\documentclass[12pt]{article}
\usepackage[utf8]{inputenc}
\usepackage{charter}
\usepackage{fullpage}
\usepackage{hyperref}
\usepackage{graphicx}
\usepackage{xspace}
\usepackage{amsmath}
\usepackage{array}
\usepackage{amssymb}
\usepackage{lineno}
\usepackage{lipsum}
\usepackage[sort&compress,numbers]{natbib}
\usepackage{caption}
\hypersetup{
    colorlinks=true,
    linkcolor=black,
    urlcolor=blue
    }
\usepackage[total={6.5in,9in},top=1in,headsep=0.1in,headheight=1in]{geometry}
\usepackage{enumitem}
\newlength{\mysize}
\newcommand{\mycfs}[1]{\setlength{\mysize}{#1pt}\fontsize{\mysize}{1.2\mysize}\selectfont}
  
\newcommand\snowmass{
\begin{center}
  \rule[-0.2in]{\hsize}{0.01in}\\
  \rule{\hsize}{0.01in}\\
  \vskip 0.1in
  Submitted to the Proceedings of the US Community Study\\ 
  on the Future of Particle Physics (Snowmass 2021)\\
  \rule{\hsize}{0.01in}\\
  \rule[+0.2in]{\hsize}{0.01in}\\[-2em]
\end{center}
}

\usepackage[firstpage=true]{background}
\backgroundsetup{contents={\parbox{6.5in}{\snowmass}}, scale=1,placement=top,opacity=1,color=black,position={3.25in,1.2in}}

\usepackage{fancyhdr}
\fancypagestyle{plain}{%
  \fancyhf{}%
  \fancyhead[C]{}
  \fancyfoot[C]{\thepage}
}

\fancypagestyle{empty}{%
  \fancyhf{}%
  \fancyfoot[C]{\thepage}
}
\newcommand{\bind}[1]{$E_{\rm bind}^{\rm #1}$\xspace}

\title{Snowmass2021 Cosmic Frontier: \\
The landscape of low-threshold dark matter direct detection in the next decade}
\date{}

\usepackage{authblk}

\author[1]{\textbf{Coordinators:} \\ Rouven Essig}
\author[2]{Graham K. Giovanetti}
\author[3]{Noah Kurinsky}
\author[4,5]{Dan McKinsey}

\author[6]{\vskip 3mm\textbf{Contributors:} \\ Karthik Ramanathan}
\author[7]{Kelly Stifter}
\author[8]{Tien-Tien Yu}
 
 \author[9]{\vskip 3mm\textbf{Endorsers:} \\ A.~Aboubrahim}
\author[10]{D.~Adams}
\author[11]{D.~S.~M.~Alves}
\author[12]{T.~Aralis}
\author[13]{H.~M.~Ara\'{u}jo}
\author[14]{D.~Baxter}
\author[15]{K.~V.~Berghaus}
\author[16]{A.~Berlin}
\author[17]{C.~Blanco}
\author[18]{I.~M.~Bloch}
\author[19]{W.~M.~Bonivento}
\author[20]{R.~Bunker}
\author[21]{S.~Burdin}
\author[22]{A.~Caminata}
\author[23]{M.C.~Carmona-Benitez}
\author[24]{L.~Chaplinsky}
\author[25]{T.~Y.~Chen}
\author[26]{S.~E.~Derenzo}
\author[27]{L.~de~Viveiros}
\author[28]{R.~Dick}
\author[29]{N.~Di~Marco}
\author[30]{P.~Du}
\author[31]{B.~Dutta}
\author[32]{R.~Ebadi}
\author[33]{T.~Emken}
\author[34]{A.~Esposito}
\author[35]{E.~Etzion}
\author[36]{J.~L.~Feng}
\author[37]{N.~Fernandez}
\author[38]{S.-F.~Ge}
\author[39]{S.~Ghosh}
\author[40]{G.~Giroux}
\author[41]{L.~Hamaide}
\author[42]{S.~A.~Hertel}
\author[43]{G.~Herrera}
\author[44]{Y.~Hochberg}
\author[45]{Y. Kahn}
\author[46]{B.~J.~Kavanagh}
\author[47]{A.~N.~Khan}
\author[48]{H.~Kluck}
\author[49]{S.~Kravitz}
\author[50]{B.~von~Krosigk}
\author[51]{J.~Kumar}
\author[52]{I.~T.~Lawson}
\author[53]{B.~V.~Lehmann}
\author[54]{T.~Lin}
\author[95]{J. Liao}
\author[55]{S.~A.~Lyon}
\author[56]{P. ~M.~Majewski}
\author[57]{C.~A.~Manzari}
\author[58]{J.~Monroe}
\author[59, 60, 61]{M.~E.~Monzani}
\author[62]{D.~E.~Morrissey}
\author[63]{D.~Norcini}
\author[64]{A.~Orly}
\author[65]{A.~Parikh}
\author[66]{J.-C.~Park}
\author[67]{P.~K.~Patel}
\author[95]{S.~Paul}
\author[68]{J. P\`erez-R\'ios}
\author[69]{A.~Phipps}
\author[70]{A.~Pocar}
\author[71]{A.~Ritz}
\author[72]{Y. Sarkis}
\author[73]{P.~Schuster}
\author[74]{T.~Schwetz}
\author[75]{S.~Shaw}
\author[76]{S.~Shin}
\author[77]{A.~Singal}
\author[78]{R.~Singh}
\author[79]{O.~Slone}
\author[80]{P. Sorensen}
\author[81]{C.~Sun}
\author[82]{M.~Szydagis}
\author[83]{D.~J.~Temples}
\author[84]{G.~Testera}
\author[85]{K.~Thieme}
\author[86]{N.~Toro}
\author[87]{T.~Trickle}
\author[88]{S.~Uemura}
\author[89]{V.~Velan}
\author[90]{E.~Vitagliano}
\author[91]{F.~Wagner}
\author[92]{G.~Wang}
\author[93]{S. Westerdale}
\author[94]{K.~M.~Zurek}

\affil[1]{C.N.~Yang Institute for Theoretical Physics, Stony Brook University, Stony Brook, NY 11794}
\affil[2]{Physics Department, Williams College, Williamstown, MA 01267, USA}
\affil[3]{SLAC National Accelerator Laboratory, 2575 Sand Hill Road, Menlo Park, CA 94025, USA}
\affil[4]{University of California Berkeley, Department of Physics, Berkeley, California 94720, USA}
\affil[5]{Lawrence Berkeley National Laboratory, 1 Cyclotron Rd., Berkeley, California 94720, USA}
\affil[6]{Division of Physics, Mathematics, \& Astronomy, California Institute of Technology, Pasadena, CA 91125, USA}
\affil[7]{Fermi National Accelerator Laboratory, Batavia, Illinois 60510, USA}
\affil[8]{Department of Physics and Institute for Fundamental Science, University of Oregon, Eugene, Oregon 97403, USA}

\affil[9]{Institut f\"ur Theoretische Physik, Westf\"alische Wilhelms-Universit\"at M\"unster, Wilhelm-Klemm-Stra{\ss}e 9, 48149 M\"unster, Germany}
\affil[10]{C.N. Yang Institute for Theoretical Physics, Stony Brook University, Stony Brook, NY 11794}
\affil[11]{Los Alamos National Laboratory, Los Alamos, NM 87544, U.S.A.}
\affil[12]{Division of Physics, Mathematics, \& Astronomy, California Institute of Technology, Pasadena, CA 91125, USA}
\affil[13]{High Energy Physics Group, Department of Physics, Imperial College London, U.K.}
\affil[14]{Fermi National Accelerator Laboratory (FNAL), Batavia, IL 60510-5011, USA}
\affil[15]{C.N. Yang Institute for Theoretical Physics, Stony Brook University, Stony Brook, NY 11794}
\affil[16]{Theoretical Physics Division, Fermi National Accelerator Laboratory, Batavia, IL 60510, USA}
\affil[17]{Department of Physics, Princeton University, Princeton, NJ USA}
\affil[18]{Department of Physics, Tel Aviv University, Tel Aviv, Israel}
\affil[19]{Istituto Nazionale di Fisica Nucleare, Cagliari Division, Italy}
\affil[20]{Pacific Northwest National Laboratory, Richland, WA 99352, USA}
\affil[21]{Department of Physics, Oliver Lodge Laboratory, University of Liverpool, Liverpool, L69 7ZE, UK}
\affil[22]{Istituto Nazionale di Fisica Nucleare, Sezione di Genova, 16146 Genoa, Italy}
\affil[23]{Pennsylvania State University, Department of Physics, University Park, PA 16802-6300, USA}
\affil[24]{Amherst Center for Fundamental Interactions and Physics Department, UMass, Amherst, MA 01003, USA}
\affil[25]{Columbia University, New York, NY 10027, USA}
\affil[26]{Lawrence Berkeley National Laboratory, 1 Cyclotron Rd., Berkeley, CA 94720, USA}
\affil[27]{Pennsylvania State University, Department of Physics, University Park, PA 16802-6300, USA}
\affil[28]{Department of Physics and Engineering Physics, University of Saskatchewan}
\affil[29]{Gran Sasso Science Institute , 67100 L'Aquila - Italy \& INFN - Laboratori Nazionali del Gran Sasso, 67010 Assergi - Italy }
\affil[30]{C.N. Yang Institute for Theoretical Physics, Stony Brook University, Stony Brook, NY 11794, USA}
\affil[31]{Mitchell Institute for Fundamental Physics and Astronomy, Department of Physics  and Astronomy, Texas A$\&$M University, College Station, Texas 77843,  USA}
\affil[32]{Department of Physics, University of Maryland, College Park, Maryland 20742, USA}
\affil[33]{Stockholm University}
\affil[34]{School of Natural Sciences, Institute for Advanced Study, Princeton, NJ 08540, US}
\affil[35]{School of Physics \& Astronomy, Tel Aviv University, Israel}
\affil[36]{Department of Physics and Astronomy, University of California, Irvine, CA 92697-4575, USA}
\affil[37]{Department of Physics, University of Illinois at Urbana-Champaign, Urbana, IL 61801, USA}
\affil[38]{Tsung-Dao Lee Institute (TDLI) \& School of Physics and Astronomy (SPA), Shanghai Jiao Tong University (SJTU), Shanghai 200240, China}
\affil[39]{Mitchell Institute for Fundamental Physics and Astronomy, Department of Physics  and Astronomy, Texas A$\&$M University, College Station, Texas 77843,  USA; School of Physics, Korea Institute for Advanced Study, Seoul 02455, Korea}
\affil[40]{Department of Physics, Queen’s University, Kingston, K7L 3N6, Canada}
\affil[41]{Theoretical Particle Physics and Cosmology Group, Department of Physics, King's College London, Strand, London, WC2R 2LS, UK}
\affil[42]{Department of Physics, University of Massachusetts at Amherst, Amherst, MA, USA 02139}
\affil[43]{Physik-Department, Technische Universität München, James-Franck-Strasse, 85748 Garching, Germany}
\affil[44]{Racah Institute of Physics, Hebrew University of Jerusalem, Jerusalem 91904, Israel}
\affil[45]{Department of Physics, University of Illinois at Urbana-Champaign, Urbana, IL 61801, USA}
\affil[46]{Instituto de F\'isica de Cantabria (IFCA, UC-CSIC), Avenida de Los Castros s/n, 39005 Santander, Spain}
\affil[47]{Max-Planck-Institut f\"ur Kernphysik, Postfach 103980, D-69029 Heidelberg, Germany}
\affil[48]{Institut f\"ur Hochenergiephysik der \"Osterreichischen Akademie der Wissenschaften, A-1050 Wien, Austria}
\affil[49]{Lawrence Berkeley National Laboratory, 1 Cyclotron Rd., Berkeley, CA 94720, USA}
\affil[50]{Institute for Astroparticle Physics (IAP), Karlsruhe Institute of Technology (KIT), 76344 Eggenstein-Leopoldshafen, Germany}
\affil[51]{Department of Physics and Astronomy, University of Hawaii,  Honolulu, Hawaii 96822, USA}
\affil[52]{SNOLAB, Lively, Ontario, Canada}
\affil[53]{Department of Physics, University of California, Santa Cruz, Santa Cruz, CA 95064, USA}
\affil[54]{Department of Physics, University of California, San Diego, CA 92093, USA}
\affil[55]{Department of Electrical and Computer Engineering, Princeton University, Princeton, NJ 08544}
\affil[56]{Particle Physics Department, STFC Rutherford Appleton Laboratory, UK}
\affil[57]{Physik-Institut, Universit\"at Z\"urich, Winterthurerstrasse 190, CH-8057 Z\"urich, Switzerland}
\affil[58]{Royal Holloway, University of London}
\affil[59]{SLAC National Accelerator Laboratory, Menlo Park, CA 94025, USA}
\affil[60]{Kavli Institute for Particle Astrophysics and Cosmology, Stanford University, Stanford, CA 94305 USA}
\affil[61]{Vatican Observatory, Castel Gandolfo, V-00120, Vatican City State}
\affil[62]{TRIUMF}
\affil[63]{Kavli Institute for Cosmological Physics, Department of Physics, Enrico Fermi Institute, University of Chicago ,  Chicago, IL 60637 USA}
\affil[64]{School of Physics and Astronomy, Tel-Aviv University, Tel-Aviv 69978, Israel}
\affil[65]{Department of Physics, Harvard University, Cambridge, MA 02138, USA}
\affil[66]{Department of Physics, Chungnam National University, Daejeon 34134, Republic of Korea}
\affil[67]{Amherst Center for Fundamental Interactions and Physics Department, UMass, Amherst, MA 01003, USA}
\affil[68]{Department of Physics and Astronomy, Stony Brook University, 11794 Stony Brook, NY (USA)}
\affil[69]{Department of Physics, California State University -- East Bay, Hayward, California 94542-3084, USA}
\affil[70]{Amherst Center for Fundamental Interactions and Physics Department, University of Massachusetts, Amherst, MA 01003}
\affil[71]{Department of Physics and Astronomy, University of Victoria, Victoria, BC, V8P 5C2, Canada}
\affil[72]{Instituto de Ciencias Nucleares, Universidad Nacional Autónoma de M\'exico, 04510 CDMX, Mexico}
\affil[73]{SLAC National Accelerator Laboratory}
\affil[74]{Karlsruhe Institute of Technology, Germany}
\affil[75]{University of California, Santa Barbara, Department of Physics, Santa Barbara, CA 93106-9530, USA}
\affil[76]{Department of Physics, Jeonbuk National University, Jeonju, Jeonbuk 54896, Korea}
\affil[77]{Department of Physics \& Astronomy, Stony Brook University, NY 11790}
\affil[78]{Institute  of  Nuclear  Physics  Polish  Academy  of  Sciences,  PL-31-342  Krak\'ow,  Poland}
\affil[79]{Department of Physics, Princeton University, Princeton, NJ 08544, USA; Center for Cosmology and Particle Physics, Department of Physics, New York University, NY, 10012, USA}
\affil[80]{Lawrence Berkeley National Laboratory, 1 Cyclotron Rd., Berkeley, CA 94720, USA}
\affil[81]{School of Physics and Astronomy, Tel-Aviv University, Tel-Aviv 69978, Israel}
\affil[82]{University at Albany SUNY, Department of Physics, Albany, NY 12222-0100 USA}
\affil[83]{Fermi National Accelerator Lab (FNAL), Astrophysics Department, Batavia, IL 60510-5011, USA}
\affil[84]{Istituto Nazionale Fisica Nucleare (INFN) Genova- Italy}
\affil[85]{Department of Physics and Astronomy, University of Hawai\textquoteleft{i}, Honolulu, HI 96822, USA}
\affil[86]{SLAC National Accelerator Laboratory and Stanford University}
\affil[87]{Walter Burke Institute for Theoretical Physics, California Institute of Technology, Pasadena, CA 91125, USA}
\affil[88]{Fermi National Accelerator Laboratory, PO Box 500, Batavia IL, 60510, USA}
\affil[89]{University of California Berkeley, Department of Physics, Berkeley, CA 94720, USA}
\affil[90]{Department of Physics and Astronomy, University of California, Los Angeles, California 90095-1547, USA}
\affil[91]{Institut f{\"u}r Hochenergiephysik der {\"O}sterreichischen Akademie der Wissenschaften, 1050 Wien, Austria}
\affil[92]{Argonne National Laboratory}
\affil[93]{Department of Physics, Princeton University, Princeton, NJ 08544}
\affil[94]{California Institute of Technology}

\begin{document}

\pagenumbering{gobble}

\maketitle

\begin{abstract}
The search for particle-like dark matter with meV-to-GeV masses has developed rapidly in the past few years.  We summarize the science case for these searches, the recent progress, and the exciting upcoming opportunities.  Funding for Research and Development and a portfolio of small dark matter projects will allow the community to capitalize on the substantial recent advances in theory and experiment and probe vast regions of unexplored dark-matter parameter space in the coming decade. 

\end{abstract}

\pagebreak 
\pagenumbering{roman}
\section*{Executive Summary}

We live in a galaxy filled with dark matter particles left over from the Big Bang. These particles surround us and continuously stream through us, our laboratories, and the Earth. Yet dark matter particles have never been detected with terrestrial experiments. 

Advanced technologies employed in detectors with extremely low energy thresholds may be used to search for rare interactions of the dark matter with ordinary matter. These detectors are surrounded by low-background shielding and placed in underground laboratories to avoid backgrounds from cosmic rays. Historically, such ``direct detection'' experiments have primarily focused on detecting particles with mass of order $1-1000$~GeV$/c^2$, but a deeper understanding of dark matter theory suggests that dark matter may be part of a dark sector and have a mass below the proton. Direct detection experiments play an essential role in determining the particle nature of such dark matter.  Moreover, for many of these theoretical models, such as when dark matter couples to ordinary matter through a light or massless mediator, direct detection provides the \textit{only} opportunity for its detection. 

This white paper focuses on new experimental approaches to probe lower mass dark matter ranging from the proton mass to twelve orders of magnitude lighter (i.e. meV$/c^2$ to GeV$/c^2$ dark matter particle masses). 

The total energy density of dark matter is known. Therefore, as the hypothetical mass of the dark matter particle decreases, the flux of dark matter particles increases in proportion. In addition, the constraints on light dark matter cross-sections with ordinary matter are currently very weak. As a result, current constraints imply that light dark matter candidates could be discovered with very small detector target exposures (as small as 1~gram-day), given sufficiently low energy thresholds and background rates. 

Because dark-sector dark matter theories exist in which the dark matter interacts 1) only with nuclei, 2) only with electrons, or 3) only with coherent modes in a target material, experiments are needed to probe each of these interactions. In each case, new parameter space may be probed with cutting-edge experiments that provide lower energy thresholds and/or lower backgrounds than currently are available.   At the same time, longer-term research and development (R\&D) efforts to lower energy thresholds and lower background rates would allow an even wider range of masses and interaction strengths to be probed. 

The search for low-mass dark matter has progressed enormously in the past few years, with significant theoretical and experimental advances.  By funding R\&D towards improved detector technologies, improved signal and background calibrations and characterizations, and improved theoretical understanding of dark matter interactions in various materials, we can build on these exciting recent developments.  These improvements can then be folded into the design of several experiments that are small, relatively inexpensive, and naturally suited to a suite of small projects allowing particle physicists to probe vast regions of well-motivated but unexplored dark matter parameter space in the next decade. 
\clearpage

\newpage
\tableofcontents

\newpage
\pagenumbering{arabic}
\section{Introduction}
\label{theory}

\begin{figure}[tph]
    \centering
    \includegraphics[width=0.45\textwidth]{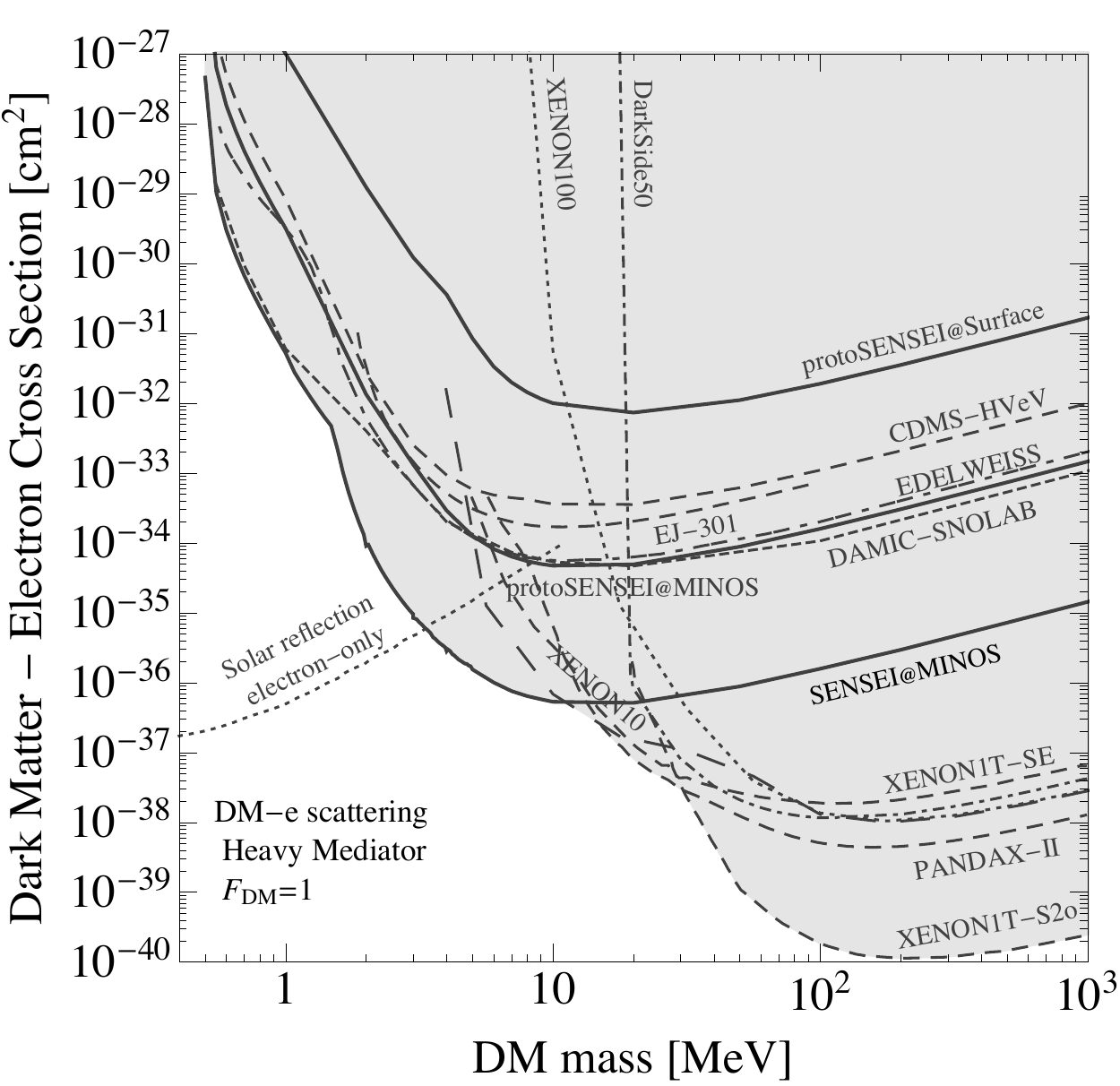}
    \includegraphics[width=0.45\textwidth]{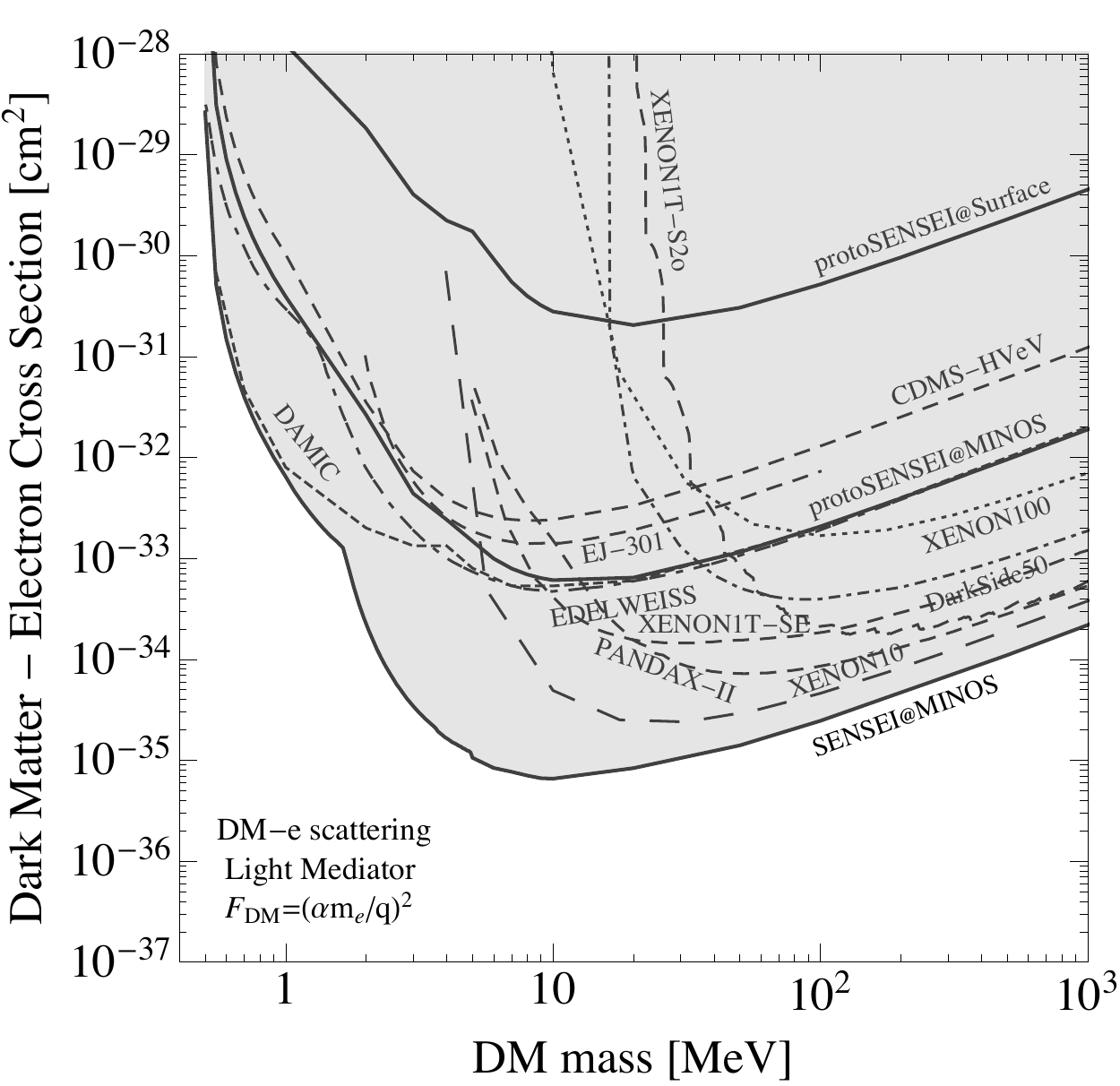}
    \includegraphics[width=0.45\textwidth]{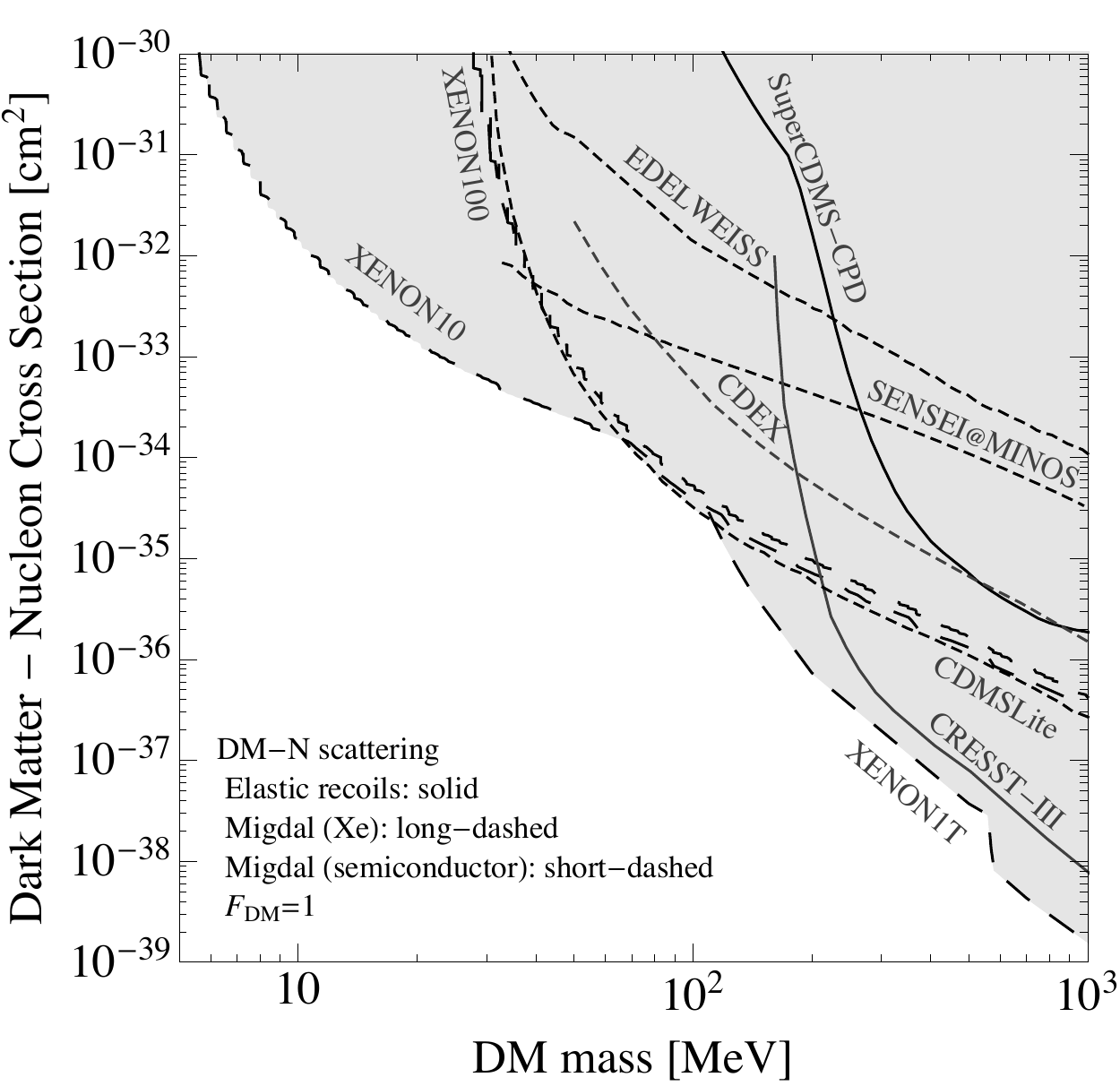}
    \includegraphics[width=0.45\textwidth]{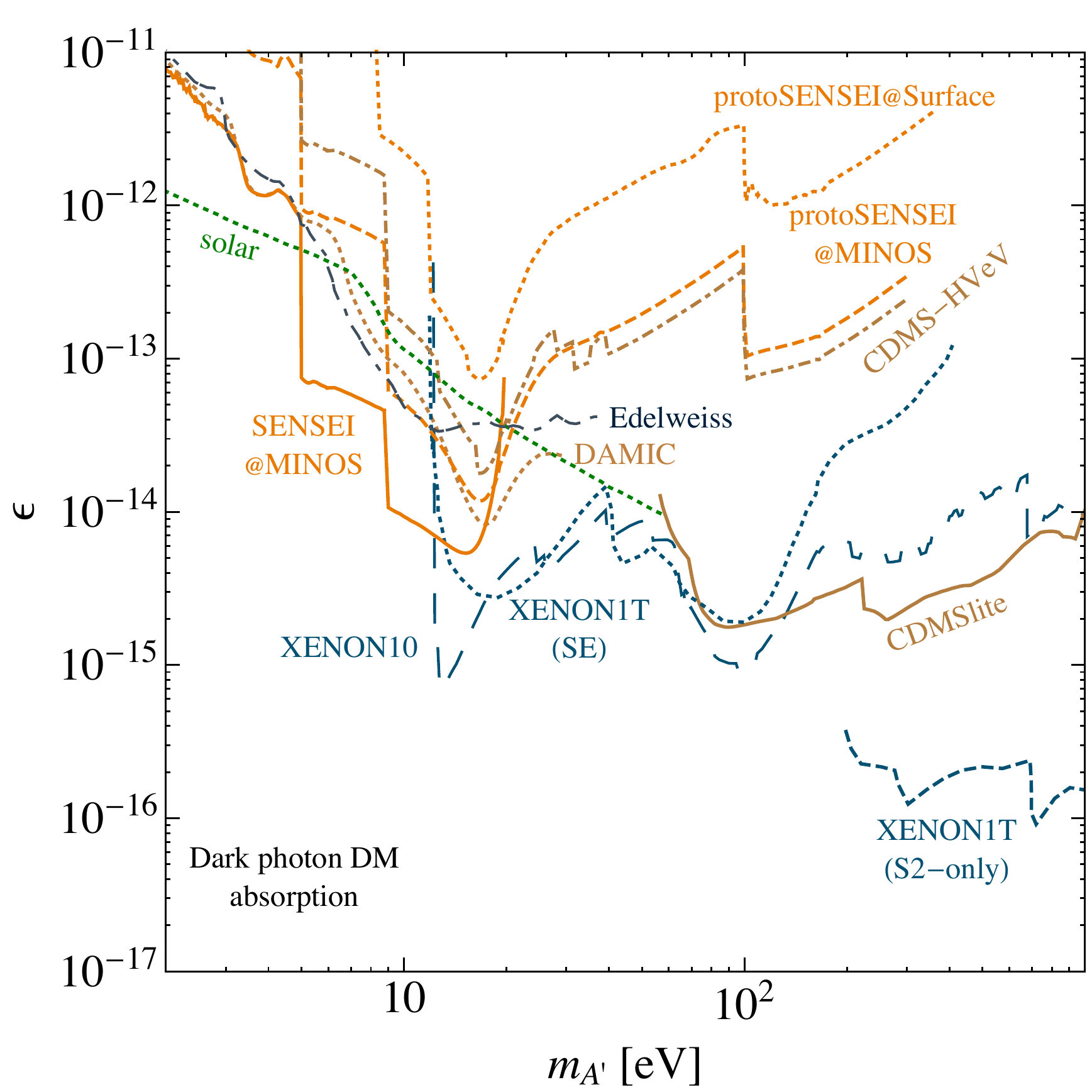}
    \caption{\textbf{Top left:} Current 90\% c.l.~limits on DM-electron scattering through a heavy mediator from SENSEI~\cite{Crisler:2018gci,Abramoff:2019dfb,SENSEI:2020dpa}, CDMS-HVeV~\cite{Agnese:2018col,Amaral:2020ryn}, DAMIC at SNOLAB~\cite{Aguilar-Arevalo:2019wdi}, EDELWEISS~\cite{Arnaud:2020svb}, DarkSide-50~\cite{Agnes:2018oej}, XENON10~\cite{Essig:2012yx,Essig:2017kqs,Angle:2011th}, XENON100~\cite{Aprile:2016wwo}, XENON1T (S2-only (``S2o'') and single-electron (``SE'') analyses)~\cite{Aprile:2019xxb,XENON:2021qze}, PandaX-II~\cite{PandaX-II:2021nsg}, and EJ-301~\cite{Blanco:2019lrf}.
 \textbf{Top right:} Same as for top left plot but assuming scattering through an ultralight mediator.  
 \textbf{Bottom left:} Current 90\% c.l.~limits on DM-nucleon scattering through a heavy mediator from searches for elastic DM-nucleus scattering from SuperCDMS-CPD~\cite{SuperCDMS:2020aus} and CRESST-III~\cite{Abdelhameed:2019hmk} (solid lines), 
 and from searches for the Migdal effect from DM-nucleus scattering from the semiconductor experiments SENSEI~\cite{SENSEI:2020dpa,Knapen:2021bwg}, CDMSlite~\cite{SuperCDMS:2022kgp}, CDEX~\cite{Liu:2019kzq}, and EDELWEISS~\cite{EDELWEISS:2022ktt} (short-dashed lines) and from the noble-liquid experiments XENON10~\cite{Essig:2019xkx}, XENON100~\cite{Essig:2019xkx}, and XENON1T~\cite{Aprile:2019jmx} (long-dashed lines).  
 \textbf{Bottom right:} Constraints on dark photon dark matter absorption are from SENSEI (orange)~\cite{Crisler:2018gci,Abramoff:2019dfb, SENSEI:2020dpa}, DAMIC (dark orange)~\cite{DAMIC:2016qck, DAMIC:2019dcn}, EDELWEISS (dark blue)~\cite{Arnaud:2020svb}, XENON10 (blue), CDMSlite (dark orange)~\cite{Bloch:2016sjj}, XENON1T (blue)~\cite{Aprile:2019xxb,XENON:2021qze}, and the Sun (green)~\cite{An:2013yfc,Redondo:2013lna,Bloch:2016sjj}. Not shown are constraints from XENON100, which are comparable to those of XENON10, as well as constraints from the ZEPLIN~\cite{Alner:2007ja,Lebedenko:2008gb}. 
}
    \label{fig:constraints}
\end{figure}

The search for particle-like dark matter (DM) interactions for DM masses below the proton (``sub-GeV DM'') is an exciting and relatively new research direction that has made enormous progress in the past few years.  In the 2013 Snowmass Community Summer Study, direct-detection searches for sub-GeV DM received no mention in the Cosmic Frontier summary report~\cite{Feng:2014uja} and only a paragraph in the dedicated white paper on direct detection~\cite{Cushman:2013zza}. Furthermore, the sole MeV-to-GeV mass DM limit at the time~\cite{Essig:2012yx} was based on a reinterpretation of data taken by XENON10 in the previous decade~\cite{Angle:2011th}. Since then, however, the number of experimental searches (see Fig.~\ref{fig:constraints}) and theoretical studies has sky-rocketed (see e.g.~\cite{Essig:2011nj,Essig:2012yx,Angle:2011th,Graham:2012su,An:2014twa,Aprile:2014eoa,Lee:2015qva,Essig:2015cda,Hochberg:2015pha,Hochberg:2015fth,Derenzo:2016fse,Aguilar-Arevalo:2016zop,Bloch:2016sjj,Cavoto:2016lqo,Hochberg:2016ntt,Hochberg:2016ajh,Hochberg:2016sqx,Kouvaris:2016afs,Robinson:2016imi,Ibe:2017yqa,Dolan:2017xbu,Tiffenberg:2017aac,Romani:2017iwi,Budnik:2017sbu,Bunting:2017net,Cavoto:2017otc,Fichet:2017bng,Knapen:2017ekk,Hochberg:2017wce,Knapen:2017xzo,Emken:2017erx,Emken:2017hnp,Emken:2017qmp,Emken:2018run,LUX:2018akb,Crisler:2018gci, Agnese:2018col, Agnes:2018oej, Settimo:2018qcm,Bringmann:2018cvk,Ema:2018bih,Akerib:2018hck,Griffin:2018bjn,CDEX:2019hzn,EDELWEISS:2019vjv,Bell:2019egg,Liu:2019kzq,XENON:2019zpr,Essig:2019xkx,Baxter:2019pnz,Abramoff:2019dfb,Aguilar-Arevalo:2019wdi,Armengaud:2019kfj,Aprile:2019jmx,Emken:2019tni,Kurinsky:2019pgb,Cappiello:2019qsw,Trickle:2019ovy,Griffin:2019mvc,Trickle:2019nya,Catena:2019gfa,Berlin:2019ahk,Berlin:2019uco,Coskuner:2019odd,Lin:2019uvt,Blanco:2019lrf,Geilhufe:2019ndy,Griffin:2020lgd,Trickle:2020oki,Hochberg:2021pkt,Akerib:2021pfd,LUX:2020yym,SENSEI:2020dpa,Liang:2020ryg,GrillidiCortona:2020owp,Bernstein:2020cpc,Ma:2019lik,Nakamura:2020kex,Du:2020ldo,Collar:2021fcl,Hochberg:2021ymx,Mitridate:2021ctr,Griffin:2021znd,Coskuner:2021qxo,Kahn:2021ttr,Berghaus:2021wrp,Blanco:2021hlm,Aguilar-Arevalo:2022kqd,SuperCDMS:2022kgp,Alexander:2016aln,Battaglieri:2017aum,BRNreport,BRNreport-detector,Essig:Physics2020,Bell:2021zkr,Bell:2021xff}), showing the growing importance of this nascent research field.  With appropriate levels of funding for experiment and theory, we have the exciting opportunity to capitalize on the recent theoretical and experimental advances and probe vast regions of unexplored DM parameter space in the coming decade.

The opportunities in low-threshold direct-detection have been explored in several recent community reports, notably the \textit{Dark Sectors workshop 2016}~\cite{Alexander:2016aln}, the \textit{US Cosmic Visions: New Ideas in Dark Matter 2017}~\cite{Battaglieri:2017aum}, the \textit{DOE Basic Research Needs for Dark Matter Small Projects New
Initiatives 2018}~\cite{BRNreport}, and the \textit{DOE Basic Research Needs Study on High Energy Physics Detector Research and Development 2019}~\cite{BRNreport-detector}. These reports paint a compelling picture of increased funding for a ``small-projects'' portfolio and for increased funding for R\&D towards higher-sensitivity (lower threshold) detectors, improved background reduction, and for improved understanding of materials. At the same time, funding for theoretical research is crucial to {\it e.g.}~increase our understanding of DM interactions in materials, calculate novel background processes, and develop additional science opportunities for low-threshold detectors. 
Given the extensive available literature that describe the science opportunities and instrumental progress in the field of low-threshold DM detection, the purpose of this whitepaper (WP) is modest, aiming to provide a high-level summary of the science opportunities and basics of low-threshold detection (Section~\ref{sec:science}), of the recent experimental advances (Section~\ref{sec:experimental}), and of the experimental challenges facing the sub-GeV DM direct-detection program in the next decade (Section~\ref{ssec:EC}). 

We conclude the introduction with a short discussion of how this WP fits in with other solicited WPs. In contrast to CF1 WP1 on ``Dark matter Direct Detection to the Neutrino Floor"~\cite{SnowmassCF1WP1}, which focuses on direct-detection opportunities for DM with masses above 1~GeV, here we focus on sub-GeV particle-like DM down to masses $\mathcal{O}({\rm meV})$. We do not consider wavelike DM, which is covered by CF2.  
There is also an overlap with CF1 WP3 on ``Calibrations and Backgrounds for Dark Matter Direct Detection"~\cite{SnowmassCF1WP3}, which focuses on backgrounds and calibrations for direct detection; here, we will focus on the efforts specifically needed for sub-GeV DM direct detection. Finally, our description of the recent incredible advances and future sensitivity goals of the ultrasensitive detectors needed for sub-GeV DM detection will be brief to avoid too much overlap with the Instrumentation Frontier's WP1 on ``Quantum Calorimeters and Single Electronic Excitation Detectors"~\cite{SnowmassIF1WP1} and WP2 on ``Photon Counting in Visible and near-IR"~\cite{SnowmassIF1WP2}. 

\section{Science Opportunities and Basics of Low-Threshold Detection}\label{sec:science}

\subsection{Sub-GeV Dark Matter and The Need for New Experiments}

The main focus of the direct-detection program in the past few decades has been the search for Weakly Interacting Massive Particles (WIMPs) (e.g.~\cite{Lee:1977,Kolb:1990vq,Jungman:1995df,Bergstrom:2000pn,Bertone:2004pz}), with masses above the proton ($\sim$1~GeV).  Several experiments have focused on WIMPs and have now reached multi-ton-scale target masses; a clear path exists for even larger detectors to reach the neutrino fog (see CF1 WP1~\cite{SnowmassCF1WP1}). 
However, a deeper understanding of DM theory and the absence of an unambiguous positive laboratory signal suggest that the DM dynamics may be more complex and that DM may be part of a  dark sector, which consists of particles that are not directly charged under the Standard Model (SM) gauge interactions.  
DM in a dark sector can span a vast mass range, including the meV-to-GeV mass range that is the focus of this whitepaper.  

\subsubsection{Sub-GeV DM Models} 

\begin{figure}[t!]
    \centering
    \includegraphics[width=0.40\textwidth]{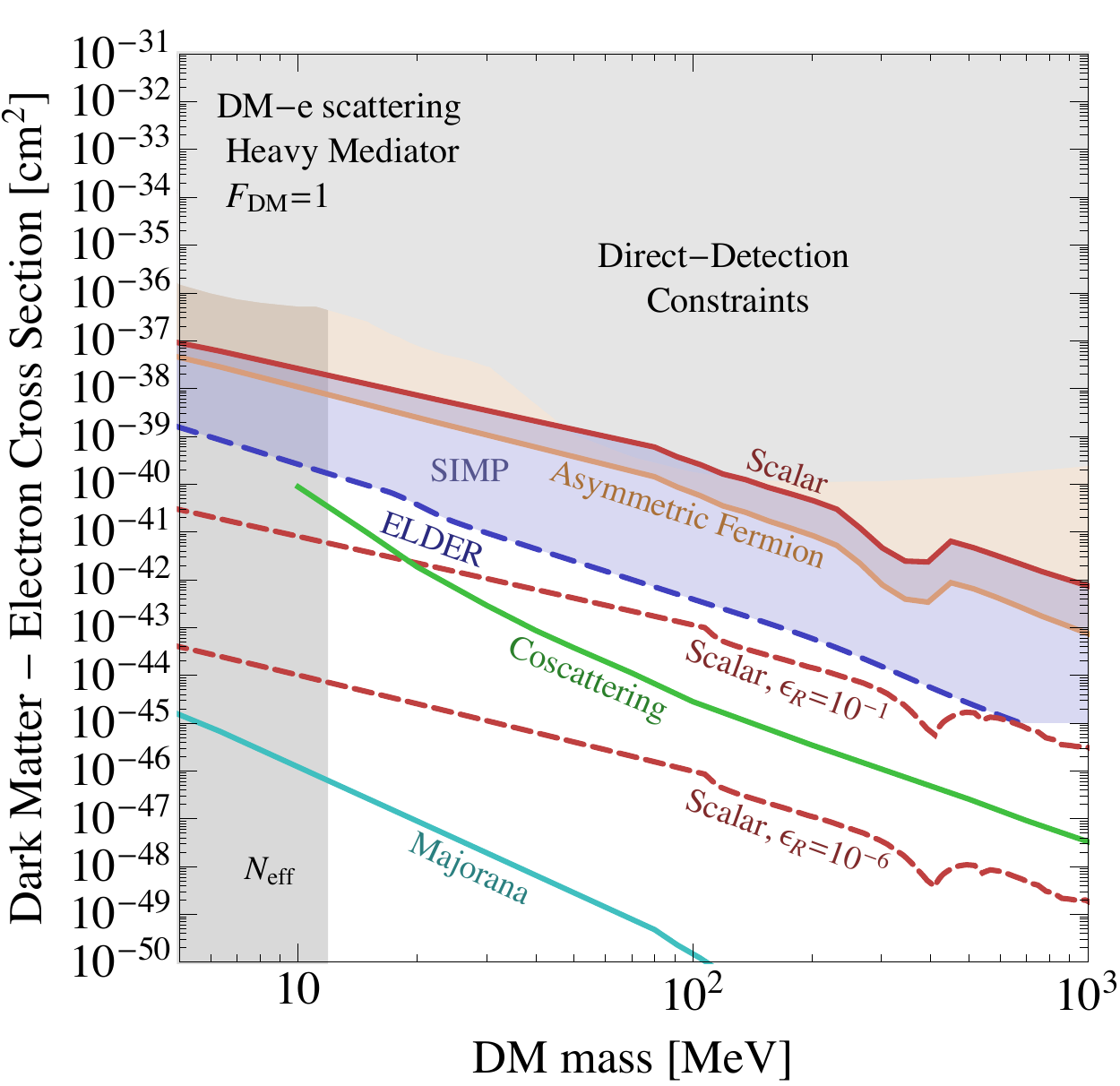}~~~
    \includegraphics[width=0.40\textwidth]{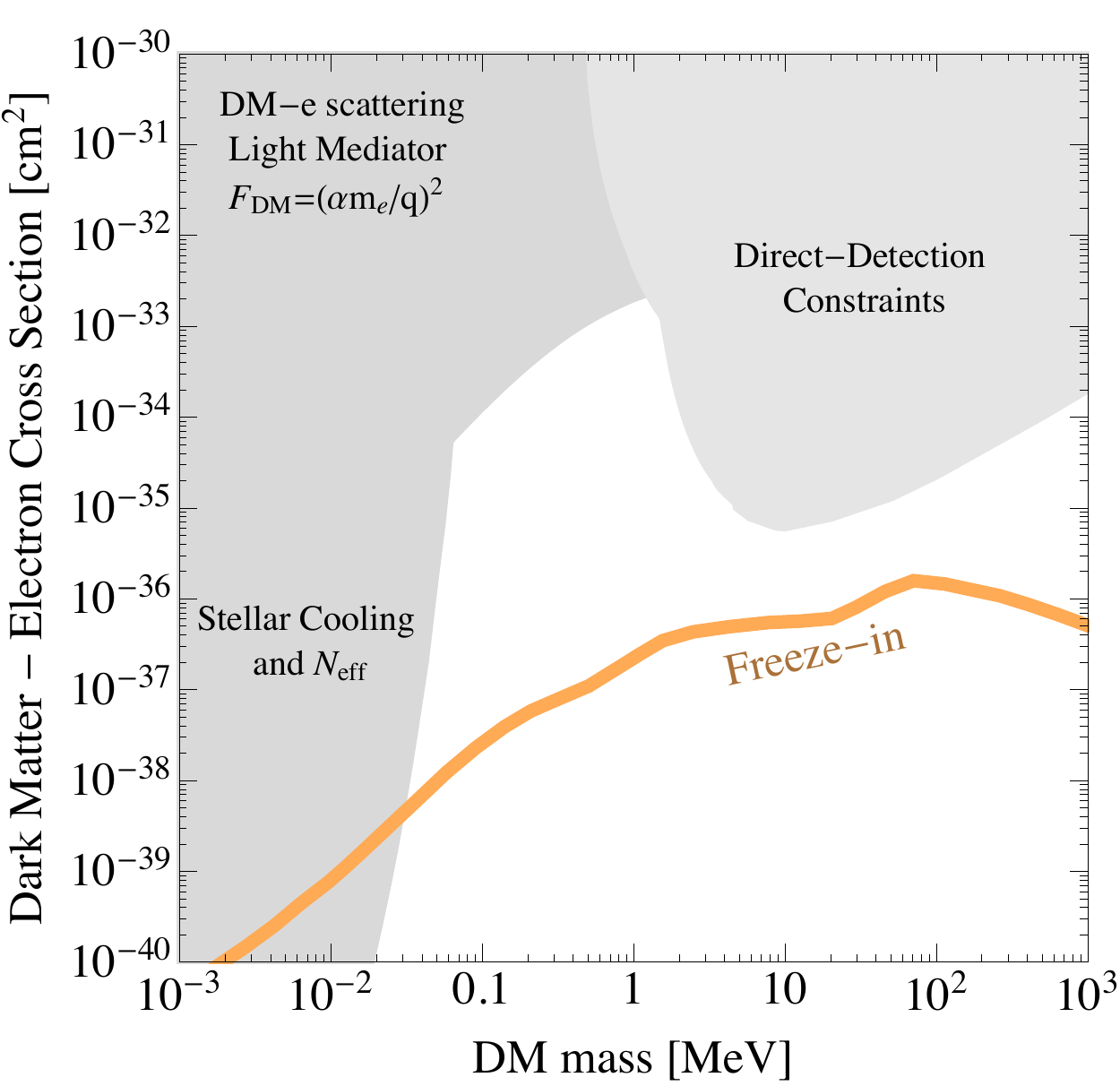}
    \includegraphics[width=0.40\textwidth]{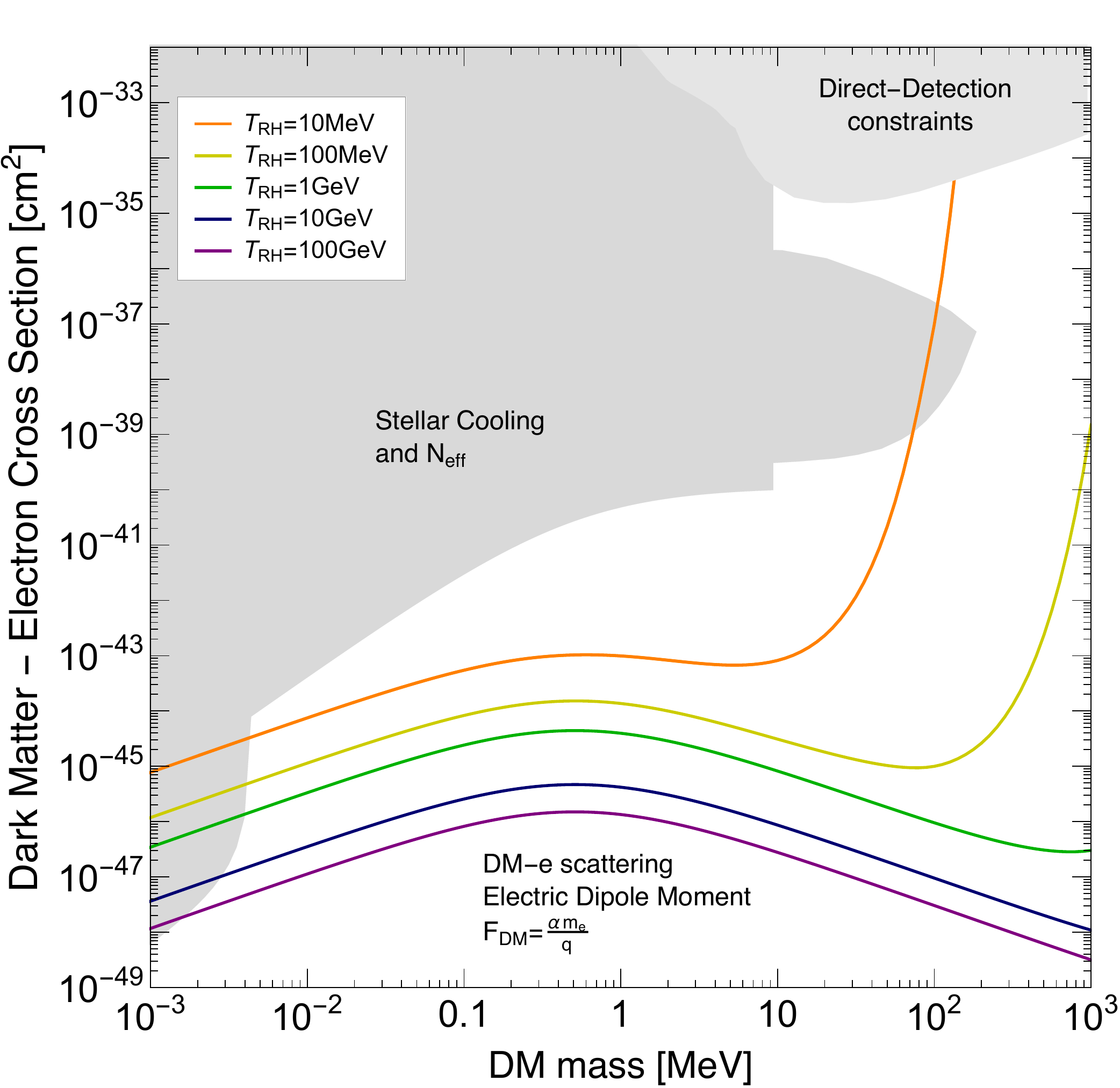}~~~
    \includegraphics[width=0.40\textwidth]{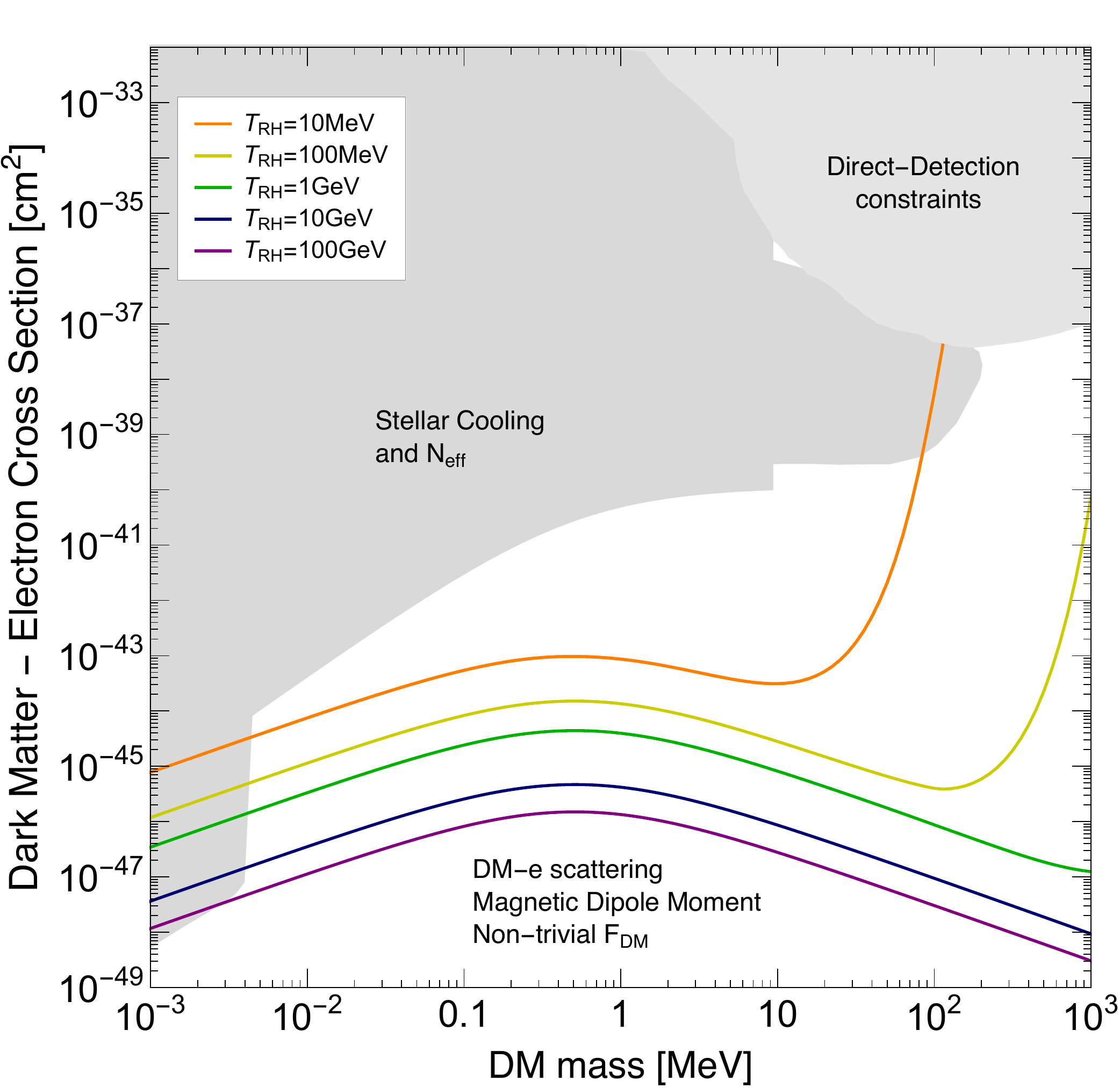}
    \caption{\mycfs{9.7}{Specific target regions in the DM-electron cross section 
    versus DM mass plane in which DM is produced with the observed relic abundance.  For these benchmark models, DM interacts with both nuclei and electrons.  
    When applicable, we show existing direct-detection bounds (light gray) (see Fig.~\ref{fig:constraints} and~\cite{Chang:2019xva}) and constraints from stellar cooling and the effective number of relativistic species ($N_{\rm eff}$) (darker gray) (see e.g.~\cite{Essig:2015cda,Knapen:2017xzo,Chang:2018rso,Chu:2018qrm,Chang:2019xva}).  There are no existing collider bounds in these parameter spaces except in the top left plot, but the constraints are model dependent and not shown (but see e.g.~\cite{Battaglieri:2017aum,Izaguirre:2015yja,Essig:2015cda,BRNreport}). 
    \textbf{Top left:} DM-electron scattering via a heavy mediator, leading to a DM form factor ($F_{\rm DM}=1$).  (i) A complex scalar produced through thermal freeze-out via $s$-channel annihilation through a massive dark photon off-resonance~\cite{Boehm:2003hm,Boehm:2003ha,Fayet:2004bw,Essig:2015cda,Izaguirre:2015yja} (\textbf{red solid line}) or near resonance~\cite{Feng:2017drg} (\textbf{red dashed lines} ($\epsilon_R$ is related to the kinetic energy needed to be on resonance); (ii) a Dirac fermion produced through an initial particle-antiparticle asymmetry and interacting with the SM through a massive dark photon~\cite{Lin:2011gj,Essig:2015cda} (\textbf{orange region/line}); (iii) a strongly interacting massive particle (SIMP) produced through $3\to2$ interactions while remaining at the SM-sector temperature through elastic scattering with SM particles~\cite{Hochberg:2014dra,Hochberg:2014kqa} (\textbf{blue region}); (iv) an elastically decoupling relic (ELDER) whose abundance is determined by its elastic scattering off SM particles~\cite{Kuflik:2015isi,Kuflik:2017iqs} (\textbf{blue dashed line}); (v) a Majorana fermion produced via thermal freeze-out through a vector mediator (\textbf{cyan line}); and (vi) a scalar DM particle produced by inelastic scattering against a lighter particle from the thermal bath (``coscattering'')~\cite{DAgnolo:2018wcn} (\textbf{green line}).  
    \textbf{Top right:} DM-electron scattering via an ultralight dark photon ($F_{\rm DM}=(\alpha m_e/q)^2$); the observed relic abundance is obtained via freeze-in along the \textbf{orange line}~\cite{Essig:2011nj,Chu:2011be,Essig:2015cda,Dvorkin:2019zdi,Chang:2019xva}. 
    \textbf{Bottom left:} DM-electron scattering through an electric dipole moment ($F_{\rm DM}=\alpha m_e/q$), the observed relic abundance is obtained via freeze-in along the colored lines for different re-heating temperatures $T_{\rm RH}$~\cite{Chang:2019xva}.  
    \textbf{Bottom right:} Same as bottom left plot but for a magnetic dipole moment interaction~\cite{Chang:2019xva}. 
    Here $F_{\rm DM}$ is non-trivial~\cite{Chang:2019xva,Catena:2019gfa}.
    }
}
    \label{fig:models}
\end{figure}

Various ``portals''---mediated by dark-sector particles---are possible and allow for several different types of DM interactions with Standard Model (SM) particles, e.g.~\cite{Hewett:2012ns,Essig:2013lka,Alexander:2016aln,Battaglieri:2017aum,Jaeckel:2010ni}. 
These interactions allow for DM to be produced in the early Universe with the observed relic abundance through new mechanisms that are distinct from the thermal freeze-out through electroweak-scale mediators typically associated with WIMPs. 
Indeed, many production mechanisms accommodate sub-GeV DM, see e.g.~\cite{Boehm:2003hm,Boehm:2003ha,McDonald:2001vt,Fayet:2004bw,Sigurdson:2004zp,Strassler:2006im,Sikivie:2006ni,Nussinov:1985xr, Kaplan:1991ah,Zurek:2008qg,Hooper:2008im,Cholis:2008vb,ArkaniHamed:2008qn,Pospelov:2008jd,Feng:2008ya,Feng:2008dz,Pospelov:2008jk,Hall:2009bx,Morrissey:2009ur,Kaplan:2009ag,Essig:2010ye,Cohen:2010kn,Essig:2011nj,Chu:2011be,Falkowski:2011xh,Lin:2011gj,Feng:2011ik,Nelson:2011sf,MarchRussell:2012hi,Arias:2012az,Graham:2012su,Kaplinghat:2013yxa,Hochberg:2014dra,Hochberg:2014kqa, Boddy:2014yra,Boddy:2014qxa,Essig:2015cda,Hochberg:2015vrg,Izaguirre:2015yja,Kuflik:2015isi,Graham:2015rva,Marsh:2015xka,D'Agnolo:2015koa,DAgnolo:2015ujb,Pappadopulo:2016pkp,Farina:2016llk,Dror:2016rxc,Feng:2017drg,Kuflik:2017iqs,Choi:2017zww,DAgnolo:2017dbv,Falkowski:2017uya,DAgnolo:2018wcn,Chu:2018qrm,Dvorkin:2019zdi,DAgnolo:2019zkf,Hambye:2019dwd,Heeba:2019jho,Evans:2019vxr,Koren:2019iuv,Chu:2019rok,Chang:2019xva,Darme:2017glc,March-Russell:2020nun,Dodelson:1993je,Shi:1998km,Abazajian:2001nj,Abazajian:2012ys,Abazajian:2017tcc,Boyarsky:2018tvu,Hochberg:2018vdo,Hochberg:2018rjs,Smirnov:2020zwf,Dutta:2019fxn,Dutta:2020enk,Dutta:2022qvn,Fernandez:2021iti}. 
Several specific predictive benchmark models that can be probed by the next-generation of sub-GeV DM direct-detection experiments discussed in this whitepaper are shown in Fig.~\ref{fig:models} in the parameter space of DM-electron scattering cross section ($\overline\sigma_e$) versus DM mass ($m_\chi$). We emphasize that these predictive target regions are also relevant for DM scattering with nuclei or DM scattering that produces collective excitations.  However, it is also possible for the DM to interact with the SM through mediators that couple, for example, only to baryons or only to leptons, necessitating experiments that probe  a range of possible interactions.  Moreover, there are many other models that are not as predictive as those shown in Fig.~\ref{fig:models}, but could nevertheless have direct-detection signals that are observable with low-threshold experiments; examples include bosonic DM (dark photons, pseudoscalars, or scalars) (e.g.,~\cite{Graham:2015rva}). 

\subsubsection{The need for a broad low-threshold direct-detection program} 

Direct-detection experiments provide a crucial avenue to understand the particle nature of DM.  
\begin{itemize}[leftmargin=0.4cm,itemindent=0.5cm,labelwidth=\itemindent,labelsep=0cm,align=left]\addtolength{\itemsep}{-0.65\baselineskip}
\item Discovering a new particle at an underground direct-detection experiment would constitute evidence that such a particle constitutes all or at least part of the DM. 
\item It is important to note that the interactions between DM and the SM sector depend on the DM and dark-sector properties. It is therefore imperative to pursue a range of direct-detection concepts and technologies that allow us to probe DM interactions with both nucleons (quarks) and with electrons in order to elucidate fully the DM properties. The community has proposed several ideas that probe both types of interactions. 
\item It is important to have a strong program for DM that includes direct-detection experiments in addition to accelerator-based probes.  Indeed, while there are several dark-sector models that can be investigated with both accelerator-based probes and direct-detection experiments, which would allow for an exciting investigation of the DM properties from multiple angles, there are some models that can only be discovered with one of these probes.  Dark-sector models that contain velocity-suppressed interactions between the DM and SM or consist of multiple dark-sector states in which the DM in DM-SM scattering is required to up-scatter into a heavier state are best probed with accelerators, while dark sectors in which the DM-SM interaction is mediated by a very light or massless particle (leading to a scattering cross section that is enhanced at low momentum transfer) are best probed by direct-detection experiments.  For example, direct-detection experiments provide the {\it only} possibility to test the freeze-in benchmark model shown in Fig.~\ref{fig:models} (top right).  In particular, although the DM was never in thermal equilibrium with ordinary matter in this scenario and the interactions between DM and SM particles 
are therefore necessarily tiny, a light mediator ($\ll\ $keV) leads to a large enhancement of the direct-detection cross section at low momentum transfers.  Schematically, this cross section is given by 
\begin{equation}
\overline\sigma_e \sim 4 \pi \alpha_D \epsilon^2 \alpha \frac{\mu_{\chi,e}^2}{q^4}\,,
\end{equation}
where the momentum transfer $q$ is at most $q_{\rm max} \sim \mu_{\chi,e} v_\chi$, $\epsilon$ is the kinetic mixing parameter, $\alpha$ ($\alpha_D$) is the usual (dark) fine-structure constant, and $\mu_{\chi,e}$ is the DM-electron reduced mass.  The small momentum transfer in direct-detection experiments provide a parametric enhancement relative to higher energy experiments, providing a unique probe of this scenario. 
\item The interactions between sub-keV bosonic DM and the SM are already strongly constrained to avoid an anomalously large stellar cooling rate. The remaining viable parameter space can only be constrained by direct-detection experiments.  
\end{itemize}

\subsection{Basics of Direct Detection} 

\subsubsection{DM scattering and DM absorption} 

There are two main types of interactions between the DM and a detector target material, which depend on the DM properties, and lead to a drastically different signal shape: absorption and scattering.  If the DM is a boson and has a direct coupling to SM particles, it can be absorbed (by, {\it e.g.}, a target electron); if instead the DM interacts with the SM particles through a new mediator, the DM (either a boson or a fermion) can scatter off the target material.  For the absorption process, the entire rest-mass energy of  the (non-relativistic) DM is deposited into the target material (usually absorbed by an electron), and the electron recoil spectrum consists of a delta-function smeared by the detector resolution.  For the scattering process, only a fraction of the DM's kinetic energy is deposited, and the signal shape is often much harder to calculate correctly.  Current technologies can probe DM absorption for DM masses as low as $m_\chi\sim 1$~eV (the silicon band gap) and DM scattering as low as $m_\chi \sim 500$~keV (for which the kinetic energy is $\frac{1}{2}m_\chi v_\chi^2 \sim 1$~eV) (see Fig.~\ref{fig:constraints}).  Detection concepts exist that could allow future technologies and experiments to probe DM absorption for masses as low as $\sim$1--10~meV and DM scattering as low as $\sim$1--10~keV (see Fig.~\ref{fig:future}). The kinematics of the absorption process are straightforward, but we will discuss in more detail the scattering process. 

\subsubsection{Kinematics of sub-GeV DM Scattering} 
\begin{figure}[t!]
    \centering
    \includegraphics[width=0.65\textwidth]{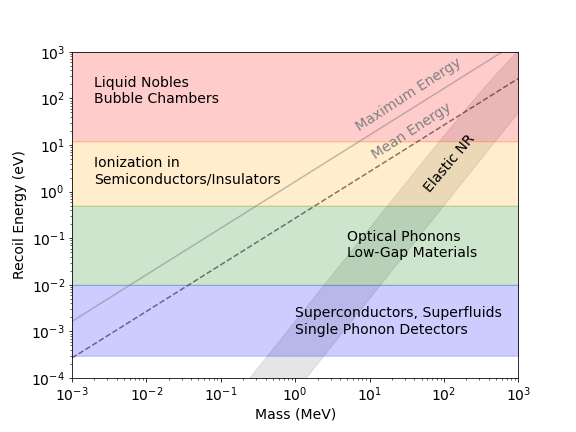}
    \caption{
    Illustration of regions of DM mass kinematically accessible with different detection concepts, detector target materials, and technologies. Elastic nuclear recoils will have an energy that lies in the ``Elastic NR'' band (the width of the band is set by scattering off hydrogen and xenon, respectively).  In contrast, inelastic scattering allow for the extraction of an $\mathcal{O}(1)$ amount of the DM kinetic energy, that lies along the lines labelled ``Maximum Energy'' and ``Mean Energy'' corresponding to DM moving at the galactic escape velocity $v_{\rm esc}\sim 544$~km/s or the mean velocity $v_{\rm mean}\sim 220$~km/s, respectively.  Recently developed technologies can probe DM masses as low as $\sim$1~MeV, while novel detectors are needed to achieve sensitivity to DM down to near the fermionic DM bound at 10~keV. 
    }
    \label{fig:materials}
\end{figure}

WIMP direct-detection experiments typically search for nuclear recoils from elastic DM-nucleus scattering.  Indeed, for many DM models, DM-nucleus scattering is the dominant interaction between the DM and SM sector.  This is because there is often a coherent enhancement across the nucleons or protons so that the DM scattering cross section scales as $Z^2$ or $A^2$. 
Although the expected DM flux increases for lower DM masses, inversely proportional to the DM mass, \begin{equation}
\textrm{DM flux} \sim 7 \times 10^9~\textrm{cm}^{-2}\textrm{s}^{-1}~\left(\frac{1~{\rm MeV}}{m_\chi}\right)\,,
\end{equation}
the DM's kinetic energy $E_{\rm kin}$, 
\begin{equation}\label{eq:Ekin}
E_{\rm kin} \lesssim 10 \mathrm{~eV} ~\left(\frac{m_\chi}{6~{\rm MeV}}\right)\,,
\end{equation}
decreases for lower DM masses (here we took the DM speed to be the galactic escape velocity, $\sim$544~km/s).  Moreover, in elastic scatters only a small fraction of the DM's kinetic energy is transferred to the recoil energy $E_{\rm NR}$ of the nucleus for DM masses $m_\chi \ll 1$~GeV,
\begin{equation}
E_{\rm NR} \lesssim 5 \mathrm{~eV} \times\left(\frac{m_\chi}{100\mathrm{~MeV}}\right)^2\left(\frac{130\mathrm{~GeV}}{m_N}\right) \,,
\label{eq:E_NR}
\end{equation}
where we took the maximum relative speed between the DM and detector to be the galactic escape velocity plus the Earth velocity ($v_{\rm esc}+v_{\rm mean}\sim (544+220)$~km/s) 
to estimate the maximum nuclear recoil energy.  
We see that the energy transfer to a nucleus, as the DM mass is lowered 
below the GeV scale, quickly falls below the threshold of the most sensitive current generation DM experiments. 
Still, several ideas exist to enhance the signal from elastic sub-GeV DM-nucleus scatters, which we will discuss in Sec.~\ref{sec:experimental}.  These are important to pursue, since in general searches for nuclear recoil signals have lower radioactive backgrounds (since the dominant radioactive background sources produce electronic recoils).
The kinetic energy and maximum nuclear recoil energies are shown in Fig.~\ref{fig:materials} for various target materials. 

In contrast to elastic DM-nucleus scatters, processes in which the DM scatters inelastically with a detector target material allow for the transfer of a large fraction of the DM's kinetic energy to the target material. 
Examples of inelastic processes are (i) DM scattering with bound electrons, (ii) DM scattering with nuclei through the Migdal effect or accompanied by a photon from bremsstrahlung, and (iii) DM-target scattering that produces a collective excitation.  
For these processes to occur, the DM's kinetic energy must be larger than the energy needed to create the desired excitation, which depends on the properties of the target material.  For example, the binding energy of the least-bound electron in atoms (e.g.~Xe) is of $\mathcal{O}$(10~eV), in insulators (e.g.~diamond, NaI) of $\mathcal{O}$(5~eV), in semiconductors (e.g., Si, Ge, GaAs) of $\mathcal{O}$(1~eV), and in low-gap materials (Dirac materials, doped semiconductors, and superconductors) of $\mathcal{O}$(few~meV).  This allows DM as light as 6~MeV ($\mathcal{O}$(keV)) to excite an electron in an atom (low-gap material) (see Eq.~(\ref{eq:Ekin})).  In addition,  solid-state targets (e.g., Si, Ge, GaAs) and superfluid He have phonon modes of $\mathcal{O}$(meV) that could be excited by the scattering of DM as light as $\mathcal{O}$(keV).  Fig.~\ref{fig:materials} shows several target materials, their excitation energy thresholds, and the DM kinetic energy. 

It is important to note that the ``kinematic matching'' for the inelastic processes is excellent in several materials, allowing DM to produce the excitation with a large rate.  As an example, we consider DM with masses in the MeV-to-GeV range that scatters off an electron in a semiconductor.  While the electron momentum is not fixed, its \textit{typical} momentum is of $\mathcal{O}(\alpha m_e) \sim \textrm{few keV}$.  Since the electron is lighter and moving much faster than the DM ($v_e\sim \mathcal{O}(\alpha)$ versus $v_\chi \sim 10^{-3}$), the momentum transfer from the DM to the electron is also $\sim\alpha m_e$, and hence the recoil energy of the electron is $\sim\alpha m_e v_\chi \sim \textrm{few eV}$. This recoil energy is above the semiconductor band gap $\mathcal{O}$(eV), and hence there is no suppression in the rate from, \textit{e.g.}, requiring the initial electron momentum to be different from its typical momentum.  The same argument is often applicable also to lower-gap materials, as well as (in various materials) to DM creating collective excitations. 

We note that the kinematics of the signal may be different than those for the typical scenarios  discussed above, e.g., in certain DM models~\cite{Smirnov:2020zwf} or if the DM (often only a small component of it) is boosted from interactions with e.g.~cosmic rays or in the Sun (see e.g.~\cite{Bringmann:2018cvk,An:2021qdl,Emken:2021lgc}) or is unbound~\cite{Herrera:2021puj}.

\subsubsection{Signals from Sub-GeV DM Scattering and Absorption}

The signal from sub-GeV DM scattering off electrons, individual nuclei (possibly with a bremsstrahlung photon or a Migdal electron), or a condensed-matter system consists of one or a few electrons, one or a few photons, or a collective excitation (such as phonons, rotons, plasmons, or magnons).  In Sec.~\ref{subsec:progress} and Sec.~\ref{sec:experimental}, we discuss recent progress in detecting these small signals. Significant theoretical advances also allow one to compute these signals rates with public codes, see e.g.~QEDark~\cite{Essig:2015cda}, EXCEED-DM~\cite{Griffin:2021znd,Trickle:2022fwt},  DarkELF~\cite{Knapen:2021bwg}, and QCDark~\cite{Dreyer:2023}, with more tools being developed. 

Given  the current constraints on sub-GeV DM, exposures of $\mathcal{O}$(gram-day) can probe new parameter space if backgrounds are under control.  In addition, near-term experiments will likely probe parameter space in which the DM interaction with ordinary matter is sufficiently large to allow for DM to scatter in the Earth before reaching the detector; this leads to a diurnal modulation in the signal rate (see, \textit{e.g.},~\cite{Collar:1992qc,Collar:1993ss,Hasenbalg:1997hs} and the DMSQUARE experiment~\cite{Avalos:2021fxm}), and could help distinguish a DM signal from backgrounds. 

It is also worth noting that it may be possible to exploit the anisotropies of condensed matter systems to allow directional detection of sub-GeV DM~\cite{Essig:2011nj,Hochberg:2016ntt,Budnik:2017sbu,Cavoto:2017otc,Griffin:2018bjn,Coskuner:2019odd,Geilhufe:2019ndy,Blanco:2019lrf,Blanco:2021hlm,Coskuner:2021qxo}, which would also lead to a daily modulation, again potentially helping to distinguish a DM signal from backgrounds. 

While the majority of efforts in sub-GeV particle DM detection are focusing on detecting electrons, photons, or collective excitations from DM, other novel possibilities exist.  Since sub-GeV DM needs to interact with ordinary matter through new light mediators, it is possible for it to have long-range interactions if the mediator is sufficiently light (this is the case for, \textit{e.g.}, the freeze-in benchmark scenario).  These interactions could lead to novel signals~\cite{Berlin:2019uco}. 

\subsection{Recent Progress and Next Goals} \label{subsec:progress}

\subsubsection{Recent Progress in Sensor Development} The past few years has seen rapid progress in detectors capable of sensing such small energy depositions. 
In particular, devices capable of detecting single electrons include two-phase xenon and argon time projection chambers (TPCs, used by XENON10/100/1T/nT, ZEPLIN-II/III/LUX/LZ, DarkSide, and PandaX), silicon Skipper-charged-coupled-devices (Skipper-CCDs, used by SENSEI, DAMIC-M, and Oscura), and high-voltage charge amplification with transition edge sensor (TESs, used by SuperCDMS HVeV) or neutron- transmutation-doped (NTD, used by EDELWEISS) readout, with TESs and NTDs also capable of sensing low-energy heat deposition from DM scattering (SuperCDMS CPD, CRESST-III, and EDELWEISS).  These technologies have led the first experimental searches for sub-GeV DM and have produced so far the leading set of constraints on sub-GeV DM interactions shown in Fig.~\ref{fig:constraints}. 
Some of these technologies are actively being scaled up to increase the target mass and exposure. 
Another technology capable of sensing single electrons are depleted $p$-channel field effect transistors (DEPFETs).  Single photons can be detected with, e.g., TESs or superconducting nanowire single photon detectors (SNSPDs).  An active R\&D program exists to push the TES technology to eventually sensing even single phonons.  We will discuss some of these technologies further in Sec.~\ref{sec:experimental}. 

Besides the phenomenal improvements in the sensors, significant progress has also been made in characterizing and understanding potential low-threshold backgrounds.  We will discuss this further in Sec.~\ref{ssec:EC}. 

\begin{figure}[t!]
    \centering
    \includegraphics[width=0.45\textwidth]{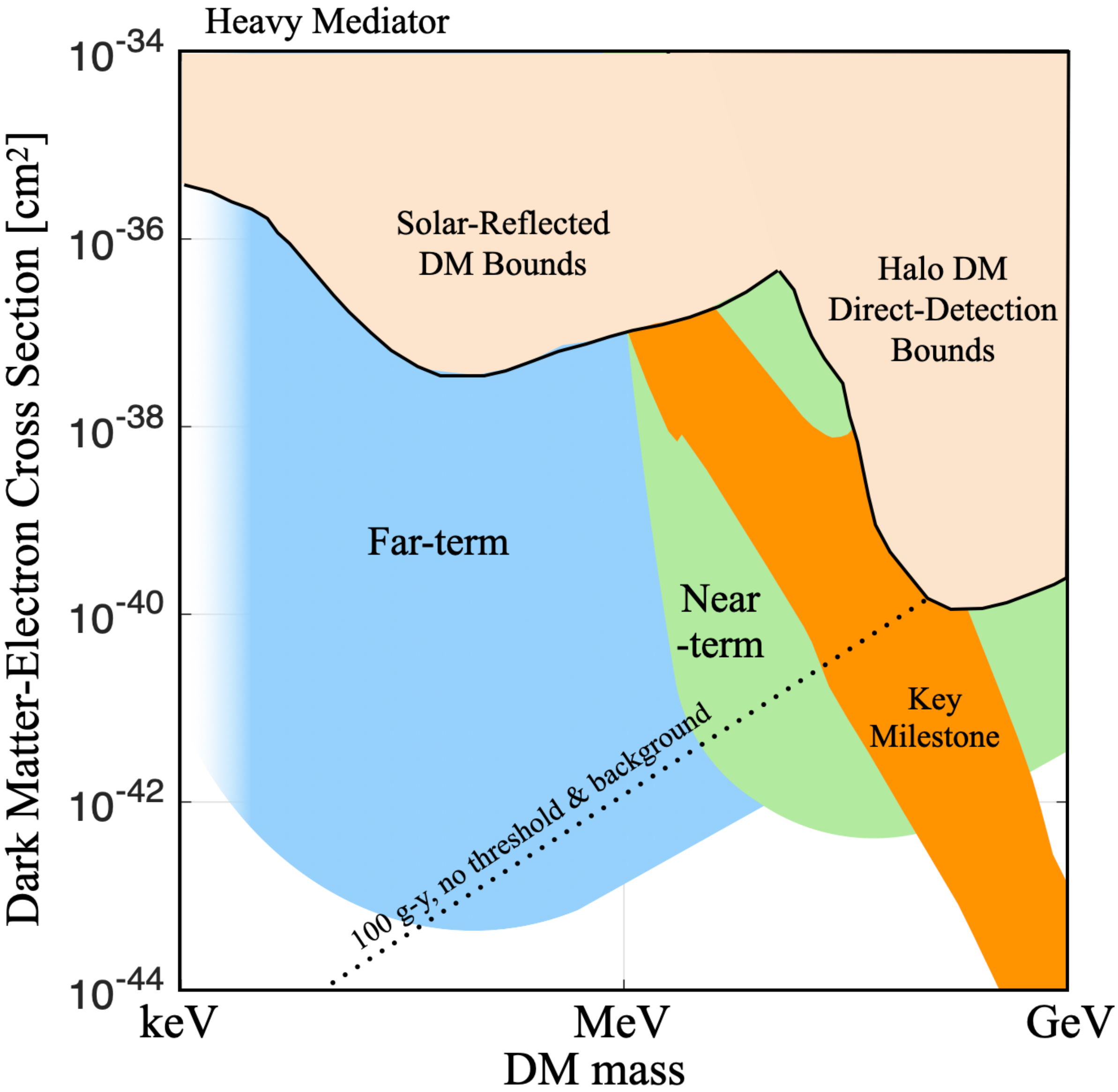}
    \includegraphics[width=0.45\textwidth]{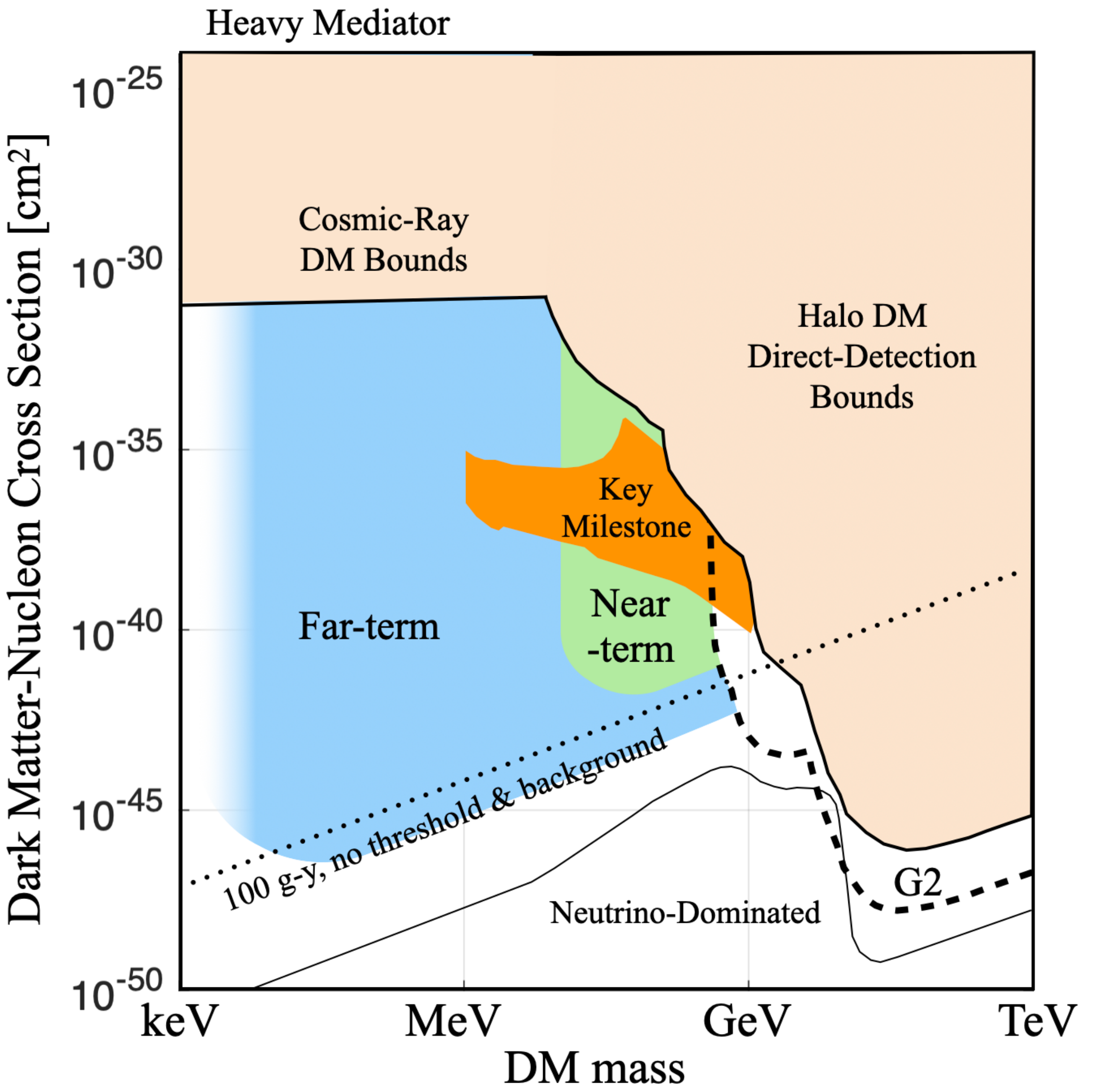}
    \includegraphics[width=0.45\textwidth]{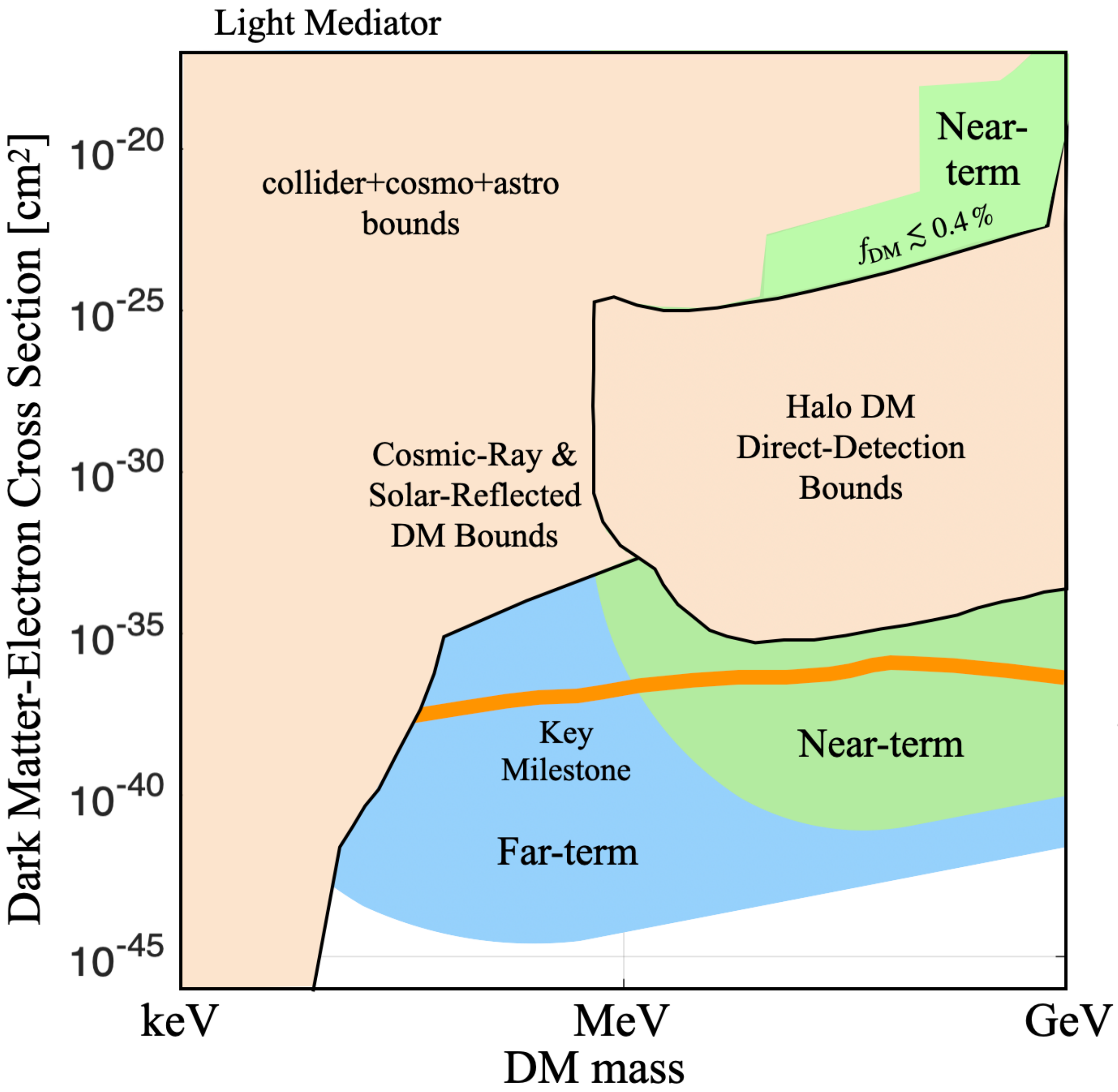}
    \includegraphics[width=0.45\textwidth]{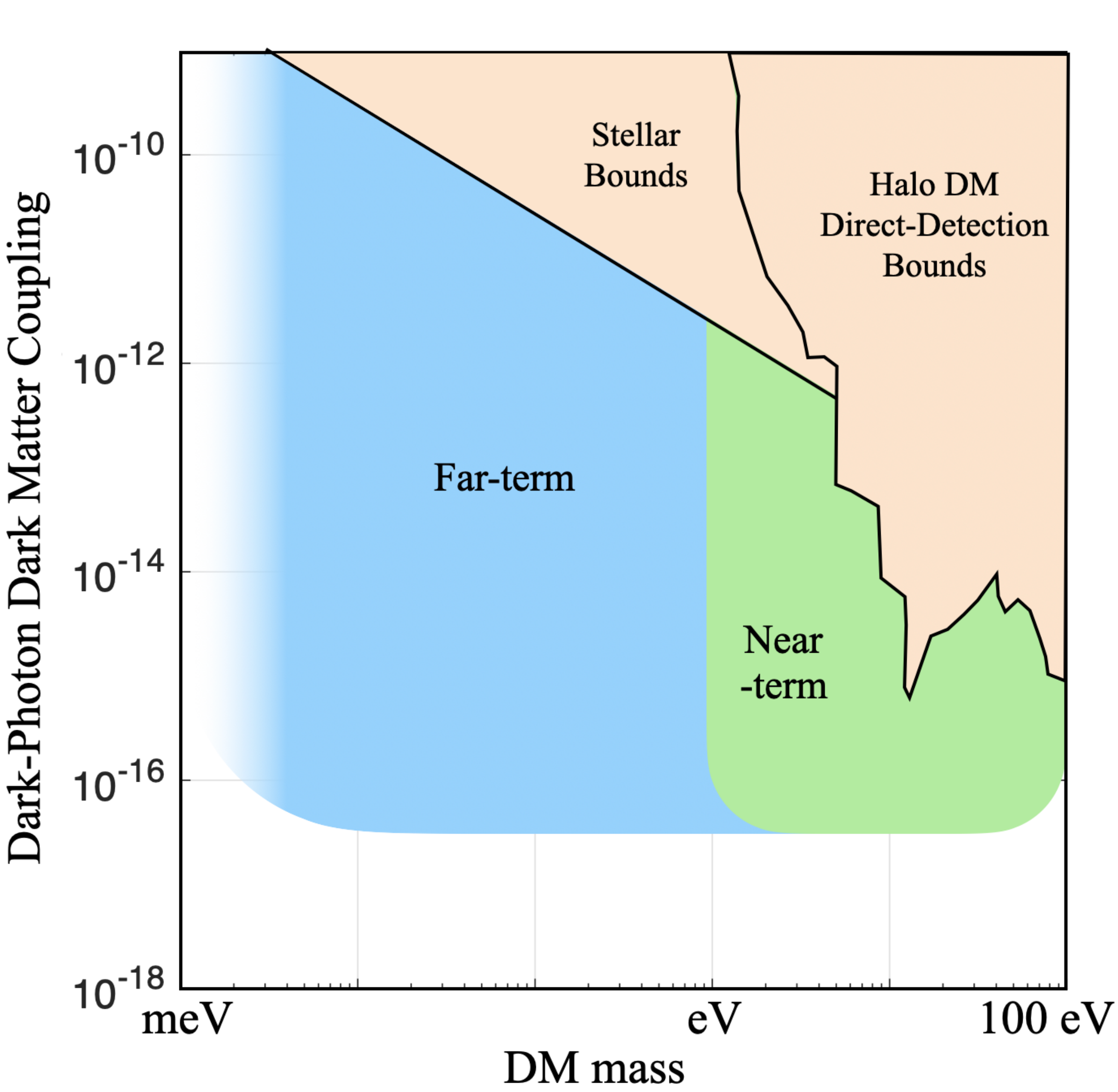}
    \caption{Figures are adapted and updated from BRN report~\cite{BRNreport}. \textbf{Top left:} Current 90\% c.l.~constraints on DM-electron scattering through a heavy mediator from direct-detection experiments (including bounds on the solar-reflected DM component) (beige, as in Fig.~\ref{fig:constraints} and from~\cite{Emken:2021lgc}, but see also~\cite{An:2021qdl,An:2017ojc}) together with approximate regions in parameter space that can be explored in the next $\sim$5 years (``near-term'', green) and on longer timescales (``far-term'', blue). Orange regions labelled “Key Milestone” represent concrete dark-matter benchmark models and are the same as in the BRN report~\cite{BRNreport}. Along the dotted line DM would produce about three events in an exposure of 100 gram-year, assuming scattering off electrons in a hypothetical target material with zero threshold.    
 \textbf{Top right:} As for left plot, but assuming DM-nuclear scattering; direct-detection bounds are from~\cite{XENON:2018voc,DarkSide:2018bpj,XENON:2019zpr,Essig:2019xkx,Aprile:2019jmx}, while the cosmic-ray accelerated DM bounds are from~\cite{Bringmann:2018cvk,Cappiello:2019qsw}. 
 \textbf{Bottom left:} As for top-left plot, but assuming scattering through an ultralight mediator.  Direct-detection bounds are as in Fig.~\ref{fig:constraints}, while other bounds are collected in~\cite{Essig:2015cda,Emken:2019tni,An:2021qdl}.  Green region at large cross section values is allowed for a subdominant DM component~\cite{Emken:2019tni}. 
 \textbf{Bottom right:} As for top-left plot, but for the case of dark-photon dark matter absorption (bounds are as in Fig.~\ref{fig:constraints}). 
}
    \label{fig:future}
\end{figure}

\subsubsection{The Next 10 Years}

The goals of the low-threshold direct-detection program over the next $\sim$10~years are multifold.  Several experiments are attempting to scale up the target mass and/or reduce  and understand low-energy backgrounds.  US-funded experiments specifically designed to search for sub-GeV DM include SENSEI, DAMIC-M, and SuperCDMS, while LZ, XENONnT, and DarkSide-20k are also expected to probe sub-GeV DM.  The recent DOE Basic Research Needs workshop on small-scale DM experiments funded the R\&D, engineering, and project development of Oscura (using Skipper-CCDs) and TESSERACT (using TES readout for various targets that include superfluid He, GaAs, sapphire, and silica).  

Several collaborations have ongoing R\&D efforts in sensor development, in reducing backgrounds, as well as in characterizing and calibrating the expected DM signal and backgrounds~\cite{BRNreport,BRNreport-detector}. Assuming funding is forthcoming, the next few years should see significant progress in probing DM scattering down to DM masses of about 1~MeV and DM absorption down to DM masses of 1~eV. At the same time, the development of improved sensors to low energy depositions should allow us to probe DM well below these masses in the latter part of this decade. Continued funding is needed in sensor development, background reduction and calibration, material characterization, and in the theoretical characterization of DM signals and backgrounds; this also includes interdisciplinary research between particle theory and experiment, condensed matter theory and experiment, AMO physics, and quantum sensor technologies. 
The outcome of this would be the exploration of many orders of magnitude of well-motivated and unexplored DM parameter space, as shown in Fig.~\ref{fig:future}.

\section{Detection Techniques}\label{sec:experimental}

In general, technology development for sub-GeV DM searches centers on detector improvement. Depending on where the improvement is made ({\it i.e.}, in threshold or discrimination power), the subsequent yield will change by different amounts for different DM search channels. Current detector R\&D for sub-GeV DM searches can generally be placed in one of 3 categories based on the energy threshold of the detector: 20~eV scale, eV-scale, and meV-scale, and we will break down this section in accordance with these thresholds. The particular mapping onto ``electron recoils'' (ER) or ``nuclear recoils'' (NR) DM sensitivity will then depend both on the nature of the interaction with the target, and on the sensitivity of the readout. The former is covered in Sec.~\ref{sec:science}, and we note that sensitivity to ER also implies sensitivity to DM-nucleus interactions via Bremsstrahlung and the Migdal effect. In the last section, we cover interactions that are collective in nature or rely on non-traditional paradigms, including deflection of dark matter and single phonon emission in collective media. In this section we focus on developments that enable lower thresholds or improved event discrimination and discuss the implications of R\&D on these developments.

To guide the discussion in this section, we refer the reader to Table~\ref{tab:expTechniques}, which lists the threshold regimes, the minimum mass that can be probed, and the relevant techniques discussed in the subsequent sections. Fig.~\ref{fig:materials} shows the breakdown of mass reach versus threshold to help map specific technologies onto science goals discussed in Sec.~\ref{sec:science}. 

\begin{table}[]
    \centering
    \begin{tabular}{|c|c|p{3cm}|p{6cm}|}
    \hline
        Section & Threshold & Lowest DM Mass Probed & Relevant Techniques, Technologies, and Materials  \\
        \hline
        3.1 & 20~eV & 20~MeV (ER)\newline 100~MeV (NR) & Noble Elements (TPCs \& SPCs) \newline Solid-State Charge Detectors \newline Phonon Detectors \newline Threshold Detectors \\
        \hline
        3.2 & 500~meV & 1~MeV (ER/NR) \newline 500~meV (Abs.) & Semiconductor Detectors \newline Athermal Phonon Detectors \newline  Scintillators \newline NIR Photon Detectors \\
        \hline
        3.3 & 5~meV & 10~keV (CE) \newline 5~meV (Abs.) & Superconductors \newline Low-Gap Materials \newline Athermal Phonon Detectors \newline Polar Materials \newline Superfluids \newline FIR Photon Detectors \newline Magnetic Bubble Chambers \newline Other new ideas\\
        \hline
    \end{tabular}
    \caption{Overview of main sub-GeV thrusts categorized by energy threshold. The three thresholds correspond very roughly to current thresholds for liquid Noble TPCs and G2 experiments (20 eV); next-generation detector thresholds, including single-charge semiconductor detector and athermal phonon detectors (0.5~eV); and the lowest threshold required to probe down to the bound on Fermionic DM (5~meV, required to probe 10~keV scattering). These are based on an experiment's potential to exclude DM candidates; in many cases, discovery potential near the lowest masses will require further improvements in energy threshold. We have also listed relevant technologies for each regime. In the column ``lowest mass probed,'' we list the lowest mass achievable assuming Electron Recoils (ER), Nuclear Recoils (NR), Absorption (Abs.), and a catch-all for other processes we refer to as collective excitations (CE). The mass bounds are based on the assumption that the given target can extract the maximum kinetic energy of the DM at galactic escape velocity; real mass bounds for specific techniques are not likely to achieve these exact targets but will in principle be well matched enough with the DM to come within an order of magnitude. The time-line for technologies that go below 5~meV in threshold are beyond a decade and we do not consider them in this whitepaper.}
    \label{tab:expTechniques}
\end{table}

\subsection{20~eV Thresholds and Higher}\label{ssec:20ev}

As described in CF1 WP1~\cite{SnowmassCF1WP1}, the current slate of DOE-funded G2 experiments and their counterparts plan to explore the parameter space for spin-independent galactic DM-nucleon couplings to within an order of magnitude of the neutrino fog for DM masses above the proton. These mature detector technologies have the potential to reach into the sub-GeV DM mass range given additional dedicated R\&D at the relevant energy scales. A natural benchmark for these detectors is energy thresholds near the $\sim20$~eV-scale, which corresponds to the energy required to produce single quanta in many of these target materials, and therefore the fundamental limit that they can probe. These detectors have the potential to explore many of the key milestone models highlighted in Fig.~\ref{fig:future}.

Significant strides have been made in recent years to enhance sensitivity in this mass range, but as energy thresholds decrease, many detectors lose their ability to discriminate against backgrounds, necessitating novel detector ideas, more stringent radiopurity requirements, dedicated calibration experiments, and an improved understanding of intrinsic detector backgrounds. Additionally, these improvements must be maintained as experiments scale to larger target masses. This section will highlight the current status of detector technologies capable of improving sensitivity to DM masses above 20~MeV for ERs and 100~MeV for NRs.
 
\subsubsection{Noble Elements}

Noble liquid detectors, particularly liquid argon (LAr) and liquid xenon (LXe) time projection chambers (TPCs), have historically been powerful probes for DM candidates with masses in the range of a few~GeV/c$^2$ to 10~GeV/c$^2$, with efforts concentrated in several science collaborations. This includes ArDM~\cite{Marchionni:2011kg}, DarkSide-50~\cite{Agnes:2015ftt}, DEAP-3600~\cite{Ajaj:2019jk}, and MiniCLEAN~\cite{Hime:2011tt} in the LAr community, which have recently come together to form the Global Argon Dark Matter Collaboration (GADMC), and PandaX~\cite{PandaX2017_DM}, LZ~\cite{Akerib:2020it}, and XENON/DARWIN~\cite{Aalbers:2016ex,Aalbers:2022dzr} in the LXe community. The members of the LZ and XENON/ DARWIN collaborations have signaled their intent to work together on a next-generation LXe experiment~\cite{LXeMOU}. Recently, a liquid-helium TPC project (ALETHEIA) has also been proposed, taking advantage of the low mass of the helium nucleus to probe lower DM masses~\cite{https://doi.org/10.48550/arxiv.2203.07901}.

To date, the lowest energy thresholds achieved by dual-phase noble liquid TPCs were obtained in studies focusing solely on the ionization signal (S2), rather than the scintillation signal (S1), as shown in Fig.~\ref{fig:constraints}. Due to the large amplification of the S2 signal in such detectors, individual electrons can be detected with high efficiency, resulting in energy thresholds below 1~keV for nuclear recoils, approaching the single quanta limit of this technology. Achieving such thresholds by analyzing only S2 signals comes with drawbacks, including the loss of discrimination between ERs and NRs, loss of time resolution, and a class of pathological single- and few-electron backgrounds that dominate the lowest energy bins~\cite{XENON:2021myl,DS50_2018_S2Only,LUX:2020vbj}. Still, this class of detector has the lowest demonstrated background rate per unit target mass, even in S2-only mode, allowing the most stringent DM cross-section limits to date over a wide range of DM masses and interaction models (see Fig.~\ref{fig:constraints}).

By scaling up noble liquid detector mass or reducing single electron backgrounds, there is the opportunity to probe lower DM cross-sections through the S2-only channel. Challenges associated with scaling a noble liquid experiment are fairly well-understood, due in large part to the current generation of tonne-scale TPCs, but methods for reducing the background rates and the detector response at the relevant energies have not been comprehensively studied. The loss of discrimination between ERs and NRs in S2-only analyses elevates the importance of mitigating electromagnetic backgrounds that arise from beta-emitters in the bulk liquid and gamma-emitting impurities within detector materials, which leads to more stringent requirements on the radiopurity of detector components and the liquid targets. Another major contribution is spurious electron backgrounds, which are likely due to some combination of delayed re-emission of electrons captured on trace electronegative impurities, photoionization of detector components, and charge buildup at the liquid-gas interface~\cite{LUX:2020vbj,Kopec:2021ccm,EDWARDS200854,Santos:2011ju,Aprile_2014,Sorensen:2017ymt,Sorensen_2018,Tomas:2018pny,Akimov:2019ogx,XENON:2021myl}. These processes are sensitive to trace levels of chemical impurities in the bulk liquid, outgassing, and the electric field configuration. Dedicated studies of spurious electron backgrounds could lead to mitigation techniques, including optimizations of drift and extraction fields and improved methods for removing electronegative impurities. Even with background improvements, significant uncertainties in the ionization yields of noble liquids to low-energy ERs and NRs hinder the analysis of data at very low energies. Significant R\&D to tackle many of the issues outlined here has been proposed and is being pursued, as detailed in~\cite{LowMassTPC_LOI},~\cite{LargeScaleLXe_LOI}, and~\cite{Bernstein:2020cpc}. 

If current and future generations of TPCs can combine the benefits of a mature, scalable technology with the low-energy sensitivity gained through these improvements, they will be capable of competitive sensitivity to a wide variety of existing low-mass models, including the ER and relatively less-studied absorption channels. Suggestions for these studies can be found in~\cite{LZG3_LOI},~\cite{DARWIN_LOI}, and~\cite{PhysRevD.104.092009}. 

Finally, the sensitivity of noble liquid detectors can be extended by doping the target medium. At low concentrations, additives with low ionization energies can increase the scintillation and ionization yield by allowing energy that would otherwise be lost as heat to translate into additional ionization and excitation. At higher concentrations, additives with low-A nuclei can serve as targets with stronger kinematic couplings to light DM particles, and additives with odd-A nuclei offer sensitivity to spin-dependent interactions. To pursue this route, dedicated calibration experiments are needed to measure yields and intrinsic distributions with the inclusion of these dopants. These ideas, as well as the requisite R\&D to bring them to fruition, are described further in~\cite{LowMassTPC_LOI} and~\cite{HydroX_LOI}. 

The NEWS-G collaboration searches for sub-GeV mass particles using spherical proportional counters (SPCs) filled with light gases, including neon, methane, and helium. Similar to their liquid counterparts, SPCs must continue to lower background rates as they scale to larger mass. Highlighted in~\cite{NEWSG_LOI}, the goal of the NEWS-G collaboration is the construction of DarkSPHERE, a 300\,cm diameter SPC made from electroformed copper that would operate at the Boulby underground laboratory. DarkSPHERE would hold up to 5-bar of helium/iso-butane (10\%) mixture, increasing exposure, lowering backgrounds, and using a lighter target nucleus than the current SNOLAB NEWS-G experiment. Additionally, the use of a novel multi-anode sensor within a low-pressure gas target could allow for directional measurement of nuclear recoils, enabling powerful event discrimination and a path into the neutrino fog.

\subsubsection{Solid-State Charge Detectors}
Echoing the continuing evolution of noble liquid TPCs described above, charge detectors have seen substantial improvements over the last several years, resulting in significantly reduced energy thresholds. Solid-state charge detectors, including charge-coupled devices (CCDs) and semiconductor crystals, can now efficiently detect single ionized electrons with the potential to push nuclear recoil sensitivity down to DM masses near 1~MeV, approximately an order of magnitude lower in mass than noble liquid detectors.  These detectors must contend with a similar set of background issues, most notably internal radiological backgrounds, external gamma-rays, and dark currents, and methods for minimizing these will become increasingly critical as these detectors scale in target mass.

CCDs have low noise, a small bandgap, and can be assembled within a low-background package, making them excellent targets in which to detect single charges produced by DM interactions. The DAMIC experiment has deployed an array of Si CCDs, achieving competitive sensitivity for GeV-scale WIMPs~\cite{DAMIC:2016qck}. CCD detector performance can be further improved by performing repeated, non-destructive readout of charges from a single pixel, a technology known as skipper-CCDs~\cite{Tiffenberg:2017aac}. These skipper-CCDs have significantly reduced readout noise, allowing for precise charge counting of single electrons. Planned and future experiments employing skipper-CCDs, including SENSEI, DAMIC-M, and Oscura, are described in Section~\ref{ssec:MidEnergy}. A significant uncertainty in applying these techniques to NRs is the calibration of ionization yield from NRs below keV energies, which is an active area of research~\cite{SnowmassCF1WP3}.

The previous generation of DM experiments at the GeV-scale also employed the Neganov-Trofimov-Luke (NTL) effect~\cite{Neganov:1985,Luke:1990ir,Agnese:2013jaa,Romani:2017iwi} to convert charge to phonon energy, which can then be read out with a conventional phonon detector (see \S~\ref{subsec3.1.3}). Upcoming experiments SuperCDMS SNOLAB~\cite{supercdms2017} and EDELWEISS~\cite{EdelweissSensitivity} will use 20~eV (Si/Ge) and 100~eV (Ge) resolution phonon detectors, equivalent to 0.1 -- 1 electron-hole pair at 100~V bias. The fiducial mass of these detectors will be $\mathcal{O}$(kg), making them competitive with DAMIC-M and SENSEI in charge-only channels at higher thresholds. Both experiments will also operate a comparable mass of detectors in low-voltage mode with slightly worse phonon resolutions in conjunction with discrete charge electronics, which will allow for ER/NR discrimination, as well as in-situ measurement of the charge yield of backgrounds relevant to sub-GeV DM searches. More detail on phonon-based charge-readout R\&D is described in Section~\ref{ssec:MidEnergy}.

Large-area silicon photomultipliers (SiPM) arrays with low dark count and cross-talk rates, such as those developed by the GADMC~\cite{DIncecco:2018fx}, could be re-purposed as stand-alone DM detectors as described in~\cite{SIPM_LOI}. In standard SiPMs, the sensitive volume is a shallow depletion region that extends 5-10~$\mu$m from the sensor surface. In back-side illuminated SiPMs (B-SiPMs), the depletion zone extends across the full thickness of the silicon substrate, with typical thicknesses on the order of 10-50~$\mu$m  and the possibility of developing B-SiPMs with thicknesses up to 500~$\mu$m. A B-SiPM-based detector using a 1~kg radio-pure silicon target operating immersed within a liquid neon veto bath, which dramatically reduces the dark count rate in the detector, and read-out with integrated CMOS electronics could achieve a threshold of $\approx$150~eV and exposure comparable to CCDs. While the ultimate energy threshold of this techniques is limited, thanks to the considerable pace of SiPM development, the low number of readout channels per wafer, and the relatively low cost of wafer production, this technology is potentially rapidly scalable. 

\subsubsection{Phonon Detectors}\label{subsec3.1.3}

Large-mass, cryogenic crystalline phonon detectors are a proven technology currently deployed by several experimental collaborations, including CRESST~\cite{Abdelhameed:2019hmk},
SuperCDMS~\cite{supercdmsWP22}, and EDELWEISS~\cite{Arnaud:2020svb}. These detectors generally consist of a crystal target outfitted with a temperature sensor capable of detecting the temperature rise that follows a particle interaction. These detectors must operate at cryogenic temperatures, generally below 50~mK, and are typically read out via one of two techniques:
\begin{itemize}
    \item Thermal readout - the temperature of the crystal is precisely monitored, and deviations from the base temperature with timescales consistent with particle interactions are interpreted as discrete events. Neutron Transmutation Doped (NTD) Ge thermistors are one technology commonly used in this approach. Transition edge sensors (TES) can also be used to monitor crystal temperature. Thermal readout has a resolution scaling with crystal volume, and thus achieving thresholds much below the 20~eV scale has not been proposed.
    \item Athermal readout - phonons are collected in superconducting sensors before they can thermalize. For example, TES-based QETs employ superconducting fins to collect phonons, effectively increasing the sensitive area of the sensor~\cite{Fink:2020noh,Ren:2020gaq}, but other technologies exist to read athermal phonons. The last several years have seen significant improvements in the detector resolution and energy thresholds of this technology, as described in detail in Section~\ref{ssec:MidEnergy}. 
\end{itemize}
The status of current-generation experiments is described below.

The SuperCDMS experiment will operate 24 detectors in a 15\,mK cryostat at SNOLAB. The detectors are fabricated from either Si or Ge in one of two configurations. iZIP detectors are sensitive to both ionization and phonons, allowing for ER/NR discrimination, which is advantageous for NR searches. HV detectors have no ionization sensors, and therefore no ER discrimination, but are operated at $\sim$100~V bias to take advantage of the NTL effect to amplify the phonon signal as described in the previous section. SuperCDMS SNOLAB expects to have sensitivity to NR masses between 0.5-5~GeV to within a decade of the neutrino fog, and to electron scattering of MeV-scale DM and/or absorption of eV-scale dark photons and/or axion-like particles (ALPs). The current experiment expects to be background limited by cosmogenically activated isotopes ($^3$H) and $^{32}$Si, $^{210}$Pb, and dark counts. SuperCDMS has established paths to upgrade the experiment, with provisions for several scenarios, including a positive detection by another G2 experiment. The SuperCDMS collaboration is undertaking a multi-stage program to improve detector performance, as described in~\cite{supercdmsWP22}, and discussed in more detail in Section~\ref{ssec:MidEnergy}. The current generation of detectors is expected to achieve resolutions on the order of 10--20~eV based on initial demonstrations~\cite{supercdmsWP22}.

CRESST-III operated an array of CaWO$_4$ cryogenic scintillating calorimeters, capable of simultaneously measuring scintillation photons and phonons via TES, at the LNGS through early 2018.  CRESST achieved 30~eV$_\text{nr}$ thresholds in a 23.6~g crystal, resulting in a sensitivity down to NR masses of 160~MeV~\cite{Abdelhameed:2019hmk} (see Fig.~\ref{fig:constraints}). The background rate in this crystal increases significantly near the detector threshold, and the origin of these events is under active study\cite{EXCESSworkshop,Proceedings:2022hmu}. Similarly, the EDELWEISS collaboration operated a 33.4~g high purity Ge crystal as an ionization and NTL enhanced phonon detector with sensitivity to eV-scale electron signals~\cite{Arnaud:2020svb} and a phonon resolution of 60~eV. This detector demonstrated the potential of cryogenic Ge detectors for sub-GeV ER searches (see Fig.~\ref{fig:constraints}). 

\subsubsection{Threshold Detectors}
Bubble chambers operate by super-heating a target fluid to a point where the localized energy deposition from a nuclear recoil creates a single bubble in the chamber, while the more diffuse energy depositions from electron recoils do not, making these detectors effectively blind to the electron recoil backgrounds that dominate other DM detection techniques~\cite{PICO_LOI}. A noble liquid bubble chamber extends the sensitivity of bubble chambers to sub-keV thresholds, maintaining the electron recoil insensitivity and 3D-event reconstruction capabilities of operating detectors while adding a scintillation channel that can be used to discriminate higher energy nuclear recoils from fast neutrons. A 10~kg scintillating bubble chamber that will be used for calibration measurements is currently under construction at Fermilab. This could be followed by increasingly large chambers on the path towards a multi-tonne detector. Noble liquid bubble chambers present a scalable, affordable, and background-free method to search for GeV-mass DM to the solar neutrino fog~\cite{noblebubble_LOI}.

A pure liquid can also be placed into a meta-stable state below the freezing point. A particle interaction can then create a nucleation site, and the resulting freezing creates a signature increase in temperature and dielectric constant. This technique is similarly insensitive to electron recoil backgrounds, providing a path into the coherent neutrino scattering fog. A super-cooled liquid detector should intrinsically possess lower-energy thresholds than needed for ionization~\cite{snowball_LOI} and super-cooled water could achieve keV-scale energy thresholds. Dedicated measurements are needed to demonstrate these energy thresholds and fully characterize electron recoil discrimination capabilities of a super-cooled detector at these energies.

\subsection{500~meV-Scale Thresholds} \label{ssec:MidEnergy}

A significant technological transition occurs below the 20~eV scale, first when DM scattering energies become comparable to the ionization energy of liquid nobles, and then when the available DM scattering energy is not sufficient to produce ionization events in a liquid noble detector. The maturity of technologies in this regime also varies significantly across different DM searches; charge and photon sensing detectors are mature at eV-scale thresholds, with multiple experiments achieving sensitivity down to the bandgap in Si/Ge, but athermal phonon detectors, and detectors focused on detection of nuclear recoils, are still an order of magnitude in sensitivity from maturity at these energy scales. In this section, we will review technologies of various maturity that can probe all DM candidates down to eV-scale energies. The energy threshold for this section has been set at 0.5~eV, roughly the bandgap of Ge, as a reflection of the maturity of Ge as a DM detector.

The mature detectors in this energy regime are R\&D improvements to technologies that have been used in DM for over a decade. 

\subsubsection{Low-Threshold Athermal Phonon Detectors}

All of the experiments with a heritage of thermal and athermal phonon detection over the last few decades are engaged in active R\&D to lower phonon detection thresholds. These are based on thermal phonon readout, in which crystal temperature is monitored and abrupt temperature changes due to radiation events are identified, as well as athermal phonon readout, in which phonons are collected by superconducting sensors before they can fully thermalize. Most experiments are on the cusp of achieving sub-eV resolution, with detectors optimized to trigger below 20~eV. Ref.~\cite{Proceedings:2022hmu} gives a recent overview of DM searches with eV-scale energy resolutions based on R\&D detectors from CRESST, EDELWEISS, SuperCDMS, and MINER.

SuperCDMS has developed 10\,g~\cite{CPD:2020xvi} and 1\,g~\cite{Ren:2020gaq} Si detectors with 3\,eV resolution, and demonstrated 10\,eV thresholds with offline triggering algorithms~\cite{Ren:2020gaq}, based on W-TES athermal phonon sensors. Both have placed limits on DM models that extend into the eV-scale energy range~\cite{SuperCDMS:2020aus,Amaral:2020ryn} in NR and ER channels, though resolutions are still more than an order of magnitude from the 0.5~eV target for probing NR DM down to the ionization limit. Additional R\&D has demonstrated noise performance equivalent to 200~meV resolution~\cite{Fink:2020noh}, close to 1~eV thresholds. Si is well-matched to DM-electron scattering down to 2~MeV, and is competitive for elastic NR searches with other light elements such as sapphire ($\mathrm{Al_2O_3}$) and CaWO$_4$. SuperCDMS is planning near-term tests with 1~g detectors similar to the Si HVeV with comparable baseline resolution at the eV scale~\cite{supercdmsWP22}.

CRESST, in collaboration with the $\nu$CLEUS coherent neutrino program, has achieved 19~eV thresholds in $\mathrm{Al_2O_3}$ (sapphire)~\cite{Angloher:2017sxg} and 30~eV in CaWO$_4$ (calcium tungstate)~\cite{Abdelhameed:2019hmk} crystals, respectively, and is engaged in R\&D to pursue lower thresholds with smaller detectors for next-generation particle detectors. As discussed in the previous section, the CRESST-III result~\cite{Abdelhameed:2019hmk} sets the best current constraints down to $\sim$200~MeV on sub-GeV DM scattering elastically off nuclei.

Several experiments have near-term plans to expand to using novel substrates for phonon detection. SPICE, part of the TESSERACT program, is building on TES-based phonon sensor work in conjunction with SuperCDMS and MINER to expand to $\mathrm{SiO_2}$ (quartz) and sapphire substrates~\cite{SPICE}. SuperCDMS is expanding work to diamond~\cite{Kurinsky:2019pgb} and SiC~\cite{Griffin:2020lgd} substrates, with both substrates expected to achieve gains in phonon resolution due to higher phonon quality and low electronic backgrounds due to higher band gaps (3-5~eV). CRESST has demonstrated a diamond detector with 70~eV resolution~\cite{Canonica:2020omq} on a CVD substrate, and is also planning near-term R\&D to achieve sub-Gev reach in diamond. While diamond is targeted primarily at NR detection in this energy range, the remaining substrates are polar, which opens low-mass detection windows through dark photon mediated interactions with optical phonons~\cite{Griffin:2019mvc}.

In a complementary approach, quantum defects in diamond and SiC are also proposed to achieve directional WIMP detection at solid-state densities~\cite{Rajendran:2017ynw,Marshall:2020azl,Ebadi:2022axg,Akerib:2022ort}. In this proposal, event registration would use charge or phonon collection technologies.  The directional information, however, would be in the form of submicron tracks of lattice damage caused by WIMP nuclear recoils. Early R\&D efforts have focused on nitrogen-vacancy defects in diamond~\cite{Marshall:2020azl,Marshall:2021kjk,Marshall:2021xiu} to establish the feasibility of this approach.  If successful, the realization of such a detector would also benefit from charge or phonon collection capabilities.

While NTDs and TES-based sensors have dominated phonon-readout for the last decade, new superconducting technologies are poised to broaden options of phonon readout. Kinetic Inductance Detectors (KIDs) are superconducting thin-film absorbers constructed as an LC resonator. The breaking of Cooper pairs due to phonon absorption within the film changes their surface impedance \textemdash~an effect that can be measured and related to the initial deposition energy via an RF tone. Current energy resolutions of $<$20~eV have been demonstrated, with a roadmap down to the eV-scale~ \cite{KIDSLOI}. Improved fabrication techniques such as ``tile-and-trim" and ``post-measurement lithographic correction", and new readout techniques like ``bandwidth recycling" are calculated to allow fabrication of over a million KID pixels on a single 6-inch wafer, exploiting the simple RF multiplexability of these devices ~\cite{millionkidLOI}.

Magnetic Microcalorimeters (MMCs) are another superconducting technology that are
being developed for sub-GeV DM detection via athermal-phonon
sensing  and phonon-pulse shape discrimination~\cite{kim2020self, kim2017novel,
MMC-cevns:2021loi}. They use thin metallic-magnetic films deposited on the surface of semiconducting or scintillating crystals for fast collection and sensing of athermal phonons with timescales on the order of microseconds, which enables   phonon-pulse shape discrimination to reduce low energy backgrounds from electron-recoils in NR searches. Near-term advances to optimize for energy resolution are anticipated to achieve $\mathcal{O}$(eV) energy thresholds, and tests are planned for sapphire and diamond, specifically targeting sub-GeV DM searches.

\subsubsection{Single-Charge Semiconductor Detectors}

As described in Sec.~\ref{ssec:20ev}, silicon CCDs have recently come to the forefront of low-mass DM detection, and have become even more powerful with the advent of the skipper-CCD technology, which enables repeated, non-destructive readout of charges from a single pixel. This method has been shown to reduce readout noise to 0.068\,e$^{-}$rms/pix~\cite{Tiffenberg:2017aac}, allowing for precise charge counting down to single electrons. The first effort utilizing these skipper-CCDs for DM searches is the SENSEI experiment, which has reported sensitivity down to the expected $\sim$1\,MeV and whose goal is to deploy about 100~g of skipper-CCDs at SNOLAB~\cite{Crisler:2018gci,Abramoff:2019dfb,SENSEI:2020dpa}. The DAMIC-M collaboration will also employ skipper-CCDs, aiming for a 1~kg target mass~\cite{Castello-Mor:2020jhd}. 

Looking ahead, the SENSEI and DAMIC-M collaborations have joined to propose Oscura~\cite{Aguilar-Arevalo:2022kqd}, a 10\,kg skipper-CCD experiment that will have significant physics reach for both DM with masses $\gtrsim$1\,MeV scattering off electrons and DM with masses $\gtrsim$1\,eV being absorbed by electrons.  
Due to the ability to clearly resolve single electrons, the energy detection threshold is already at its fundamental limit. Instead, the ultimate sensitivity of these experiments will be limited by overall exposure, as well as a detailed understanding of low-energy physics phenomena and detector effects. Thus, the technological R\&D challenges are more related to scaling the detector size rather than detector threshold, and include fabrication of CCDs, low-noise readout of many channels, and reduction of backgrounds coming from detector materials.  There has been recent progress in understanding the single- and few-electron events~\cite{Du:2020ldo,SENSEI:2021hcn}, but more work is needed.  

Semiconducting crystals read out by TES-based phonon sensors have a long history in DM searches, but were recently demonstrated to have single electron resolution in silicon by the SuperCDMS (Si HVeV) and EDELWEISS (Ge) collaborations, showing RMS noise equivalent to 0.03~\cite{Ren:2020gaq} and 0.53~\cite{Arnaud:2020svb} electron-holes pairs, respectively. As the band gaps in Si and Ge are similar, these experiments demonstrate similar reach for DM--electron scattering and absorption processes as skipper-CCD experiments. Their true power comes from the ability to detect sub-\bind{elec} energy deposits in the form of phonons, allowing for a significant extension in DM mass reach. As much of the required technical R\&D is driven by this detection channel, as described in~\cite{supercdmsWP22}, details are left to Sec.~\ref{ssec:collective}. 

In addition to the NR reach through phonon-only detection discussed earlier, diamond and silicon carbide are also useful as substrates for single charge detection\cite{Kurinsky:2019pgb,Griffin:2020lgd,diamondLOI}. While these materials have higher band gaps, which limit the mass reach of ionization-based searches to $\geq$5\,MeV, they also ensure deeper impurity traps, making detectors made from these materials less susceptible to absorption of infrared photons, which are known to cause excess charge backgrounds. Their high dielectric strength will allow kV-scale voltage biases, which translates to charge resolution of better than 10$^{-3}$~e$^{-}$. Work on phonon-mediated charge readout as well as direct single-charge readout have been proposed. More details are discussed in ~\cite{supercdmsWP22,SPLENDOR}. 

\subsubsection{Single Photon Detectors with Low Dark Counts}

A particularly fruitful area of collaboration between HEP, astrophysics, and photonics has been in the domain of single photon counting. In the optical and NIR regime, many of the technologies discussed in this whitepaper are closely tied to developments aimed at survey science, such as the skipper-CCDs, which are closely related to the focal planes developed for DES and DESI. This synergy has also motivated the use of other optical imaging technologies not only as secondary readout, but as the primary DM scattering target as well. For electron-recoil targets, the main drivers are the effective threshold at low energies, fiducial mass, and dark counts. Achieving dark count rates of O(mHz/g) or lower in the optical, and (kHz/g) in the near-infrared (NIR) regime, are required for any potential technology to compare favorably to the state of the art. The requirement for high fill factor and quantum efficiency for future UVOIR focal planes will naturally make many technologies suitable for DM detection provided the dark counts can be effectively minimized. The equivalent comparison for a 100~$\mu$m pixel pitch and 500~$\mu$m in a Si detector in ambient conditions is $10^{-2}$~counts/px/s, assuming it can achieve at least $10^{-5}$~counts/px/s or lower if run in a low-background facility.

One example of this synergy is the use of superconducting nanowire single-photon detectors (SNSPDs) directly as the active target to probe DM that scatters off electrons in the SNSPDs and bosonic DM that is absorbed in the SNSPDs~\cite{hochberg2019snspd,Hochberg:2021yud}. Using different materials, energy thresholds of 0.73\,eV and 0.25\,eV have been demonstrated, which would extend the mass reach of DM detectors beyond the current $\sim$1\,MeV limits, and may be reduced further through subsequent R\&D. A benefit of this approach is that for larger energy deposits, above the eV-scale, these devices may also have sensitivity to directionality, opening a new window on background discrimination. As noted above, a key challenge to overcome as this technology matures will be fabricating them at a scale that is useful for DM experiments. Existing experiments rely on established WSi nanowire technology for DM detection at optical and near-IR wavelengths; longer-term R\&D to extend this reach lower is discussed in the next section.

\subsubsection{Scintillating Detectors}
\label{sssec:scin}

Many crystals scintillate at optical to UV wavelengths, and a number of crystal scintillators have been proposed as DM detection targets. The majority of these have not been demonstrated with low backgrounds at the single photon level or with photon efficiency adequate to detect single scintillation events. For example, CRESST has used scintillation from calcium tungstate crystals for ER/NR discrimination, but the low light yield from events at low energy precludes the use of this detection mode below 100~eV~\cite{Angloher:2017sxg}. Thus, materials which primarily scintillate as a means of energy relaxation are needed, both to allow them to be used as primary detectors and to ensure high single photon detection efficiency correlates to thresholds at the scintillation photon energy.

Organic scintillators are promising targets for sub-GeV DM detection with electron scattering~\cite{Blanco:2019lrf,Blanco:2021hlm}. Aromatic organic compounds are naturally anisotropic due to planar carbon rings, with the lowest-energy electronic wavefunctions well-approximated by linear combinations of 2p$_z$ orbitals. The typical energy of the lowest excited state is a few eV, comparable to 1~eV bandgap of silicon detectors. Organic compounds have long been used as scintillation detectors for high-energy particles~\cite{birks2013theory}, which allows the excitation to be detected through emission of a photon. In organic \emph{crystals} in particular, intermolecular forces are weak enough that the electrons are localized to each molecule, preserving the anisotropy, in contrast to the delocalized electrons in semiconductors. Ref.~\cite{Blanco:2021hlm} found that trans-stilbene had the same total rate per unit mass as silicon for $m_\chi \gtrsim 5 \ {\rm MeV}$, and featured a peak-to-trough modulation amplitude of $\sim$25\% for $m_\chi \gtrsim 10 \ {\rm MeV}$ with light mediator exchange (and $\sim$10\% for heavy mediators). The modulation increases to $\sim$200\% for $m_\chi = 2 \ {\rm MeV}$ when the maximum DM kinetic energy approaches the 4.2 eV scintillation threshold of trans-stilbene. Such detectors have strong synergy with the skipper-CCD program, because a large-mass organic scintillator target can be coupled to a few CCDs with appropriate light focusing; the energy of the scintillation photon lies between the 1- and 2-electron bins, providing some protection against dark rate backgrounds.

The TESSERACT project, part of the Dark Matter New Initiatives (DMNI) program, builds a program around coupling TES-based phonon detectors to other novel detector materials, including superfluid helium, gallium arsenide, sapphire, and silica. 
HeRALD (Helium Roton Apparatus for Light Dark Matter) is part of the TESSERACT program and proposes to instrument a vessel of superfluid $^4$He for scintillation, triplet excimer, phonon, and roton detection \cite{hertel2019helium, He4LOI}, such that the ratios of signals may be used for discrimination. 
The TESs can be used to detect scintillation photons and the quenching of triplet excimers. Alternatively, ionizing the ejected atom and then accelerating them with an electric field, the ejected ions can be calorimetrically detected by the release of the heat of adsorption. For further discussion of the phonon and roton channels, refer to Sec.~\ref{sssec:phonon}. In any case, signals from the evaporated helium atoms may be collected into multiple detectors, allowing coincidence-based rejection of instrumental backgrounds. 

Crystal scintillators have also been identified as excellent target materials for probing DM-electron interactions, with gallium arsenide (GaAs) perhaps being the most promising~\cite{Derenzo:2016fse,Derenzo:2020hfo}. TESSERACT is developing GaAs as their primary target for electromagnetically-interacting DM due to the fact that it has a bandgap at 1.42\,eV, giving it a similar intrinsic reach to Si, and has the added benefits of intrinsic radio-purity and the capability to grow it commercially in large crystals. The phonon detectors will then be sensitive to the 1.33\,eV scintillation photons produced by interactions in the crystal, while coincident detection of phonon signals enables discrimination of electronic recoil signals from nuclear recoil and heat-only backgrounds. GaAs also seems to have little to no afterglow~\cite{Derenzo:2018plr,Vasiukov:2019hwn}. 

A similar proposal is given in~\cite{GaAsLOI}, and also offers the option to instrument the GaAs with SNSPDs described in the previous section. In both cases, R\&D will be required to optimize the crystal size, surface roughness (see e.g.~\cite{Derenzo:2022cnn} for details), and dopant density in order to maximize photon production and collection in the sensor. R\&D is also necessary in the case of either sensors choice: the energy threshold of TESs would need to be lowered from the current limit of $\sim$3eV down to the requisite 1.33\,eV, and SNSPDs must be scaled up from their current mm$^{2}$-scale size to cm$^{2}$ and beyond. 

Another proposal uses Quantum Dots (QDs) as a target material.  QDs are nanocrystals of semiconducting material whose band-edge electronic properties are determined by their characteristic size. PbS QDs can be tuned to have optical gaps as low as 0.37~eV and are excellent scintillators with near unity quantum yields, which can be used to search for DM interactions~\cite{Blanco:2022cel}. Additionally, multi-exciton generation in these QDs is efficient for MeV-scale DM-electron scattering. The coincident two-photon signals of the bi-exciton in PbS QDs is an intrinsically low-background signal. 

\subsection{meV-scale thresholds} \label{ssec:collective}

Target energy thresholds of $\mathcal{O}$(meV) to probe sub-GeV DM require new approaches to go beyond the eV-scale ionization and scintillation thresholds fleshed out in the previous sections. Though some of the prior discussed technologies can be extended down in sensitivity, the difference in DM kinematics and now comparable inter-atomic length scale necessitates thinking about interactions as occurring \textit{collectively} with the excitations themselves rather than with a single nucleus or electron.

From a detector perspective, this improved sensitivity requires accessing quanta at the meV level. A non-exhaustive list of studied and proposed physical mechanisms are:
\begin{itemize}
    \item Athermal phonons: Phonons are excited states of vibrational modes in a lattice\textemdash with commonly considered crystalline targets like Si, Ge, NaI, CsI, and CaWO$_4$. Phonons here come in two flavors: acoustic, in which lattice sites move together and are gapless in energy, and optical, where they move oppositely. A relevant scale for interactions directly coupling to phonons rather than individual nuclei or electrons and then shedding their energy, is the Debye energy ($\mathcal{O}$(10)~meV in the aforementioned crystals). Phonons also exist in superfluid materials, like $^4$He. 
    \item Superfluid Rotons: High-momentum excitations of the superfluid are called rotons (though they possess no angular momentum). Importantly there is a regime in which these superfluid phonons and rotons are kinematically forbidden from decaying, resulting in ballistic propagation with macroscopically large mean-free paths and lifetimes~\cite{hertel2019helium, maris2017helium}. The phonons and rotons may then carry information from the initial track, potentially allowing discrimination against instrumental backgrounds. 
    \item Cooper-pair quasiparticles: In superconductors, the paired \textit{Cooper}-electrons can be broken due to energy deposits, resulting in quasiparticles. These quasiparticles can propagate macroscopic distances in high-quality crystals and cyclically convert to phonons as well. Cooper-pair breaking energies are approximately related to the superconducting transition temperature $T_c$ as $\Delta\approx1.76/k_B/T_C$, with common materials like aluminium and aluminum-based alloys demonstrating pair-breaking energies of $\mathcal{O}$(10\textendash100)~$\mu$eV. 
    \item Far infrared photons: The vibrational modes of molecules often lie at the $\mathcal{O}$(100)~meV, implying that excited molecular states can decay to produce a Far-Infrared (FIR) photon spectrum following the absorption or scattering of DM (see e.g. \cite{Wang:2022cyk}).
    \item Magnetic domains: These are similar to phonons in that they are lattice excitations but are instead spin ordering waves\textemdash of electron-spin, nuclear-spin, or even molecular-spin\textemdash in magnetically ordered materials, with characteristic energies around the Bohr magneton, of $\mathcal{O}$(10)~$\mu$eV. These excitations can be important for DM interactions with a spin-dependent coupling, with preliminary calculations done by~\cite{Trickle:2019ovy}.
    \item Low bandgap and novel materials: Engineered materials, like graphene (falling under the umbrella of \textit{Dirac materials}, possessing unique linear rather than parabolic dispersion relations) can have electronic bandgaps of $\mathcal{O}$(meV) allowing for ionization detectors with much reduced thresholds. The energy scale for these materials largely arises from higher-order couplings such as spin-orbit or magnetic couplings between and within unit cells, which produce semi-metals in compounds which would not ordinarily have a bandgap (see e.g. \cite{rosa2020colossal}).
    \item Plasmons and EM couplings: Plasmons are collective oscillations of charges within a bulk. In contrast to DM interacting with individual electrons, the DM here couples to the electron density within the detector. The parameter of interest for these couplings is the complex dielectric function of the target which encodes the material's response to perturbations of its charge density~\cite{Hochberg:2021pkt,Gelmini:2020xir,Knapen:2021bwg}.
    \item Millicharge Effects: In general if the DM is charged, or mixes kinematically with the photon, one can expect sensitive probes of electric or magnetic field strength modulation to reveal signatures of DM interaction as detection of a millicharged particle. These searches have been described in detail in collider literature but have only recently been explored in the context of direct detection\cite{Berlin:2019uco}.
\end{itemize}

Technologies, techniques, and proposed experiments that exploit a few of the above mechanisms are briefly explored below.

\subsubsection{Phonons and Quasiparticles} \label{sssec:phonon}

Presently, the two devices used to sense athermal phonons are TESs~\cite{phononLOI} and KIDs~\cite{KIDSLOI,millionkidLOI}, both patterned on the surface of a crystal, sensitive to quasiparticle production through phonon energy deposits. Ballistic phonon propagation within a bulk is ultimately limited by surface contact and reflection, leading to surface mediated down-conversion~\cite{leman2012invited}, which is dependent on the target material, film thickness, and acoustic impedance mismatch between target and sensor to name a few long-term concerns. 

Direct quasiparticle trapping from a superconducting target to sensor is another avenue explored in the literature~\cite{Hochberg:2015fth}. The energy resolution of TES detectors is primarily driven by the superconducting transition temperature, motivating search for low $T_c$ materials with sharp transition features. The lower bound on KID sensitivity is set by Generation-Recombination noise (which is the quasiparticle fluctuation due to the thermal Cooper-pair breaking) and Two-Level System noise due to defects and dangling bonds etc. in the dielectric-film interface. Both sets of devices will benefit, in amplifier limited operation, from quantum limited parametric amplifiers and lower gap materials, with eventual goals of $\mathcal{O}$(10) meV resolving sensors~\cite{phononLOI, KIDSLOI}.

If DM is coupled to the standard model electromagnetic current (like a dark photon kinetically mixing with an ordinary photon) then the DM will couple via a dipole moment to an ionic crystal, like Al$_2$O$_3$ (sapphire), quartz (SiO$_2$), or SiC (silicon carbide), producing optical phonons~\cite{Knapen:2017ekk,Griffin:2020lgd}. This mechanism allows for enhanced probing of dark photon parameter space by optimizing the dielectric properties of the target. Complementarily, crystals with high sound speeds like diamond, silicon carbide (SiC), and silicon oxide (SiO$_2$) can be sensitive to scalar interactions that produce $\mathcal{O}$(meV) acoustic phonons~\cite{Griffin:2020lgd, diamondLOI, Kurinsky:2019pgb}.

Major ongoing \& proposed experiments in the DM crystalline phonon coupling world have been discussed in the previous sections. Future techniques to sense athermal phonon signatures include coupling superconducting qubits to a target to sense phonons, with one example being Quantum Capacitance Detectors \cite{echternach2018single}, which have been successfully demonstrated as single quasiparticle counting photon detectors. Phonon-induced decoherence is a significant challenge facing superconducting quantum computing, and further study of phonon coupling mechanisms also promises to expand their use as phonon sensors~\cite{Wilen:2020lgg}.

The HeRALD program, as previously noted in Sec.~\ref{sssec:scin}, is part of the TESSERACT experiment. For energy depositions below the lowest-energy atomic excitation, the available signals are phonons and rotons. By instrumenting the superfluid with TESs, these excitations, down to the 1 meV level, can be detected, as the ballistic phonons and rotons, once incident on the vacuum-liquid interface (which has a 0.62 meV evaporation threshold) can eject a $^4$He atom which are sensed by TES-based detectors above. As these are long-lived and carry information away from the event site, the momentum distribution of the excitations (measured through the pulse shape observed in helium atom quantum evaporation) can be used to discriminate against instrumental background \cite{hertel2019helium}. For DM of especially low mass, momentum and energy may be efficiently transferred to virtual excitations of the superfluid helium which then decay to phonon modes. This can in principle enable detection down to keV-scale DM masses~\cite{Schutz:2016tid,Knapen:2016cue}.

Alternatively, a novel technique is being developed for sensing phonons and rotons in liquid helium based on nano-mechanical resonators with ultra-high force sensitivity~\cite{LHe_Resonators}. This requires building resonators adapted to operate within superfluid helium and capable of sensing individual excitations within the liquid caused by a particle interaction. Preliminary calculations show that an array of such devices can adequately instrument a large volume of liquid helium, potentially achieving a sensitivity a factor of ten better than approaches being pursued by other groups. The experimental design also necessitates theoretical and numerical analysis of the interaction mechanisms of low-mass DM with helium and the subsequent process of generating the detectable excitations to understand, from first principles, the scientific reaches, limitations and overall design of a liquid helium low-mass DM detector~\cite{Matchev:2021fuw}.

Another experimental proposal suggests introducing a $^3$He layer to the top of either $^4$He or a standard crystal (i.e. Si, SiO$_2$, YIG) target, depending on the required DM coupling. At $\mathcal{O}$(100) mK temperatures, the $^3$He will form Andreev bound states that will reside on the surface and phonons in the crystal or superfluid bulk will eject the $^3$He in similar fashion to the previously described $^4$He atoms, to be read out by an envisioned qubit system~\cite{He3LOI,Lyon:2022sza}. The QUEST-DMC consortium will use NEMS with SQUID readout in superfluid $^3$He at 80 $\mu$K to detect spin-dependent interactions of light dark matter, building on the low-temperature infrastructures of the European Microkelvin Platform in the UK at Royal Holloway, University of London and Lancaster University~\cite{Casey}


\subsubsection{Far-IR Single Photon Detectors as Targets} \label{sssec:fir}

Continuing the theme from the previous section, we note that many meV-eV ERDM detection technologies directly map onto requirements for new FIR photon detectors. Many of the technological challenges inherent to expanding detector sensitivity to the 10s of microns, and eventually the THz regime, require reducing thresholds and operating detectors at lower temperatures to minimize dark counts. There is substantial synergy between development of large FIR and THz focal planes with low dark counts and expansion of both dark photon and scattering searches into the sub-eV regime, given the lack of competing technologies in these areas. The majority of existing FIR sensors are based on avalanching technology, which tends to trade long-wavelength sensitivity for increased dark counts and smaller fiducial volumes.

For example, a possible expansion to the skipper-CCD DM program is by heavy degenerate doping of CCDs to reduce the effective bandgap. Dopants in semiconductors at small concentration create new energy levels of electrons that lie $\mathcal O(10-100)\,\mathrm{meV}$ below (above) the conduction (valence) band. Therefore, doped semiconductors are sensitive to energy deposit of $\mathcal O(10-100)\,\mathrm{meV}$, which can be used to detect light DM with $\mathcal O(10-100)\,\mathrm{keV}$ masses via DM-electron scattering process or $\mathcal O(10-100)\,\mathrm{meV}$ masses via absorption. It has been demonstrated that doped semiconductors have a broader sensitivity compared to many other proposed targets, and can probe the freeze-in benchmark, along which the correct relic abundance of DM is obtained, with a modest exposure~\cite{Du:2022dxf}. Moreover, since doped semiconductors are easy to fabricate and widely used in industries, one of the attractive features of this proposal is that they could be readily realized with near-term  technologies. Studies of tradeoffs between dark rate, dopant concentration, quantum efficiency, and operation temperature are needed to understand the viablility of this approach.

There is also a roadmap for expanding SNSPD DM sensitivity beyond that demonstrated in~\cite{Hochberg:2021yud}. Moving from WSi to lower-$T_c$ materials, such as NbTi and Al, will allow for lower energy sensitivity due to the lower bandgap of the nanowire material. SNSPDs have been demonstrated down to thresholds of 125 meV~\cite{Verma:2020gso}, which along with their flat stackable geometry make them attractive for low threshold directional detection~\cite{hochberg2019snspd}. Growth of nanowires to increase fiducial mass per nanowire, and expansion of total mass through multiplexed arrays of nanowires, is a promising avenue to build gram-scale superconducting targets for DM searches.

In addition, it has been reported in~\cite{Kim:2020bwm} that achieving yet another order of magnitude in mass reach, from 1\,keV to 0.1\,MeV, is possible through the use of graphene-based Josephson junction (GJJ) single-photon detectors as the active target. Through a coupling to electrons, DM can scatter off free electrons in the graphene, transferring some fraction of its incoming kinetic energy. The recoiling electron heats up and thermalizes with nearby electrons, triggering the GJJ and allowing the detection of 0.1\,meV energy deposits. This sensitivity has already been demonstrated, and DM search results from a pathfinder detector may be available shortly. If this program moves forward, as with other detection schemes using novel materials, a major R\&D challenge will be fabrication and readout of the target material at scale.

\subsubsection{Low-Gap and Novel meV-Scale Materials} \label{sssec:dirac}

Molecular vibrational modes of 100 meV can be excited by DM-nucleus interactions, for both spin-dependent and spin-independent cases, able to probe masses down to 100 keV (0.2 eV) for scattering (absorption). Proposals have investigated carbon monoxide (CO), hydrogen fluoride (HF), and scandium hydride (ScHx) as the gaseous targets with emitted photons in the $\sim$150 to 500 meV range~\cite{Essig:2019kfe,MolecularLOI}. In addition, experiments based on hydrogen-rich crystals with known excitations in the mid-IR have also been proposed~\cite{Wang:2022cyk}. These types of targets will require targeted programs to determine which are suitable for low-background searches, and to characterize the lifetime and collection efficiency of their excitations. Depending the interaction, the de-excitation can lead to either a single or multi-photon cascade, detectable by SNSPDs, KIDs/TESs or any other technology. 

In addition to materials currently used as mid-IR photon sensors, there are a number of new materials with tunable bandgaps, produced through magnetic ordering. SPLENDOR, an effort to use the narrow gap semiconductors La$_3$Cd$_2$As$_6$ and Eu$_5$In$_2$Sb$_6$~\cite{rosa2020colossal} read out via low-threshold cryogenic charge amplifiers~\cite{Phipps:2016gdx,Juillard:2019njs,SPLENDOR}, has been funded for FY22-24 by the Los Alamos National Laboratory LDRD program. This supports growth of novel substrates, characterization of electronic response with electron loss spectroscopy, and development of conventional amplifiers as well as amplifiers based on charge-sensitive qubits and related structures~\cite{Tosi19,Brock21}. These materials have O(10~meV) bandgaps, and the tunability though application of an external field may also allow these materials to have their gaps varied in-situ to achieve lower gaps than those present at zero field, allowing for further lowering of thresholds for charge producing events.

Finally, Dirac materials can exhibit electron band structures that are linear, with gapless dispersion relations, and vanishing density of states at the Fermi surface. By carefully engineering and/or doping the crystalline structure, $\mathcal{O}$(meV) band-gaps can be generated. A relevant example of this class of material is bi-layer graphene~\cite{HOCHBERG2017239}, with tunable band-gaps depending on applied voltages or zirconium-pentatelluride (ZrTe$_5$) with a 35~meV bandgap~\cite{hochberg2018detection}. Depending on the material, DM signals can manifest as either long-lived athermal phonons signals read out by traditional TESs or KIDs (thanks to disfavored electron-phonon thermalization) or ionized electrons overcoming the material work function---to be externally collected and measured.

\subsubsection{Novel Detection Schemes} \label{subsec:plasmons}

In addition to advances in well-established detection methodologies, the challenge of detecting meV-scale excitations has also begun to inspire work in new methods of particle detection. This may be the area of direct detection that expands the most over the next decade, and requires ingenuity to identify systems that are be well-suited to this challenge. These techniques will fall into two categories: novel systems with meV-scale binding energies, or techniques that make use of the higher particle density of DM at low masses to make classical scattering measurements of the DM fluid. Two examples are given here. 

Single molecular magnets (SMMs) are molecules that are effectively non-interacting nano-scale magnets. A crystalline preparation of SMMs, aligned with an external magnetic field will exist in a meta-stable state. DM interactions depositing as little as 1 meV can form a localized hot-spot and trigger a spin-flip cascade in the crystal, leading to a macroscopically detectable change in crystal magnetization potentially readable by SQUIDs or other techniques. Recent work by~\cite{bunting2017magnetic} investigated the behaviour of various transition metal and lanthanide SMMs for DM search applicability. 

Experiments can investigate the coupling of charged DM with the SM, with common examples being Dark Photons or millicharged particles. One proposed experimental pathway to hunt for millicharged DM is a `deflection' experiment~\cite{Berlin:2019uco}. The DM population passes through a shielded region, effectively a capacitor, with an applied AC EM field. As the DM passes through, it is `deflected' into positively and negatively charged particles, creating a train of alternating millicharge densities that themselves induce small oscillating EM fields detectable by a resonant structure like an antenna in the far-field. This is novel approach to detecting sub-GeV DM that applies some principles from detecting wave-like DM but in a strictly classical scattering sense, and has not yet been explored from a practical standpoint. If it can be proven to be a viable experiment it is complementary to the techniques discussed here in that the medium is a classic field rather than a large mass, and it is thus limited by the volume of the field that can be actively monitored.

\section{Experimental Challenges} \label{ssec:EC}

For the success of new experimental techniques designed to search for low-mass DM, new experimental challenges must be faced and overcome. Two of these essential challenges are 1) the mitigation and characterization of backgrounds and 2) the quantification of signals. 

\textbf{Backgrounds:} The scientific reach of a given experiment is determined by both its DM signal strength and its backgrounds. Although proposals for new experiments tend to focus on the DM-induced signals that might be detected, in practice all experiments are limited by backgrounds, and searching new DM parameter space entails understanding and reducing these backgrounds as much as possible. This is a complex undertaking, and different experimental techniques are limited by different sets of background sources. For WIMP searches, most backgrounds arise from nuclear physics processes, but for experiments looking for signals in the eV scale and below, new backgrounds sources become relevant. Intimate understanding and effective mitigation of backgrounds from chemical physics, condensed matter physics, and atomic physics processes becomes essential. 

\textbf{Calibration:} In order to demonstrate DM sensitivity, any DM direct detection experiment must quantitatively characterize the signal that might be expected from a DM particle interacting with its target material. For traditional WIMP searches for nuclear recoil (NR) signals, this has largely been accomplished using MeV-scale neutrons, such as might be produced with neutron generators or neutron sources. But with the wide variety of signal types in sub-GeV DM experiments, nuclear recoil, electronic recoil, and collective excitation calibrations must be performed, at very low energies. This necessitates the development of a new set of calibration techniques. 

In the subsections below, experimental challenges particular to sub-GeV DM detection are summarized, with emphasis on phenomena in the meV to tens of eV energy range. A broader treatment of background and calibration topics (including phenomena at keV energies and above) may be found in~\cite{SnowmassCF1WP3}. 

\subsection{Backgrounds}

In practice, DM direct detection experiments are operated until they are background-limited, and then improved or upgraded so as to reduce backgrounds. Reduction of backgrounds in DM direct detection experiments is therefore a constant activity. All current experiments are limited by backgrounds, and background reduction is an area of constant innovation. New ideas for experiments owe their strength not only to their possible DM signatures, but also to their ability to reduce or actively reject backgrounds. 
For sufficiently large exposures, solar neutrinos can eventually limit the sensitivity of experiments to DM (moreover, in minimal extensions of the SM with right handed neutrinos, active neutrinos could interact via a magnetic moment interaction, which is enhanced at low recoil energies)~\cite{Essig:2011nj,Hochberg:2015fth,Essig:2018tss,Schwemberger:2022fjl,Donchenko:2021fnf}.

1708.02248). Weak scatterings of solar neutrinos are expected to be small given current exposures, but in minimal extensions of the SM with right handed neutrinos, active neutrinos could interact via a magnetic moment interaction, which is enhanced at low recoil energies: 2202.01254, 2111.03331 

\subsubsection{Radioactive backgrounds specific to meV-eV scales} The radioactive backgrounds endemic to WIMP detectors---gamma-rays, x-rays, fast neutrons, alpha and beta decays---still create backgrounds of relevance for sub-GeV DM detection; without a sophisticated shield and extremely low-background materials, it is difficult to reach radioactive background rates below 1 events/keV/kg/day. But because the characteristic energies of these backgrounds are much higher than those of the DM interactions, radioactive backgrounds are often less significant than many other sources of background. One exception is the Thomson and Delbr\"{u}ck scattering of gamma rays, which can produce low-energy nuclear recoils as well as a phonon signal~\cite{Robinson:2016imi,Berghaus:2021wrp}.  Another concern is epithermal neutrons: fast neutrons may lose most of their energy in one part of the apparatus and then create energy deposits in the DM detector before escape or capture, thus producing an excess of events at very low energy. Reduction of these backgrounds requires careful detector design to prevent gamma-rays and epithermal neutrons from reaching the active detector. Use of active neutron or gamma-ray vetoes can also be effective.

\subsubsection{Dark current} Any experiment for which charge is a signal will have backgrounds due to dark current. While this is dominated by single electrons, pile-up of single electrons can lead to multiple-electron backgrounds, as can multiplication of electrons in high-field regions of the detector. These backgrounds may be reduced through some combination of improving single-electron resolution, reducing leakage current, improved detector speed so as to reduce electron pileup, and limiting any high-field regions where electrons might multiply.

\subsubsection{Low-energy thermal and athermal backgrounds} For detectors that collect phonon signals, fluctuations near threshold can result in backgrounds. This background source includes vibration-induced heat backgrounds. An active area of R\&D is the reduction of vibrational backgrounds. 

\subsubsection{Environmental photon backgrounds} Photons produced around and in the detectors can create unwanted backgrounds.  These include infra-red radiation, Cherenkov radiation, and transition radiation~\cite{Du:2020ldo}. These can be reduced through a combination of cold infra-red shielding and the reduction of insulating materials around the DM detector. A modeling of this background is possible by carefully characterizing the materials in and around the sensor. 

\subsubsection{Backgrounds arising from other solid-state or chemical effects} At the eV scale or below, low-energy backgrounds that were of little importance for WIMP searches begin to be relevant. These include chemical backgrounds such as relaxation of metastable states and afterglow production of single photons, as well as solid state ``little earthquake'' backgrounds from the release of mechanical stress (including stresses produced when cooling the detector). Reduction of these backgrounds is another area of active R\&D. One mitigation technique is to avoid stressing the detector; another is to keep the detector in a state that cannot easily relax.  Use of materials with little chemical or solid-state processes in the energy range of interest may allow background reduction, as may experimental techniques in which signal in a low-background target is sensed in coincidence by multiple detectors.  

\subsection{Calibrations}

\subsubsection{Neutron scattering} Production of eV-scale nuclear recoils is a particular challenge, requiring keV-scale or lower energy neutron sources. In addition it is advantageous for the neutrons to be monoenergetic, so that the resulting nuclear recoil energy spectrum may be more effectively analyzed to produce an energy-dependent signal model. One approach is to use an accelerator to produce a nuclear reaction (such as $^7Li(p,n)$ or $^{51}V(p,n)$) yielding monoenergetic neutrons. A second approach is to use a photoneutron source. A third approach is to filter neutrons using a material that passes only a selected neutron energy. In some cases, like an iron-filtered SbBe source, two methods may be used simultaneously to enable a portable source with lower intrinsic backgrounds. 

\subsubsection{Low-energy ionization and scintillation yields} Neutron sources as described above may be used to determine the ionization and scintillation response at very low nuclear recoil energies relevant to sub-GeV DM detection. Because of the small number of ionization and scintillation quanta, these calibrations can be difficult. In particular, a key experimental question is the existence of a ``kinematic cutoff” (perhaps in line with the Frenkel defect energy \cite{stein2018analysis}, which is the energy needed to dislocate an atom from its site, of $\mathcal{O}$(10) eV in Si and Ge.) below which no ionization signal can be expected, curtailing ionization detector sensitivity even if nominally of lower threshold.  This motivates multi-channel detector development, as one might still expect a populated phonon or other excitation spectrum from such interactions. In addition, calibrations at the low energies comparable to electron binding energies are needed to explore signals expected from electronic recoils from sub-GeV DM particles. All told, these measurements are critical for quantifying and thereby justifying claimed DM sensitivity for any experiments making use of light or charge signals.

\subsubsection{Thomson and Delbr\"{u}ck gamma-ray scattering} In principle, these processes (where a gamma-ray elastic scatters with a nucleus) may be used for nuclear recoil calibration. Advantages include the higher and better-collimated fluxes that may be achieved with gamma rays, and the naturally low energies deposited. The main challenge is distinguishing these nuclear recoils from the significantly higher rate of Compton scatters, which must be distinguished through energy deposition and event type discrimination. 

\subsubsection{Low-energy Compton spectrum} Understanding the response of detectors to low-energy Compton scatters is important for understanding background spectra in detail, particularly around the binding energies of the target, where step-like spectral features exist and present measurements deviate from calculations \cite{ramanathan:2017compton, Botti:2022lkm}. Detailed gamma-ray calibration experiments are needed to generate accurate background models of this type. 

\subsubsection{X-ray scattering and absorption} X-rays are highly effective for mono-energetic electronic recoil calibration, and well suited for determining energy scale and resolution at 100 eV to tens of keV energies. An area of active R\&D is the implementation of x-ray sources that may be exposed to a detector or removed on demand, which can be a challenge since x-rays are easily absorbed by intervening material. X-ray sources may also be created through neutron activation, thus allowing a source that provides early calibration, but decays away to allow low background operation as the experiment proceeds.

\subsubsection{Laser and LED photon calibration} For eV scales or lower, lasers and LEDs are essential for probing detector response. In general these must either be implemented next to the detector (thus creating potential sources of radioactive or other local backgrounds) or must be guided to the detector through optical fiber or through windows (potentially creating backgrounds guided through the fiber or window into the DM detector).  Experimental design must take into account the calibration needs, as well as ensure that calibration sources do not create undue backgrounds.  Another consideration is the need to calibrate a detector in its bulk, rather than only on its surface, which can require a careful choice of excitation wavelength. A photon calibration source that can scan over a detector may also be employed, so as to characterize its uniformity of response. 

\subsubsection{Measurement of the Migdal effect} The Migdal effect, in which sudden nuclear acceleration upon scattering can lead to direct ionization of atomic electrons,  can be used to significantly improve the sensitivity of direct detection experiments to sub-GeV DM particles. Though yet to be confirmed experimentally, this process can be calculated for both isolated atoms and in condensed media. Experimental validation of the Migdal effect rates and electron spectra would allow detailed comparison of data to quantum mechanical calculations, and determine if there are any departures from expectation (several collaborations are already attempting to measure the Migdal effect, see e.g.~\cite{Araujo:2022wjh,Adams:2022zvg}).

\subsection{Conclusion} 

For new experiments searching for event-by-event sub-GeV DM interactions, a new energy regime (meV to eV scales) must be rigorously controlled for backgrounds and precisely quantified for potential DM signals. Many techniques are being developed for this new experimental challenge. New methods for calibration or background reduction can often applied across multiple experiments, enabling a broad impact.



\bibliography{main.v2.bbl}

\begin{thebibliography}{100}
\providecommand{\url}[1]{{#1}}
\providecommand{\urlprefix}{URL }
\expandafter\ifx\csname urlstyle\endcsname\relax
  \providecommand{\doi}[1]{DOI \discretionary{}{}{}#1}\else
  \providecommand{\doi}{DOI \discretionary{}{}{}\begingroup
  \urlstyle{rm}\Url}\fi

\bibitem{Feng:2014uja}
J.L. Feng, et~al., in \emph{{Community Summer Study 2013}: {Snowmass on the
  Mississippi}} (2014)

\bibitem{Cushman:2013zza}
P.~Cushman, et~al., in \emph{{Community Summer Study 2013: Snowmass on the
  Mississippi (CSS2013) Minneapolis, MN, USA, July 29-August 6, 2013}} (2013).
\newblock \urlprefix\url{http://arxiv.org/pdf/1310.8327.pdf}

\bibitem{Essig:2012yx}
R.~Essig, A.~Manalaysay, J.~Mardon, P.~Sorensen, T.~Volansky, Phys. Rev. Lett.
  \textbf{109}, 021301 (2012).
\newblock \doi{10.1103/PhysRevLett.109.021301}

\bibitem{Angle:2011th}
J.~Angle, et~al., Phys. Rev. Lett. \textbf{107}, 051301 (2011).
\newblock \doi{10.1103/PhysRevLett.110.249901, 10.1103/PhysRevLett.107.051301}.
\newblock [Erratum: Phys. Rev. Lett.110,249901(2013)]

\bibitem{Essig:2011nj}
R.~Essig, J.~Mardon, T.~Volansky, Phys. Rev. \textbf{D85}, 076007 (2012).
\newblock \doi{10.1103/PhysRevD.85.076007}

\bibitem{Graham:2012su}
P.W. Graham, D.E. Kaplan, S.~Rajendran, M.T. Walters, Phys.Dark Univ.
  \textbf{1}, 32 (2012).
\newblock \doi{10.1016/j.dark.2012.09.001}

\bibitem{An:2014twa}
H.~An, M.~Pospelov, J.~Pradler, A.~Ritz, Phys. Lett. \textbf{B747}, 331 (2015).
\newblock \doi{10.1016/j.physletb.2015.06.018}

\bibitem{Aprile:2014eoa}
E.~Aprile, et~al., Phys. Rev. \textbf{D90}(6), 062009 (2014).
\newblock \doi{10.1103/PhysRevD.90.062009}

\bibitem{Lee:2015qva}
S.K. Lee, M.~Lisanti, S.~Mishra-Sharma, B.R. Safdi, Phys. Rev. \textbf{D92}(8),
  083517 (2015).
\newblock \doi{10.1103/PhysRevD.92.083517}

\bibitem{Essig:2015cda}
R.~Essig, M.~Fernandez-Serra, J.~Mardon, A.~Soto, T.~Volansky, T.T. Yu, JHEP
  \textbf{05}, 046 (2016).
\newblock \doi{10.1007/JHEP05(2016)046}

\bibitem{Hochberg:2015pha}
Y.~Hochberg, Y.~Zhao, K.M. Zurek, Phys. Rev. Lett. \textbf{116}(1), 011301
  (2016).
\newblock \doi{10.1103/PhysRevLett.116.011301}

\bibitem{Hochberg:2015fth}
Y.~Hochberg, M.~Pyle, Y.~Zhao, K.M. Zurek, JHEP \textbf{08}, 057 (2016).
\newblock \doi{10.1007/JHEP08(2016)057}

\bibitem{Derenzo:2016fse}
S.~Derenzo, R.~Essig, A.~Massari, A.~Soto, T.T. Yu, Phys. Rev. \textbf{D96}(1),
  016026 (2017).
\newblock \doi{10.1103/PhysRevD.96.016026}

\bibitem{Aguilar-Arevalo:2016zop}
A.~Aguilar-Arevalo, et~al., Phys. Rev. Lett. \textbf{118}(14), 141803 (2017).
\newblock \doi{10.1103/PhysRevLett.118.141803}

\bibitem{Bloch:2016sjj}
I.M. Bloch, R.~Essig, K.~Tobioka, T.~Volansky, T.T. Yu, JHEP \textbf{06}, 087
  (2017).
\newblock \doi{10.1007/JHEP06(2017)087}

\bibitem{Cavoto:2016lqo}
G.~Cavoto, E.N.M. Cirillo, F.~Cocina, J.~Ferretti, A.D. Polosa, Eur. Phys. J.
  \textbf{C76}(6), 349 (2016).
\newblock \doi{10.1140/epjc/s10052-016-4193-7}

\bibitem{Hochberg:2016ntt}
Y.~Hochberg, Y.~Kahn, M.~Lisanti, C.G. Tully, K.M. Zurek, Phys. Lett.
  \textbf{B772}, 239 (2017).
\newblock \doi{10.1016/j.physletb.2017.06.051}

\bibitem{Hochberg:2016ajh}
Y.~Hochberg, T.~Lin, K.M. Zurek, Phys. Rev. \textbf{D94}(1), 015019 (2016).
\newblock \doi{10.1103/PhysRevD.94.015019}

\bibitem{Hochberg:2016sqx}
Y.~Hochberg, T.~Lin, K.M. Zurek, Phys. Rev. \textbf{D95}(2), 023013 (2017).
\newblock \doi{10.1103/PhysRevD.95.023013}

\bibitem{Kouvaris:2016afs}
C.~Kouvaris, J.~Pradler, Phys. Rev. Lett. \textbf{118}(3), 031803 (2017).
\newblock \doi{10.1103/PhysRevLett.118.031803}

\bibitem{Robinson:2016imi}
A.E. Robinson, Phys. Rev. D \textbf{95}(2), 021301 (2017).
\newblock \doi{10.1103/PhysRevD.95.021301}.
\newblock [Erratum: Phys.Rev.D 95, 069907 (2017)]

\bibitem{Ibe:2017yqa}
M.~Ibe, W.~Nakano, Y.~Shoji, K.~Suzuki, JHEP \textbf{03}, 194 (2018).
\newblock \doi{10.1007/JHEP03(2018)194}

\bibitem{Dolan:2017xbu}
M.J. Dolan, F.~Kahlhoefer, C.~McCabe, Phys. Rev. Lett. \textbf{121}(10), 101801
  (2018).
\newblock \doi{10.1103/PhysRevLett.121.101801}

\bibitem{Tiffenberg:2017aac}
J.~Tiffenberg, M.~Sofo-Haro, A.~Drlica-Wagner, R.~Essig, Y.~Guardincerri,
  S.~Holland, T.~Volansky, T.T. Yu, Phys. Rev. Lett. \textbf{119}(13), 131802
  (2017).
\newblock \doi{10.1103/PhysRevLett.119.131802}

\bibitem{Romani:2017iwi}
R.K. Romani, et~al., Appl. Phys. Lett. \textbf{112}, 043501 (2018).
\newblock \doi{10.1063/1.5010699}

\bibitem{Budnik:2017sbu}
R.~Budnik, O.~Chesnovsky, O.~Slone, T.~Volansky, Phys. Lett. \textbf{B782}, 242
  (2018).
\newblock \doi{10.1016/j.physletb.2018.04.063}

\bibitem{Bunting:2017net}
P.C. Bunting, G.~Gratta, T.~Melia, S.~Rajendran, Phys. Rev. \textbf{D95}(9),
  095001 (2017).
\newblock \doi{10.1103/PhysRevD.95.095001}

\bibitem{Cavoto:2017otc}
G.~Cavoto, F.~Luchetta, A.D. Polosa, Phys. Lett. \textbf{B776}, 338 (2018).
\newblock \doi{10.1016/j.physletb.2017.11.064}

\bibitem{Fichet:2017bng}
S.~Fichet, Phys. Rev. Lett. \textbf{120}(13), 131801 (2018).
\newblock \doi{10.1103/PhysRevLett.120.131801}

\bibitem{Knapen:2017ekk}
S.~Knapen, T.~Lin, M.~Pyle, K.M. Zurek, Phys. Lett. \textbf{B785}, 386 (2018).
\newblock \doi{10.1016/j.physletb.2018.08.064}

\bibitem{Hochberg:2017wce}
Y.~Hochberg, Y.~Kahn, M.~Lisanti, K.M. Zurek, A.~Grushin, R.~Ilan, S.M.
  Griffin, Z.F. Liu, S.F. Weber, Phys. Rev. \textbf{D97}(1), 015004 (2017).
\newblock \doi{10.1103/PhysRevD.97.015004}

\bibitem{Knapen:2017xzo}
S.~Knapen, T.~Lin, K.M. Zurek, Phys. Rev. D \textbf{96}(11), 115021 (2017).
\newblock \doi{10.1103/PhysRevD.96.115021}

\bibitem{Emken:2017erx}
T.~Emken, C.~Kouvaris, I.M. Shoemaker, Phys. Rev. \textbf{D96}(1), 015018
  (2017).
\newblock \doi{10.1103/PhysRevD.96.015018}

\bibitem{Emken:2017hnp}
T.~Emken, C.~Kouvaris, N.G. Nielsen, Phys. Rev. \textbf{D97}(6), 063007 (2018).
\newblock \doi{10.1103/PhysRevD.97.063007}

\bibitem{Emken:2017qmp}
T.~Emken, C.~Kouvaris, JCAP \textbf{1710}(10), 031 (2017).
\newblock \doi{10.1088/1475-7516/2017/10/031}

\bibitem{Emken:2018run}
T.~Emken, C.~Kouvaris, Phys. Rev. \textbf{D97}(11), 115047 (2018).
\newblock \doi{10.1103/PhysRevD.97.115047}

\bibitem{LUX:2018akb}
D.S. Akerib, et~al., Phys. Rev. Lett. \textbf{122}(13), 131301 (2019).
\newblock \doi{10.1103/PhysRevLett.122.131301}

\bibitem{Crisler:2018gci}
M.~Crisler, R.~Essig, J.~Estrada, G.~Fernandez, J.~Tiffenberg, M.~Sofo~haro,
  T.~Volansky, T.T. Yu, Phys. Rev. Lett. \textbf{121}(6), 061803 (2018).
\newblock \doi{10.1103/PhysRevLett.121.061803}

\bibitem{Agnese:2018col}
R.~Agnese, et~al., Phys. Rev. Lett. \textbf{121}(5), 051301 (2018).
\newblock \doi{10.1103/PhysRevLett.121.051301}.
\newblock [Erratum: Phys.Rev.Lett. 122, 069901 (2019)]

\bibitem{Agnes:2018oej}
P.~Agnes, et~al., Phys. Rev. Lett. \textbf{121}(11), 111303 (2018).
\newblock \doi{10.1103/PhysRevLett.121.111303}

\bibitem{Settimo:2018qcm}
M.~Settimo, in \emph{{Proceedings, 53Rd Rencontres De Moriond on Cosmology: La
  Thuile, Italy, March 17-24, 2018}} (2018)

\bibitem{Bringmann:2018cvk}
T.~Bringmann, M.~Pospelov, Phys. Rev. Lett. \textbf{122}(17), 171801 (2019).
\newblock \doi{10.1103/PhysRevLett.122.171801}

\bibitem{Ema:2018bih}
Y.~Ema, F.~Sala, R.~Sato, Phys. Rev. Lett. \textbf{122}(18), 181802 (2019).
\newblock \doi{10.1103/PhysRevLett.122.181802}

\bibitem{Akerib:2018hck}
D.S. Akerib, et~al., Phys. Rev. Lett. \textbf{122}(13), 131301 (2019).
\newblock \doi{10.1103/PhysRevLett.122.131301}

\bibitem{Griffin:2018bjn}
S.~Griffin, S.~Knapen, T.~Lin, K.M. Zurek, Phys. Rev. D \textbf{98}(11), 115034
  (2018).
\newblock \doi{10.1103/PhysRevD.98.115034}

\bibitem{CDEX:2019hzn}
Z.Z. Liu, et~al., Phys. Rev. Lett. \textbf{123}(16), 161301 (2019).
\newblock \doi{10.1103/PhysRevLett.123.161301}

\bibitem{EDELWEISS:2019vjv}
E.~Armengaud, et~al., Phys. Rev. D \textbf{99}(8), 082003 (2019).
\newblock \doi{10.1103/PhysRevD.99.082003}

\bibitem{Bell:2019egg}
N.F. Bell, J.B. Dent, J.L. Newstead, S.~Sabharwale, T.J. Weiler, Phys. Rev. D
  \textbf{101}(1), 015012 (2019).
\newblock \doi{10.1103/PhysRevD.101.015012}

\bibitem{Liu:2019kzq}
Z.Z. Liu, et~al., Phys. Rev. Lett. \textbf{123}(16), 161301 (2019).
\newblock \doi{10.1103/PhysRevLett.123.161301}

\bibitem{XENON:2019zpr}
E.~Aprile, et~al., Phys. Rev. Lett. \textbf{123}(24), 241803 (2019).
\newblock \doi{10.1103/PhysRevLett.123.241803}

\bibitem{Essig:2019xkx}
R.~Essig, J.~Pradler, M.~Sholapurkar, T.T. Yu, Phys. Rev. Lett.
  \textbf{124}(2), 021801 (2019).
\newblock \doi{10.1103/PhysRevLett.124.021801}

\bibitem{Baxter:2019pnz}
D.~Baxter, Y.~Kahn, G.~Krnjaic, Phys. Rev. D \textbf{101}(7), 076014 (2019).
\newblock \doi{10.1103/PhysRevD.101.076014}

\bibitem{Abramoff:2019dfb}
O.~Abramoff, et~al., Phys. Rev. Lett. \textbf{122}(16), 161801 (2019).
\newblock \doi{10.1103/PhysRevLett.122.161801}

\bibitem{Aguilar-Arevalo:2019wdi}
A.~Aguilar-Arevalo, et~al., Phys. Rev. Lett. \textbf{123}(18), 181802 (2019).
\newblock \doi{10.1103/PhysRevLett.123.181802}

\bibitem{Armengaud:2019kfj}
E.~Armengaud, et~al., Phys. Rev. \textbf{D99}(8), 082003 (2019).
\newblock \doi{10.1103/PhysRevD.99.082003}

\bibitem{Aprile:2019jmx}
E.~Aprile, et~al., Phys. Rev. Lett. \textbf{123}(24), 241803 (2019).
\newblock \doi{10.1103/PhysRevLett.123.241803}

\bibitem{Emken:2019tni}
T.~Emken, R.~Essig, C.~Kouvaris, M.~Sholapurkar, JCAP \textbf{1909}(09), 070
  (2019).
\newblock \doi{10.1088/1475-7516/2019/09/070}

\bibitem{Kurinsky:2019pgb}
N.A. Kurinsky, T.C. Yu, Y.~Hochberg, B.~Cabrera, Phys. Rev. D \textbf{99}(12),
  123005 (2019).
\newblock \doi{10.1103/PhysRevD.99.123005}

\bibitem{Cappiello:2019qsw}
C.~Cappiello, J.F. Beacom, Phys. Rev. \textbf{D100}(10), 103011 (2019).
\newblock \doi{10.1103/PhysRevD.100.103011}

\bibitem{Trickle:2019ovy}
T.~Trickle, Z.~Zhang, K.M. Zurek, Phys. Rev. Lett. \textbf{124}(20), 201801
  (2019).
\newblock \doi{10.1103/PhysRevLett.124.201801}

\bibitem{Griffin:2019mvc}
S.M. Griffin, K.~Inzani, T.~Trickle, Z.~Zhang, K.M. Zurek, Phys. Rev. D
  \textbf{101}(5), 055004 (2019).
\newblock \doi{10.1103/PhysRevD.101.055004}

\bibitem{Trickle:2019nya}
T.~Trickle, Z.~Zhang, K.M. Zurek, K.~Inzani, S.~Griffin, JHEP \textbf{03}, 036
  (2019).
\newblock \doi{10.1007/JHEP03(2020)036}

\bibitem{Catena:2019gfa}
R.~Catena, T.~Emken, N.~Spaldin, W.~Tarantino, Phys. Rev. Res. \textbf{2}(3),
  033195 (2019).
\newblock \doi{10.1103/PhysRevResearch.2.033195}

\bibitem{Berlin:2019ahk}
A.~Berlin, R.T. D'Agnolo, S.A.R. Ellis, C.~Nantista, J.~Neilson, P.~Schuster,
  S.~Tantawi, N.~Toro, K.~Zhou, JHEP \textbf{07}(07), 088 (2020).
\newblock \doi{10.1007/JHEP07(2020)088}

\bibitem{Berlin:2019uco}
A.~Berlin, R.T. D'Agnolo, S.A.R. Ellis, P.~Schuster, N.~Toro, Phys. Rev. Lett.
  \textbf{124}(1), 011801 (2020).
\newblock \doi{10.1103/PhysRevLett.124.011801}

\bibitem{Coskuner:2019odd}
A.~Coskuner, A.~Mitridate, A.~Olivares, K.M. Zurek, Phys. Rev. D
  \textbf{103}(1), 016006 (2021).
\newblock \doi{10.1103/PhysRevD.103.016006}

\bibitem{Lin:2019uvt}
T.~Lin, PoS \textbf{333}, 009 (2019).
\newblock \doi{10.22323/1.333.0009}

\bibitem{Blanco:2019lrf}
C.~Blanco, J.I. Collar, Y.~Kahn, B.~Lillard, Phys. Rev. D \textbf{101}(5),
  056001 (2020).
\newblock \doi{10.1103/PhysRevD.101.056001}

\bibitem{Geilhufe:2019ndy}
R.M. Geilhufe, F.~Kahlhoefer, M.W. Winkler, Phys. Rev. D \textbf{101}(5),
  055005 (2020).
\newblock \doi{10.1103/PhysRevD.101.055005}

\bibitem{Griffin:2020lgd}
S.M. Griffin, Y.~Hochberg, K.~Inzani, N.~Kurinsky, T.~Lin, T.~Chin, Phys. Rev.
  D \textbf{103}(7), 075002 (2021).
\newblock \doi{10.1103/PhysRevD.103.075002}

\bibitem{Trickle:2020oki}
T.~Trickle, Z.~Zhang, K.M. Zurek.
\newblock {Effective Field Theory of Dark Matter Direct Detection With
  Collective Excitations}.
\newblock arXiv:2009.13534 [hep-ph] (2020)

\bibitem{Hochberg:2021pkt}
Y.~Hochberg, Y.~Kahn, N.~Kurinsky, B.V. Lehmann, T.C. Yu, K.K. Berggren, Phys.
  Rev. Lett. \textbf{127}(15), 151802 (2021).
\newblock \doi{10.1103/PhysRevLett.127.151802}

\bibitem{Akerib:2021pfd}
D.S. Akerib, et~al.
\newblock {Enhancing the sensitivity of the LUX-ZEPLIN (LZ) dark matter
  experiment to low energy signals}.
\newblock arXiv:2101.08753 [astro-ph.IM] (2021)

\bibitem{LUX:2020yym}
D.S. Akerib, et~al., Phys. Rev. D \textbf{104}(1), 012011 (2021).
\newblock \doi{10.1103/PhysRevD.104.012011}

\bibitem{SENSEI:2020dpa}
{L. Barak {\it et al.} [SENSEI Collaboration]}, Phys. Rev. Lett.
  \textbf{125}(17), 171802 (2020).
\newblock \doi{10.1103/PhysRevLett.125.171802}

\bibitem{Liang:2020ryg}
Z.L. Liang, C.~Mo, F.~Zheng, P.~Zhang, Phys. Rev. D \textbf{104}(5), 056009
  (2021).
\newblock \doi{10.1103/PhysRevD.104.056009}

\bibitem{GrillidiCortona:2020owp}
G.~Grilli~di Cortona, A.~Messina, S.~Piacentini, JHEP \textbf{11}, 034 (2020).
\newblock \doi{10.1007/JHEP11(2020)034}

\bibitem{Bernstein:2020cpc}
A.~Bernstein, et~al., J. Phys. Conf. Ser. \textbf{1468}(1), 012035 (2020).
\newblock \doi{10.1088/1742-6596/1468/1/012035}

\bibitem{Ma:2019lik}
H.~Ma, Z.~She, Z.~Liu, L.~Yang, Q.~Yue, Z.~Zeng, T.~Xue, J. Phys. Conf. Ser.
  \textbf{1468}(1), 012070 (2020).
\newblock \doi{10.1088/1742-6596/1468/1/012070}

\bibitem{Nakamura:2020kex}
K.D. Nakamura, K.~Miuchi, S.~Kazama, Y.~Shoji, M.~Ibe, W.~Nakano, PTEP
  \textbf{2021}(1), 013C01 (2021).
\newblock \doi{10.1093/ptep/ptaa162}

\bibitem{Du:2020ldo}
P.~Du, D.~Egana-Ugrinovic, R.~Essig, M.~Sholapurkar, Phys. Rev. X
  \textbf{12}(1), 011009 (2022).
\newblock \doi{10.1103/PhysRevX.12.011009}

\bibitem{Collar:2021fcl}
J.I. Collar, A.R.L. Kavner, C.M. Lewis, Phys. Rev. D \textbf{103}(12), 122003
  (2021).
\newblock \doi{10.1103/PhysRevD.103.122003}

\bibitem{Hochberg:2021ymx}
Y.~Hochberg, E.D. Kramer, N.~Kurinsky, B.V. Lehmann.
\newblock {Directional Detection of Light Dark Matter in Superconductors}.
\newblock arXiv:2109.04473 [hep-ph] (2021)

\bibitem{Mitridate:2021ctr}
A.~Mitridate, T.~Trickle, Z.~Zhang, K.M. Zurek, JHEP \textbf{09}, 123 (2021).
\newblock \doi{10.1007/JHEP09(2021)123}

\bibitem{Griffin:2021znd}
S.M. Griffin, K.~Inzani, T.~Trickle, Z.~Zhang, K.M. Zurek, Phys. Rev. D
  \textbf{104}(9), 095015 (2021).
\newblock \doi{10.1103/PhysRevD.104.095015}

\bibitem{Coskuner:2021qxo}
A.~Coskuner, T.~Trickle, Z.~Zhang, K.M. Zurek.
\newblock {Directional Detectability of Dark Matter With Single Phonon
  Excitations: Target Comparison}.
\newblock arXiv:2102.09567 [hep-ph] (2021)

\bibitem{Kahn:2021ttr}
Y.~Kahn, T.~Lin.
\newblock {Searches for light dark matter using condensed matter systems}.
\newblock arXiv:2108.03239 [hep-ph] (2021)

\bibitem{Berghaus:2021wrp}
K.V. Berghaus, R.~Essig, Y.~Hochberg, Y.~Shoji, M.~Sholapurkar, Phys. Rev. D
  \textbf{106}(2), 023026 (2022).
\newblock \doi{10.1103/PhysRevD.106.023026}

\bibitem{Blanco:2021hlm}
C.~Blanco, Y.~Kahn, B.~Lillard, S.D. McDermott, Phys. Rev. D \textbf{104},
  036011 (2021).
\newblock \doi{10.1103/PhysRevD.104.036011}

\bibitem{Aguilar-Arevalo:2022kqd}
A.~Aguilar-Arevalo, et~al.,   (2022)

\bibitem{SuperCDMS:2022kgp}
M.~Al-Bakry, et~al.,   (2022)

\bibitem{Alexander:2016aln}
J.~Alexander, et~al.,  (2016).
\newblock
  \urlprefix\url{http://lss.fnal.gov/archive/2016/conf/fermilab-conf-16-421.pdf}

\bibitem{Battaglieri:2017aum}
M.~Battaglieri, et~al., in \emph{{U.S. Cosmic Visions: New Ideas in Dark Matter
  College Park, MD, USA, March 23-25, 2017}} (2017).
\newblock
  \urlprefix\url{http://lss.fnal.gov/archive/2017/conf/fermilab-conf-17-282-ae-ppd-t.pdf}

\bibitem{BRNreport}
{Department of Energy},  (2018).
\newblock
  \urlprefix\url{https://science.osti.gov/-/media/hep/pdf/Reports/Dark_Matter_New_Initiatives_rpt.pdf}

\bibitem{BRNreport-detector}
{Department of Energy},  (2019).
\newblock
  \urlprefix\url{https://science.osti.gov/-/media/hep/pdf/Reports/2020/DOE_Basic_Research_Needs_Study_on_High_Energy_Physics.pdf}

\bibitem{Essig:Physics2020}
R.~Essig, Physics 13, 172  (2020)

\bibitem{Bell:2021zkr}
N.F. Bell, J.B. Dent, B.~Dutta, S.~Ghosh, J.~Kumar, J.L. Newstead, Phys. Rev. D
  \textbf{104}(7), 076013 (2021).
\newblock \doi{10.1103/PhysRevD.104.076013}

\bibitem{Bell:2021xff}
N.F. Bell, J.B. Dent, B.~Dutta, S.~Ghosh, J.~Kumar, J.L. Newstead, I.M.
  Shoemaker, Phys. Rev. D \textbf{104}, 076020 (2021)

\bibitem{SnowmassCF1WP1}
P.~Cushman, R.~Gaitskell, C.~Galbiati, B.~Loer, {Snowmass2021 Cosmic Frontier
  White Paper: Dark matter direct detection to the neutrino floor} (2022)

\bibitem{SnowmassCF1WP3}
D.~Baxter, R.~Bunker, S.~Shaw, S.~Westerdale, {Snowmass2021 Cosmic Frontier
  White Paper: Calibrations and backgrounds for dark matter direct detection}
  (2022)

\bibitem{SnowmassIF1WP1}
T.~Cecil, K.~Irwin, M.~Pyle, R.~Maruyama, {Snowmass2021 Instrumentation
  Frontier White Paper: Quantum Calorimeters and Single Electronic Excitation
  Detectors} (2022)

\bibitem{SnowmassIF1WP2}
C.~Escobar, C.~Rogan, J.~Estrada, {Snowmass2021 Instrumentation Frontier White
  Paper: Photon Counting in Visible and near-IR} (2022)

\bibitem{Amaral:2020ryn}
D.W. Amaral, et~al., Phys. Rev. D \textbf{102}(9), 091101 (2020).
\newblock \doi{10.1103/PhysRevD.102.091101}

\bibitem{Arnaud:2020svb}
Q.~Arnaud, et~al., Phys. Rev. Lett. \textbf{125}(14), 141301 (2020).
\newblock \doi{10.1103/PhysRevLett.125.141301}

\bibitem{Essig:2017kqs}
R.~Essig, T.~Volansky, T.T. Yu, Phys. Rev. \textbf{D96}(4), 043017 (2017).
\newblock \doi{10.1103/PhysRevD.96.043017}

\bibitem{Aprile:2016wwo}
E.~Aprile, et~al., Phys. Rev. \textbf{D94}(9), 092001 (2016).
\newblock \doi{10.1103/PhysRevD.94.092001, 10.1103/PhysRevD.95.059901}.
\newblock [Erratum: Phys. Rev.D95,no.5,059901(2017)]

\bibitem{Aprile:2019xxb}
E.~Aprile, et~al., Phys. Rev. Lett. \textbf{123}(25), 251801 (2019).
\newblock \doi{10.1103/PhysRevLett.123.251801}

\bibitem{XENON:2021qze}
E.~Aprile, et~al., Phys. Rev. D \textbf{106}(2), 022001 (2022).
\newblock \doi{10.1103/PhysRevD.106.022001}

\bibitem{PandaX-II:2021nsg}
C.~Cheng, et~al., Phys. Rev. Lett. \textbf{126}(21), 211803 (2021).
\newblock \doi{10.1103/PhysRevLett.126.211803}

\bibitem{SuperCDMS:2020aus}
I.~Alkhatib, et~al., Phys. Rev. Lett. \textbf{127}, 061801 (2021).
\newblock \doi{10.1103/PhysRevLett.127.061801}

\bibitem{Abdelhameed:2019hmk}
A.~Abdelhameed, et~al., Phys. Rev. D \textbf{100}(10), 102002 (2019).
\newblock \doi{10.1103/PhysRevD.100.102002}

\bibitem{Knapen:2021bwg}
S.~Knapen, J.~Kozaczuk, T.~Lin, Phys. Rev. D \textbf{105}(1), 015014 (2022).
\newblock \doi{10.1103/PhysRevD.105.015014}

\bibitem{EDELWEISS:2022ktt}
E.~Armengaud, et~al.,   (2022)

\bibitem{DAMIC:2016qck}
A.~Aguilar-Arevalo, et~al., Phys. Rev. Lett. \textbf{118}(14), 141803 (2017).
\newblock \doi{10.1103/PhysRevLett.118.141803}

\bibitem{DAMIC:2019dcn}
A.~Aguilar-Arevalo, et~al., Phys. Rev. Lett. \textbf{123}(18), 181802 (2019).
\newblock \doi{10.1103/PhysRevLett.123.181802}

\bibitem{An:2013yfc}
H.~An, M.~Pospelov, J.~Pradler, Phys. Lett. B \textbf{725}, 190 (2013).
\newblock \doi{10.1016/j.physletb.2013.07.008}

\bibitem{Redondo:2013lna}
J.~Redondo, G.~Raffelt, JCAP \textbf{08}, 034 (2013).
\newblock \doi{10.1088/1475-7516/2013/08/034}

\bibitem{Alner:2007ja}
G.J. Alner, et~al., Astropart. Phys. \textbf{28}, 287 (2007).
\newblock \doi{10.1016/j.astropartphys.2007.06.002}

\bibitem{Lebedenko:2008gb}
V.N. Lebedenko, et~al., Phys. Rev. D \textbf{80}, 052010 (2009).
\newblock \doi{10.1103/PhysRevD.80.052010}

\bibitem{Lee:1977}
B.W. {Lee}, S.~{Weinberg}, Phys. Rev. Lett. \textbf{39}(4), 165 (1977).
\newblock \doi{10.1103/PhysRevLett.39.165}

\bibitem{Kolb:1990vq}
E.W. Kolb, M.S. Turner, Front. Phys. \textbf{69}, 1 (1990)

\bibitem{Jungman:1995df}
G.~Jungman, M.~Kamionkowski, K.~Griest, Phys. Rept. \textbf{267}, 195 (1996).
\newblock \doi{10.1016/0370-1573(95)00058-5}

\bibitem{Bergstrom:2000pn}
L.~Bergstrom, Rept. Prog. Phys. \textbf{63}, 793 (2000).
\newblock \doi{10.1088/0034-4885/63/5/2r3}

\bibitem{Bertone:2004pz}
G.~Bertone, D.~Hooper, J.~Silk, Phys. Rept. \textbf{405}, 279 (2005).
\newblock \doi{10.1016/j.physrep.2004.08.031}

\bibitem{Hewett:2012ns}
\emph{{Fundamental Physics at the Intensity Frontier}}.
\newblock \doi{10.2172/1042577}.
\newblock
  \urlprefix\url{http://lss.fnal.gov/archive/preprint/fermilab-conf-12-879-ppd.shtml}

\bibitem{Essig:2013lka}
R.~Essig, et~al., in \emph{{Proceedings, 2013 Community Summer Study on the
  Future of U.S. Particle Physics: Snowmass on the Mississippi (Cs$S^2$013):
  Minneapolis, Mn, Usa, July 29-August 6, 2013}} (2013).
\newblock
  \urlprefix\url{http://www.slac.stanford.edu/econf/C1307292/docs/IntensityFrontier/NewLight-17.pdf}

\bibitem{Jaeckel:2010ni}
J.~Jaeckel, A.~Ringwald, Ann. Rev. Nucl. Part. Sci. \textbf{60}, 405 (2010).
\newblock \doi{10.1146/annurev.nucl.012809.104433}

\bibitem{Boehm:2003hm}
C.~Boehm, P.~Fayet, Nucl. Phys. \textbf{B683}, 219 (2004).
\newblock \doi{10.1016/j.nuclphysb.2004.01.015}

\bibitem{Boehm:2003ha}
C.~Boehm, P.~Fayet, J.~Silk, Phys.Rev. \textbf{D69}, 101302 (2004).
\newblock \doi{10.1103/PhysRevD.69.101302}

\bibitem{McDonald:2001vt}
J.~McDonald, Phys. Rev. Lett. \textbf{88}, 091304 (2002).
\newblock \doi{10.1103/PhysRevLett.88.091304}

\bibitem{Fayet:2004bw}
P.~Fayet, Phys. Rev. \textbf{D70}, 023514 (2004).
\newblock \doi{10.1103/PhysRevD.70.023514}

\bibitem{Sigurdson:2004zp}
K.~Sigurdson, M.~Doran, A.~Kurylov, R.R. Caldwell, M.~Kamionkowski, Phys. Rev.
  \textbf{D70}, 083501 (2004).
\newblock \doi{10.1103/PhysRevD.70.083501, 10.1103/PhysRevD.73.089903}.
\newblock [Erratum: Phys. Rev.D73,089903(2006)]

\bibitem{Strassler:2006im}
M.J. Strassler, K.M. Zurek, Phys.Lett. \textbf{B651}, 374 (2007).
\newblock \doi{10.1016/j.physletb.2007.06.055}

\bibitem{Sikivie:2006ni}
P.~Sikivie, Lect. Notes Phys. \textbf{741}, 19 (2008).
\newblock \doi{10.1007/978-3-540-73518-2_2}.
\newblock [,19(2006)]

\bibitem{Nussinov:1985xr}
S.~Nussinov, Phys.Lett. \textbf{B165}, 55 (1985).
\newblock \doi{10.1016/0370-2693(85)90689-6}

\bibitem{Kaplan:1991ah}
D.B. Kaplan, Phys.Rev.Lett. \textbf{68}, 741 (1992).
\newblock \doi{10.1103/PhysRevLett.68.741}

\bibitem{Zurek:2008qg}
K.M. Zurek, Phys. Rev. \textbf{D79}, 115002 (2009).
\newblock \doi{10.1103/PhysRevD.79.115002}

\bibitem{Hooper:2008im}
D.~Hooper, K.M. Zurek, Phys. Rev. \textbf{D77}, 087302 (2008).
\newblock \doi{10.1103/PhysRevD.77.087302}

\bibitem{Cholis:2008vb}
I.~Cholis, L.~Goodenough, N.~Weiner, Phys. Rev. \textbf{D79}, 123505 (2009).
\newblock \doi{10.1103/PhysRevD.79.123505}

\bibitem{ArkaniHamed:2008qn}
N.~Arkani-Hamed, D.P. Finkbeiner, T.R. Slatyer, N.~Weiner, Phys. Rev.
  \textbf{D79}, 015014 (2009).
\newblock \doi{10.1103/PhysRevD.79.015014}

\bibitem{Pospelov:2008jd}
M.~Pospelov, A.~Ritz, Phys. Lett. \textbf{B671}, 391 (2009).
\newblock \doi{10.1016/j.physletb.2008.12.012}

\bibitem{Feng:2008ya}
J.L. Feng, J.~Kumar, Phys. Rev. Lett. \textbf{101}, 231301 (2008).
\newblock \doi{10.1103/PhysRevLett.101.231301}

\bibitem{Feng:2008dz}
J.L. Feng, J.~Kumar, L.E. Strigari, Phys. Lett. \textbf{B670}, 37 (2008).
\newblock \doi{10.1016/j.physletb.2008.10.038}

\bibitem{Pospelov:2008jk}
M.~Pospelov, A.~Ritz, M.B. Voloshin, Phys. Rev. \textbf{D78}, 115012 (2008).
\newblock \doi{10.1103/PhysRevD.78.115012}

\bibitem{Hall:2009bx}
L.J. Hall, K.~Jedamzik, J.~March-Russell, S.M. West, JHEP \textbf{1003}, 080
  (2010).
\newblock \doi{10.1007/JHEP03(2010)080}

\bibitem{Morrissey:2009ur}
D.E. Morrissey, D.~Poland, K.M. Zurek, JHEP \textbf{0907}, 050 (2009).
\newblock \doi{10.1088/1126-6708/2009/07/050}

\bibitem{Kaplan:2009ag}
D.E. Kaplan, M.A. Luty, K.M. Zurek, Phys.Rev. \textbf{D79}, 115016 (2009).
\newblock \doi{10.1103/PhysRevD.79.115016}

\bibitem{Essig:2010ye}
R.~Essig, J.~Kaplan, P.~Schuster, N.~Toro, Submitted to: Physical Review D
  (2010)

\bibitem{Cohen:2010kn}
T.~Cohen, D.J. Phalen, A.~Pierce, K.M. Zurek, Phys. Rev. \textbf{D82}, 056001
  (2010).
\newblock \doi{10.1103/PhysRevD.82.056001}

\bibitem{Chu:2011be}
X.~Chu, T.~Hambye, M.H.G. Tytgat, JCAP \textbf{1205}, 034 (2012).
\newblock \doi{10.1088/1475-7516/2012/05/034}

\bibitem{Falkowski:2011xh}
A.~Falkowski, J.T. Ruderman, T.~Volansky, JHEP \textbf{05}, 106 (2011).
\newblock \doi{10.1007/JHEP05(2011)106}

\bibitem{Lin:2011gj}
T.~Lin, H.B. Yu, K.M. Zurek, Phys.Rev. \textbf{D85}, 063503 (2012).
\newblock \doi{10.1103/PhysRevD.85.063503}

\bibitem{Feng:2011ik}
J.L. Feng, Y.~Shadmi, Phys. Rev. \textbf{D83}, 095011 (2011).
\newblock \doi{10.1103/PhysRevD.83.095011}

\bibitem{Nelson:2011sf}
A.E. Nelson, J.~Scholtz, Phys. Rev. \textbf{D84}, 103501 (2011).
\newblock \doi{10.1103/PhysRevD.84.103501}

\bibitem{MarchRussell:2012hi}
J.~March-Russell, J.~Unwin, S.M. West, JHEP \textbf{08}, 029 (2012).
\newblock \doi{10.1007/JHEP08(2012)029}

\bibitem{Arias:2012az}
P.~Arias, D.~Cadamuro, M.~Goodsell, J.~Jaeckel, J.~Redondo, A.~Ringwald, JCAP
  \textbf{1206}, 013 (2012).
\newblock \doi{10.1088/1475-7516/2012/06/013}

\bibitem{Kaplinghat:2013yxa}
M.~Kaplinghat, S.~Tulin, H.B. Yu, Phys. Rev. \textbf{D89}(3), 035009 (2014).
\newblock \doi{10.1103/PhysRevD.89.035009}

\bibitem{Hochberg:2014dra}
Y.~Hochberg, E.~Kuflik, T.~Volansky, J.G. Wacker, Phys. Rev. Lett.
  \textbf{113}, 171301 (2014).
\newblock \doi{10.1103/PhysRevLett.113.171301}

\bibitem{Hochberg:2014kqa}
Y.~Hochberg, E.~Kuflik, H.~Murayama, T.~Volansky, J.G. Wacker, Phys. Rev. Lett.
  \textbf{115}(2), 021301 (2015).
\newblock \doi{10.1103/PhysRevLett.115.021301}

\bibitem{Boddy:2014yra}
K.K. Boddy, J.L. Feng, M.~Kaplinghat, T.M.P. Tait, Phys.Rev. \textbf{D89}(11),
  115017 (2014).
\newblock \doi{10.1103/PhysRevD.89.115017}

\bibitem{Boddy:2014qxa}
K.K. Boddy, J.L. Feng, M.~Kaplinghat, Y.~Shadmi, T.M.P. Tait, Phys.Rev.
  \textbf{D90}(9), 095016 (2014).
\newblock \doi{10.1103/PhysRevD.90.095016}

\bibitem{Hochberg:2015vrg}
Y.~Hochberg, E.~Kuflik, H.~Murayama, JHEP \textbf{05}, 090 (2016).
\newblock \doi{10.1007/JHEP05(2016)090}

\bibitem{Izaguirre:2015yja}
E.~Izaguirre, G.~Krnjaic, P.~Schuster, N.~Toro, Phys. Rev. Lett.
  \textbf{115}(25), 251301 (2015).
\newblock \doi{10.1103/PhysRevLett.115.251301}

\bibitem{Kuflik:2015isi}
E.~Kuflik, M.~Perelstein, N.R.L. Lorier, Y.D. Tsai, Phys. Rev. Lett.
  \textbf{116}(22), 221302 (2016).
\newblock \doi{10.1103/PhysRevLett.116.221302}

\bibitem{Graham:2015rva}
P.W. Graham, J.~Mardon, S.~Rajendran, Phys. Rev. \textbf{D93}(10), 103520
  (2016).
\newblock \doi{10.1103/PhysRevD.93.103520}

\bibitem{Marsh:2015xka}
D.J.E. Marsh, Phys. Rept. \textbf{643}, 1 (2016).
\newblock \doi{10.1016/j.physrep.2016.06.005}

\bibitem{D'Agnolo:2015koa}
R.T. D'Agnolo, J.T. Ruderman, Phys. Rev. Lett. \textbf{115}(6), 061301 (2015).
\newblock \doi{10.1103/PhysRevLett.115.061301}

\bibitem{DAgnolo:2015ujb}
R.T. D'Agnolo, J.T. Ruderman, Phys. Rev. Lett. \textbf{115}(6), 061301 (2015).
\newblock \doi{10.1103/PhysRevLett.115.061301}

\bibitem{Pappadopulo:2016pkp}
D.~Pappadopulo, J.T. Ruderman, G.~Trevisan, Phys. Rev. \textbf{D94}(3), 035005
  (2016).
\newblock \doi{10.1103/PhysRevD.94.035005}

\bibitem{Farina:2016llk}
M.~Farina, D.~Pappadopulo, J.T. Ruderman, G.~Trevisan, JHEP \textbf{12}, 039
  (2016).
\newblock \doi{10.1007/JHEP12(2016)039}

\bibitem{Dror:2016rxc}
J.A. Dror, E.~Kuflik, W.H. Ng, Phys. Rev. Lett. \textbf{117}(21), 211801
  (2016).
\newblock \doi{10.1103/PhysRevLett.117.211801}

\bibitem{Feng:2017drg}
J.L. Feng, J.~Smolinsky, Phys. Rev. D \textbf{96}(9), 095022 (2017).
\newblock \doi{10.1103/PhysRevD.96.095022}

\bibitem{Kuflik:2017iqs}
E.~Kuflik, M.~Perelstein, N.R.L. Lorier, Y.D. Tsai, JHEP \textbf{08}, 078
  (2017).
\newblock \doi{10.1007/JHEP08(2017)078}

\bibitem{Choi:2017zww}
S.M. Choi, Y.~Hochberg, E.~Kuflik, H.M. Lee, Y.~Mambrini, H.~Murayama,
  M.~Pierre, JHEP \textbf{10}, 162 (2017).
\newblock \doi{10.1007/JHEP10(2017)162}

\bibitem{DAgnolo:2017dbv}
R.T. D'Agnolo, D.~Pappadopulo, J.T. Ruderman, Phys. Rev. Lett. \textbf{119}(6),
  061102 (2017).
\newblock \doi{10.1103/PhysRevLett.119.061102}

\bibitem{Falkowski:2017uya}
A.~Falkowski, E.~Kuflik, N.~Levi, T.~Volansky, Phys. Rev. D \textbf{99}(1),
  015022 (2019).
\newblock \doi{10.1103/PhysRevD.99.015022}

\bibitem{DAgnolo:2018wcn}
R.T. D'Agnolo, C.~Mondino, J.T. Ruderman, P.J. Wang, JHEP \textbf{08}, 079
  (2018).
\newblock \doi{10.1007/JHEP08(2018)079}

\bibitem{Chu:2018qrm}
X.~Chu, J.~Pradler, L.~Semmelrock, Phys. Rev. \textbf{D99}(1), 015040 (2019).
\newblock \doi{10.1103/PhysRevD.99.015040}

\bibitem{Dvorkin:2019zdi}
C.~Dvorkin, T.~Lin, K.~Schutz, Phys. Rev. \textbf{D99}(11), 115009 (2019).
\newblock \doi{10.1103/PhysRevD.99.115009}

\bibitem{DAgnolo:2019zkf}
R.T. D'Agnolo, D.~Pappadopulo, J.T. Ruderman, P.J. Wang, Phys. Rev. Lett.
  \textbf{124}(15), 151801 (2019).
\newblock \doi{10.1103/PhysRevLett.124.151801}

\bibitem{Hambye:2019dwd}
T.~Hambye, M.H.G. Tytgat, J.~Vandecasteele, L.~Vanderheyden, Phys. Rev.
  \textbf{D100}(9), 095018 (2019).
\newblock \doi{10.1103/PhysRevD.100.095018}

\bibitem{Heeba:2019jho}
S.~Heeba, F.~Kahlhoefer, Phys. Rev. D \textbf{101}(3), 035043 (2019).
\newblock \doi{10.1103/PhysRevD.101.035043}

\bibitem{Evans:2019vxr}
J.A. Evans, C.~Gaidau, J.~Shelton, JHEP \textbf{01}, 032 (2019).
\newblock \doi{10.1007/JHEP01(2020)032}

\bibitem{Koren:2019iuv}
S.~Koren, R.~McGehee, Phys. Rev. D \textbf{101}(5), 055024 (2019).
\newblock \doi{10.1103/PhysRevD.101.055024}

\bibitem{Chu:2019rok}
X.~Chu, J.L. Kuo, J.~Pradler, L.~Semmelrock, Phys. Rev. \textbf{D100}(8),
  083002 (2019).
\newblock \doi{10.1103/PhysRevD.100.083002}

\bibitem{Chang:2019xva}
J.H. Chang, R.~Essig, A.~Reinert, JHEP \textbf{03}, 141 (2019).
\newblock \doi{10.1007/JHEP03(2021)141}

\bibitem{Darme:2017glc}
L.~Darmé, S.~Rao, L.~Roszkowski, JHEP \textbf{03}, 084 (2018).
\newblock \doi{10.1007/JHEP03(2018)084}

\bibitem{March-Russell:2020nun}
J.~March-Russell, H.~Tillim, S.M. West, Phys. Rev. D \textbf{102}(8), 083018
  (2020).
\newblock \doi{10.1103/PhysRevD.102.083018}

\bibitem{Dodelson:1993je}
S.~Dodelson, L.M. Widrow, Phys. Rev. Lett. \textbf{72}, 17 (1994).
\newblock \doi{10.1103/PhysRevLett.72.17}

\bibitem{Shi:1998km}
X.D. Shi, G.M. Fuller, Phys. Rev. Lett. \textbf{82}, 2832 (1999).
\newblock \doi{10.1103/PhysRevLett.82.2832}

\bibitem{Abazajian:2001nj}
K.~Abazajian, G.M. Fuller, M.~Patel, Phys. Rev. D \textbf{64}, 023501 (2001).
\newblock \doi{10.1103/PhysRevD.64.023501}

\bibitem{Abazajian:2012ys}
K.N. Abazajian, et~al.
\newblock {Light Sterile Neutrinos: A White Paper}.
\newblock arXiv:1204.5379 [hep-ph] (2012)

\bibitem{Abazajian:2017tcc}
K.N. Abazajian, Phys. Rept. \textbf{711-712}, 1 (2017).
\newblock \doi{10.1016/j.physrep.2017.10.003}

\bibitem{Boyarsky:2018tvu}
A.~Boyarsky, M.~Drewes, T.~Lasserre, S.~Mertens, O.~Ruchayskiy, Prog. Part.
  Nucl. Phys. \textbf{104}, 1 (2019).
\newblock \doi{10.1016/j.ppnp.2018.07.004}

\bibitem{Hochberg:2018vdo}
Y.~Hochberg, E.~Kuflik, H.~Murayama, Phys. Rev. D \textbf{99}(1), 015005
  (2019).
\newblock \doi{10.1103/PhysRevD.99.015005}

\bibitem{Hochberg:2018rjs}
Y.~Hochberg, E.~Kuflik, R.~Mcgehee, H.~Murayama, K.~Schutz, Phys. Rev. D
  \textbf{98}(11), 115031 (2018).
\newblock \doi{10.1103/PhysRevD.98.115031}

\bibitem{Smirnov:2020zwf}
J.~Smirnov, J.F. Beacom, Phys. Rev. Lett. \textbf{125}(13), 131301 (2020).
\newblock \doi{10.1103/PhysRevLett.125.131301}

\bibitem{Dutta:2019fxn}
B.~Dutta, S.~Ghosh, J.~Kumar, Phys. Rev. D \textbf{100}, 075028 (2019).
\newblock \doi{10.1103/PhysRevD.100.075028}

\bibitem{Dutta:2020enk}
B.~Dutta, S.~Ghosh, J.~Kumar, Phys. Rev. D \textbf{102}(7), 075041 (2020).
\newblock \doi{10.1103/PhysRevD.102.075041}

\bibitem{Dutta:2022qvn}
B.~Dutta, S.~Ghosh, J.~Kumar, in \emph{{2022 Snowmass Summer Study}} (2022)

\bibitem{Fernandez:2021iti}
N.~Fernandez, Y.~Kahn, J.~Shelton, JHEP \textbf{07}, 044 (2022).
\newblock \doi{10.1007/JHEP07(2022)044}

\bibitem{Chang:2018rso}
J.H. Chang, R.~Essig, S.D. McDermott, JHEP \textbf{09}, 051 (2018).
\newblock \doi{10.1007/JHEP09(2018)051}

\bibitem{An:2021qdl}
H.~An, H.~Nie, M.~Pospelov, J.~Pradler, A.~Ritz, Phys. Rev. D \textbf{104}(10),
  103026 (2021).
\newblock \doi{10.1103/PhysRevD.104.103026}

\bibitem{Emken:2021lgc}
T.~Emken, Phys. Rev. D \textbf{105}(6), 063020 (2022).
\newblock \doi{10.1103/PhysRevD.105.063020}

\bibitem{Herrera:2021puj}
G.~Herrera, A.~Ibarra, Phys. Lett. B \textbf{820}, 136551 (2021).
\newblock \doi{10.1016/j.physletb.2021.136551}

\bibitem{Trickle:2022fwt}
T.~Trickle, Phys. Rev. D \textbf{107}(3), 035035 (2023).
\newblock \doi{10.1103/PhysRevD.107.035035}

\bibitem{Dreyer:2023}
C.~Dreyer, R.~Essig, M.~Fernandez-Serra, A.~Singal, C.~Zhen,   (to appear)

\bibitem{Collar:1992qc}
J.I. Collar, F.T. Avignone, Phys. Lett. B \textbf{275}, 181 (1992).
\newblock \doi{10.1016/0370-2693(92)90873-3}

\bibitem{Collar:1993ss}
J.I. Collar, F.T. Avignone, III, Phys. Rev. \textbf{D47}, 5238 (1993).
\newblock \doi{10.1103/PhysRevD.47.5238}

\bibitem{Hasenbalg:1997hs}
F.~Hasenbalg, D.~Abriola, F.T. Avignone, J.I. Collar, D.E. Di~Gregorio, A.O.
  Gattone, H.~Huck, D.~Tomasi, I.~Urteaga, Phys. Rev. \textbf{D55}, 7350
  (1997).
\newblock \doi{10.1103/PhysRevD.55.7350}

\bibitem{Avalos:2021fxm}
N.~\'Avalos, et~al., J. Phys. Conf. Ser. \textbf{2156}, 012074 (2021).
\newblock \doi{10.1088/1742-6596/2156/1/012074}

\bibitem{An:2017ojc}
H.~An, M.~Pospelov, J.~Pradler, A.~Ritz, Phys. Rev. Lett. \textbf{120}(14),
  141801 (2018).
\newblock \doi{10.1103/PhysRevLett.120.141801}

\bibitem{XENON:2018voc}
E.~Aprile, et~al., Phys. Rev. Lett. \textbf{121}(11), 111302 (2018).
\newblock \doi{10.1103/PhysRevLett.121.111302}

\bibitem{DarkSide:2018bpj}
P.~Agnes, et~al., Phys. Rev. Lett. \textbf{121}(8), 081307 (2018).
\newblock \doi{10.1103/PhysRevLett.121.081307}

\bibitem{Marchionni:2011kg}
A.~Marchionni, et~al., J. Phys. Conf. Ser. \textbf{308}(1), 012006 (2011).
\newblock \doi{10.1088/1742-6596/308/1/012006}.
\newblock
  \urlprefix\url{http://stacks.iop.org/1742-6596/308/i=1/a=012006?key=crossref.3b0daaef0024d93c7c6ff7c46b315935}

\bibitem{Agnes:2015ftt}
P.~Agnes, et~al., Phys. Rev. \textbf{D93}(8), 081101 (2016).
\newblock \doi{10.1103/PhysRevD.93.081101}

\bibitem{Ajaj:2019jk}
R.~Ajaj, et~al., Phys. Rev. D \textbf{100}(2), 022004 (2019).
\newblock \doi{10.1103/PhysRevD.100.022004}.
\newblock \urlprefix\url{https://link.aps.org/doi/10.1103/PhysRevD.100.022004}

\bibitem{Hime:2011tt}
A.~Hime, arXiv  (2011).
\newblock \urlprefix\url{http://arxiv.org/abs/1110.1005v1}

\bibitem{PandaX2017_DM}
X.~Cui, et~al., Phys. Rev. Lett. \textbf{119}, 181302 (2017).
\newblock \doi{10.1103/PhysRevLett.119.181302}.
\newblock
  \urlprefix\url{https://link.aps.org/doi/10.1103/PhysRevLett.119.181302}

\bibitem{Akerib:2020it}
D.S. Akerib, et~al., Phys. Rev. D \textbf{101}(5), 351 (2020).
\newblock \doi{10.1103/PhysRevD.101.052002}.
\newblock \urlprefix\url{https://link.aps.org/doi/10.1103/PhysRevD.101.052002}

\bibitem{Aalbers:2016ex}
J.~Aalbers, et~al., JCAP \textbf{2016}(11), 017 (2016).
\newblock \doi{10.1088/1475-7516/2016/11/017}.
\newblock
  \urlprefix\url{http://stacks.iop.org/1475-7516/2016/i=11/a=017?key=crossref.f31089b6aea3d9592e77c4ae35cda8f1}

\bibitem{Aalbers:2022dzr}
J.~Aalbers, et~al., J. Phys. G \textbf{50}(1), 013001 (2023).
\newblock \doi{10.1088/1361-6471/ac841a}

\bibitem{LXeMOU}
{Leading Xenon Researchers unite to build next-generation Dark Matter Detector}
  (2021).
\newblock
  \urlprefix\url{https://www.sanfordlab.org/press-release/leading-xenon-researchers-unite-build-next-generation-dark-matter-detector}

\bibitem{https://doi.org/10.48550/arxiv.2203.07901}
J.~Liao, Y.~Gao, Z.~Liang, Z.~Ouyang, Z.~Peng, L.~Zhang, L.~Zhang, J.~Zheng,
  J.~Zhou.
\newblock Introduction to a low-mass dark matter project, aletheia: A liquid
  helium time projection chamber in dark matter (2022).
\newblock \doi{10.48550/ARXIV.2203.07901}.
\newblock \urlprefix\url{https://arxiv.org/abs/2203.07901}

\bibitem{XENON:2021myl}
E.~Aprile, et~al.,   (2021)

\bibitem{DS50_2018_S2Only}
P.~Agnes, et~al., Phys. Rev. Lett. \textbf{121}, 081307 (2018).
\newblock \doi{10.1103/PhysRevLett.121.081307}.
\newblock
  \urlprefix\url{https://link.aps.org/doi/10.1103/PhysRevLett.121.081307}

\bibitem{LUX:2020vbj}
{D.\ S.\ Akerib {\it et al.} [LUX Collaboration]}, Phys. Rev. D
  \textbf{102}(9), 092004 (2020).
\newblock \doi{10.1103/PhysRevD.102.092004}

\bibitem{Kopec:2021ccm}
A.~Kopec, A.L. Baxter, M.~Clark, R.F. Lang, S.~Li, J.~Qin, R.~Singh, JINST
  \textbf{16}(07), P07014 (2021).
\newblock \doi{10.1088/1748-0221/16/07/P07014}

\bibitem{EDWARDS200854}
B.~Edwards, H.~Arajo, V.~Chepel, D.~Cline, T.~Durkin, J.~Gao, C.~Ghag,
  E.~Korolkova, V.~Lebedenko, A.~Lindote, M.~Lopes, R.~Loscher, A.~Murphy,
  F.~Neves, W.~Ooi, J.P. {da Cunha}, R.~Preece, G.~Salinas, C.~Silva,
  V.~Solovov, N.~Smith, P.~Smith, T.~Sumner, C.~Thorne, R.~Walker, H.~Wang,
  J.~White, F.~Wolfs, Astroparticle Physics \textbf{30}(2), 54 (2008).
\newblock \doi{https://doi.org/10.1016/j.astropartphys.2008.06.006}.
\newblock
  \urlprefix\url{https://www.sciencedirect.com/science/article/pii/S0927650508000789}

\bibitem{Santos:2011ju}
E.~Santos, et~al., JHEP \textbf{12}, 115 (2011).
\newblock \doi{10.1007/JHEP12(2011)115}

\bibitem{Aprile_2014}
Journal of Physics G: Nuclear and Particle Physics \textbf{41}(3), 035201
  (2014).
\newblock \doi{10.1088/0954-3899/41/3/035201}.
\newblock \urlprefix\url{https://doi.org/10.1088%2F0954-3899%2F41%2F3%2F035201}

\bibitem{Sorensen:2017ymt}
P.~Sorensen.
\newblock {Electron train backgrounds in liquid xenon dark matter search
  detectors are indeed due to thermalization and trapping}.
\newblock arXiv:1702.04805 [physics.ins-det] (2017)

\bibitem{Sorensen_2018}
P.~Sorensen, K.~Kamdin, Journal of Instrumentation \textbf{13}(02), P02032
  (2018).
\newblock \doi{10.1088/1748-0221/13/02/p02032}.
\newblock
  \urlprefix\url{https://doi.org/10.1088%2F1748-0221%2F13%2F02%2Fp02032}

\bibitem{Tomas:2018pny}
A.~Tom\'as, H.~Ara\'ujo, A.~Bailey, A.~Bayer, E.~Chen, B.~L\'opez~Paredes,
  T.~Sumner, Astropart. Phys. \textbf{103}, 49 (2018).
\newblock \doi{10.1016/j.astropartphys.2018.07.001}

\bibitem{Akimov:2019ogx}
D.Y. Akimov, et~al., JINST \textbf{15}(02), P02020 (2020).
\newblock \doi{10.1088/1748-0221/15/02/P02020}

\bibitem{LowMassTPC_LOI}
S.~Westerdale, et~al.
\newblock R\&d for low-threshold noble liquid detectors.
\newblock Snowmass2021 LOI (2020)

\bibitem{LargeScaleLXe_LOI}
R.~Gaitskell, et~al.
\newblock The exploitation of xe large scale detector technology for a range of
  future rare event physics searches.
\newblock Snowmass2021 LOI (2020)

\bibitem{LZG3_LOI}
S.~Shaw, et~al.
\newblock Particle dark matter searches with a g3 liquid-xenon detector.
\newblock Snowmass2021 LOI (2020)

\bibitem{DARWIN_LOI}
L.~Baudis, et~al.
\newblock Dark matter physics with the darwin observatory.
\newblock Snowmass2021 LOI (2020)

\bibitem{PhysRevD.104.092009}
D.S. Akerib, et~al., Phys. Rev. D \textbf{104}, 092009 (2021).
\newblock \doi{10.1103/PhysRevD.104.092009}.
\newblock \urlprefix\url{https://link.aps.org/doi/10.1103/PhysRevD.104.092009}

\bibitem{HydroX_LOI}
H.~Lippincott, et~al.
\newblock Hydrox- using hydrogen doped in liquid xenon to search for dark
  matter.
\newblock Snowmass2021 LOI (2020)

\bibitem{NEWSG_LOI}
G.~Giroux, et~al.
\newblock Search for low mass wimps with spherical proportional counters.
\newblock Snowmass2021 LOI (2020)

\bibitem{Neganov:1985}
B.~Neganov, V.~Trofimov, Otkrytiya, Izobret \textbf{146}, 215 (1985)

\bibitem{Luke:1990ir}
P.~Luke, J.~Beeman, F.~Goulding, S.~Labov, E.~Silver, Nucl.Instrum.Meth.
  \textbf{A289}, 406 (1990).
\newblock \doi{10.1016/0168-9002(90)91510-I}

\bibitem{Agnese:2013jaa}
R.~Agnese, et~al., Phys.Rev.Lett. \textbf{112}(4), 041302 (2014).
\newblock \doi{10.1103/PhysRevLett.112.041302}

\bibitem{supercdms2017}
R.~Agnese, et~al., Phys. Rev. D \textbf{95}, 082002 (2017).
\newblock \doi{10.1103/PhysRevD.95.082002}.
\newblock \urlprefix\url{https://link.aps.org/doi/10.1103/PhysRevD.95.082002}

\bibitem{EdelweissSensitivity}
Q.~Arnaud, et~al., Physical Review D \textbf{97}(2) (2018).
\newblock \doi{10.1103/physrevd.97.022003}.
\newblock \urlprefix\url{http://dx.doi.org/10.1103/PhysRevD.97.022003}

\bibitem{DIncecco:2018fx}
M.~D'Incecco, C.~Galbiati, G.K. Giovanetti, G.~Korga, X.~Li, A.~Mandarano,
  A.~Razeto, D.~Sablone, C.~Savarese, IEEE Trans. Nucl. Sci. \textbf{65}(4),
  1005 (2018).
\newblock \doi{10.1109/TNS.2018.2799325}.
\newblock \urlprefix\url{https://ieeexplore.ieee.org/document/8272038/}

\bibitem{SIPM_LOI}
C.~Savarese, et~al.
\newblock Silicon photomultipliers as a target for low-mass dark matter
  searches.
\newblock Snowmass2021 LOI (2020)

\bibitem{supercdmsWP22}
S.~Collaboration, {A Strategy for Low-Mass Dark Matter Searches with Cryogenic
  Detectors in the SuperCDMS SNOLAB Facility} (2022)

\bibitem{Fink:2020noh}
C.W. Fink, et~al., AIP Adv. \textbf{10}(8), 085221 (2020).
\newblock \doi{10.1063/5.0011130}

\bibitem{Ren:2020gaq}
R.~Ren, et~al., Phys. Rev. D \textbf{104}(3), 032010 (2021).
\newblock \doi{10.1103/PhysRevD.104.032010}

\bibitem{EXCESSworkshop}
{See talks at EXCESS workshop, June 2021, https://indico.cern.ch/event/1013203/
  .}

\bibitem{Proceedings:2022hmu}
A.~Fuss, M.~Kaznacheeva, F.~Reindl, F.~Wagner (eds.).
\newblock \emph{{EXCESS workshop: Descriptions of rising low-energy spectra}}
  (2022)

\bibitem{PICO_LOI}
A.~Robinson, et~al.
\newblock Multi-ton scale bubble chambers.
\newblock Snowmass2021 LOI (2020)

\bibitem{noblebubble_LOI}
E.~Dahl, et~al.
\newblock Reaching the solar cevns floor with noble liquid bubble chambers.
\newblock Snowmass2021 LOI (2020)

\bibitem{snowball_LOI}
M.~Szydagis, et~al.
\newblock Metastable water: Breakthrough technology for dark matter and
  neutrinos.
\newblock Snowmass2021 LOI (2020)

\bibitem{CPD:2020xvi}
C.W. Fink, et~al., Appl. Phys. Lett. \textbf{118}(2), 022601 (2021).
\newblock \doi{10.1063/5.0032372}

\bibitem{Angloher:2017sxg}
G.~Angloher, et~al., Eur. Phys. J. C \textbf{77}(9), 637 (2017).
\newblock \doi{10.1140/epjc/s10052-017-5223-9}

\bibitem{SPICE}
{D.\ N.\ McKinsey {\it et al.} [TESSERACT Collaboration]}.
\newblock {The TESSERACT Dark Matter Project, SNOWMASS LOI}.
\newblock Availabe
  \href{www.snowmass21.org/docs/files/summaries/CF/SNOWMASS21-CF1_CF2-IF1_IF8-120.pdf}{[Online]}.
\newblock
  \urlprefix\url{www.snowmass21.org/docs/files/summaries/CF/SNOWMASS21-CF1_CF2-IF1_IF8-120.pdf}

\bibitem{Canonica:2020omq}
L.~Canonica, et~al., J. Low Temp. Phys. \textbf{199}(3-4), 606 (2020).
\newblock \doi{10.1007/s10909-020-02350-4}

\bibitem{Rajendran:2017ynw}
S.~Rajendran, N.~Zobrist, A.O. Sushkov, R.~Walsworth, M.~Lukin, Phys. Rev. D
  \textbf{96}(3), 035009 (2017).
\newblock \doi{10.1103/PhysRevD.96.035009}

\bibitem{Marshall:2020azl}
M.C. Marshall, M.J. Turner, M.J.H. Ku, D.F. Phillips, R.L. Walsworth, Quantum
  Sci. Technol. \textbf{6}(2), 024011 (2021).
\newblock \doi{10.1088/2058-9565/abe5ed}

\bibitem{Ebadi:2022axg}
R.~Ebadi, et~al., in \emph{{2022 Snowmass Summer Study}} (2022)

\bibitem{Akerib:2022ort}
D.S. Akerib, et~al., in \emph{{2022 Snowmass Summer Study}} (2022)

\bibitem{Marshall:2021kjk}
M.C. Marshall, D.F. Phillips, M.J. Turner, M.J.H. Ku, T.~Zhou, N.~Delegan, F.J.
  Heremans, M.V. Holt, R.L. Walsworth, Phys. Rev. Applied \textbf{16}(5),
  054032 (2021).
\newblock \doi{10.1103/PhysRevApplied.16.054032}

\bibitem{Marshall:2021xiu}
M.C. Marshall, R.~Ebadi, C.~Hart, M.J. Turner, M.J.H. Ku, D.F. Phillips, R.L.
  Walsworth, Phys. Rev. Applied \textbf{17}(2), 024041 (2022).
\newblock \doi{10.1103/PhysRevApplied.17.024041}

\bibitem{KIDSLOI}
S.~Golwala, et~al.
\newblock Phonon-mediated kid-based detectors for low-mass dark matter
  detection and coherent elastic neutrino-nucleus scattering.
\newblock Snomwass2021 LOI (2020)

\bibitem{millionkidLOI}
J.~Gao, et~al.
\newblock Million-pixel kinetic inductance detector arrays for sub-gev light
  dark matter search.
\newblock Snowmass2021 LOI (2020)

\bibitem{kim2020self}
G.B. Kim, Journal of Low Temperature Physics \textbf{199}(3), 1004 (2020)

\bibitem{kim2017novel}
G.~Kim, J.~Choi, H.~Jo, C.~Kang, H.~Kim, I.~Kim, S.~Kim, Y.~Kim, C.~Lee,
  H.~Lee, et~al., Astroparticle Physics \textbf{91}, 105 (2017)

\bibitem{MMC-cevns:2021loi}
{G. Kim},
  \urlprefix\url{https://www.snowmass21.org/docs/files/summaries/NF/SNOWMASS21-
  NF10_NF7_Geon-Bo_Kim-157.pdf}.
\newblock {Snowmass 2021 Letter of Interest}

\bibitem{Castello-Mor:2020jhd}
N.~Castello-Mor,   (2020).
\newblock \doi{10.1016/j.nima.2019.162933}

\bibitem{SENSEI:2021hcn}
L.~Barak, et~al., Phys. Rev. Applied \textbf{17}(1), 014022 (2022).
\newblock \doi{10.1103/PhysRevApplied.17.014022}

\bibitem{diamondLOI}
N.~Kurinsky, et~al.
\newblock Cryogenic carbon detectors for dark matter searches.
\newblock Snowmass2021 LOI (2020)

\bibitem{SPLENDOR}
{N.~A.~Kurinsky {\it et al.} [SPLENDOR Collaboration]}.
\newblock {Low-gap charge detection for fundamental physics searches, SNOWMASS
  LOI}.
\newblock Availabe
  \href{https://www.snowmass21.org/docs/files/summaries/CF/SNOWMASS21-CF1_CF2-IF1_IF2_Kurinsky-029.pdf}{[Online]}.
\newblock
  \urlprefix\url{https://www.snowmass21.org/docs/files/summaries/CF/SNOWMASS21-CF1_CF2-IF1_IF2_Kurinsky-029.pdf}

\bibitem{hochberg2019snspd}
Y.~Hochberg, I.~Charaev, S.W. Nam, V.~Verma, M.~Colangelo, K.K. Berggren, Phys.
  Rev. Lett. \textbf{123}, 151802 (2019).
\newblock \doi{10.1103/PhysRevLett.123.151802}.
\newblock
  \urlprefix\url{https://link.aps.org/doi/10.1103/PhysRevLett.123.151802}

\bibitem{Hochberg:2021yud}
Y.~Hochberg, B.V. Lehmann, I.~Charaev, J.~Chiles, M.~Colangelo, S.W. Nam, K.K.
  Berggren, Phys. Rev. D \textbf{106}(11), 112005 (2022).
\newblock \doi{10.1103/PhysRevD.106.112005}

\bibitem{birks2013theory}
J.B. Birks, \emph{The theory and practice of scintillation counting:
  International series of monographs in electronics and instrumentation},
  vol.~27 (Elsevier, 2013)

\bibitem{hertel2019helium}
S.A. Hertel, A.~Biekert, J.~Lin, V.~Velan, D.N. McKinsey, Phys. Rev. D
  \textbf{100}, 092007 (2019).
\newblock \doi{10.1103/PhysRevD.100.092007}.
\newblock \urlprefix\url{https://link.aps.org/doi/10.1103/PhysRevD.100.092007}

\bibitem{He4LOI}
S.~Hertel, et~al.
\newblock Calorimetric readout of a superfluid 4he target mass.
\newblock Snowmass2021 LOI (2020)

\bibitem{Derenzo:2020hfo}
S.~Derenzo, E.~Bourret, C.~Frank-Rotsch, S.~Hanrahan, M.~Garcia-Sciveresa,
  Nucl. Instrum. Meth. A \textbf{989}, 164957 (2021).
\newblock \doi{10.1016/j.nima.2020.164957}

\bibitem{Derenzo:2018plr}
S.~Derenzo, E.~Bourret, S.~Hanrahan, G.~Bizarri, J. Appl. Phys.
  \textbf{123}(11), 114501 (2018).
\newblock \doi{10.1063/1.5018343}

\bibitem{Vasiukov:2019hwn}
S.~Vasiukov, F.~Chiossi, C.~Braggio, G.~Carugno, F.~Moretti, E.~Bourret,
  S.~Derenzo,   (2019)

\bibitem{GaAsLOI}
R.~Essig, et~al.
\newblock A scintillating n-type gaas detector for sub-gev dark matter direct
  detection.
\newblock Snowmass2021 LOI (2020)

\bibitem{Derenzo:2022cnn}
S.~Derenzo,   (2022)

\bibitem{Blanco:2022cel}
C.~Blanco, R.~Essig, M.~Fernandez-Serra, H.~Ramani, O.~Slone,   (2022)

\bibitem{maris2017helium}
H.J. Maris, G.M. Seidel, D.~Stein, Phys. Rev. Lett. \textbf{119}, 181303
  (2017).
\newblock \doi{10.1103/PhysRevLett.119.181303}.
\newblock
  \urlprefix\url{https://link.aps.org/doi/10.1103/PhysRevLett.119.181303}

\bibitem{Wang:2022cyk}
G.~Wang, C.L. Chang, M.~Lisovenko, V.~Novosad, V.G. Yefremenko, J.~Zhang, J.
  Low Temp. Phys. \textbf{209}(3-4), 379 (2022).
\newblock \doi{10.1007/s10909-022-02784-y}

\bibitem{rosa2020colossal}
P.~Rosa, Y.~Xu, M.~Rahn, J.~Souza, S.~Kushwaha, L.~Veiga, A.~Bombardi,
  S.~Thomas, M.~Janoschek, E.~Bauer, et~al., npj Quantum Materials
  \textbf{5}(1), 1 (2020)

\bibitem{Gelmini:2020xir}
G.B. Gelmini, V.~Takhistov, E.~Vitagliano, Phys. Lett. B \textbf{809}, 135779
  (2020).
\newblock \doi{10.1016/j.physletb.2020.135779}

\bibitem{phononLOI}
M.~Pyle, et~al.
\newblock Improving the sensitivity of athermal phonon sensors for light mass
  dark matter.
\newblock Snowmass 2021 LOI (2020)

\bibitem{leman2012invited}
S.W. Leman, Review of Scientific Instruments \textbf{83}(9), 091101 (2012)

\bibitem{echternach2018single}
P.~Echternach, B.~Pepper, T.~Reck, C.~Bradford, Nature Astronomy \textbf{2}(1),
  90 (2018)

\bibitem{Wilen:2020lgg}
C.D. Wilen, et~al., Nature \textbf{594}(7863), 369 (2021).
\newblock \doi{10.1038/s41586-021-03557-5}

\bibitem{Schutz:2016tid}
K.~Schutz, K.M. Zurek, Phys. Rev. Lett. \textbf{117}(12), 121302 (2016).
\newblock \doi{10.1103/PhysRevLett.117.121302}

\bibitem{Knapen:2016cue}
S.~Knapen, T.~Lin, K.M. Zurek, Phys. Rev. \textbf{D95}(5), 056019 (2017).
\newblock \doi{10.1103/PhysRevD.95.056019}

\bibitem{LHe_Resonators}
Y.~Lee, K.~Matchev, T.~Saab, W.~Xue, {Snowmass2021 Cosmic Frontier White Paper:
  Dark matter direct detection to the neutrino floor} (2022)

\bibitem{Matchev:2021fuw}
K.T. Matchev, J.~Smolinsky, W.~Xue, Y.~You,   (2021)

\bibitem{He3LOI}
S.A. Lyon, et~al.
\newblock Quantum sensing of 3he for low-mass dark matter detection.
\newblock Snowmass2021 LOI (2020)

\bibitem{Lyon:2022sza}
S.A. Lyon, K.~Castoria, E.~Kleinbaum, Z.~Qin, A.~Persaud, T.~Schenkel,
  K.~Zurek,   (2022)

\bibitem{Casey}
A.~Casey.
\newblock {Quantum Enhanced Superfluid Technologies for Dark Matter and
  Cosmology, QUEST–DMC} (2020).
\newblock
  \urlprefix\url{https://conference.ippp.dur.ac.uk/event/934/contributions/4941/attachments/4038/4675/QUEST-DMC_AC_HEP_upload.pdf}

\bibitem{Du:2022dxf}
P.~Du, D.~Ega\~na Ugrinovic, R.~Essig, M.~Sholapurkar,   (2212.04504)

\bibitem{Verma:2020gso}
V.B. Verma, et~al.,   (2012.09979)

\bibitem{Kim:2020bwm}
D.~Kim, J.C. Park, K.C. Fong, G.H. Lee,   (2002.07821)

\bibitem{Essig:2019kfe}
R.~Essig, J.~Pérez-Ríos, H.~Ramani, O.~Slone, Phys. Rev. Research (1) (2019).
\newblock \doi{10.1103/PhysRevResearch.1.033105}.
\newblock [Phys. Rev. Research.1,033105(2019)]

\bibitem{MolecularLOI}
K.~Berggren, R.~Essig, et~al.
\newblock Direct detection of spin-independent and spin-dependent nuclear
  scattering of sub-gev dark matter using molecular excitations and
  superconducting nanowire single-photon detectors.
\newblock Snowmass2021 LOI (2020)

\bibitem{Phipps:2016gdx}
A.~Phipps, A.~Juillard, B.~Sadoulet, B.~Serfass, Y.~Jin, Nucl. Instrum. Meth. A
  \textbf{940}, 181 (2019).
\newblock \doi{10.1016/j.nima.2019.06.022}

\bibitem{Juillard:2019njs}
A.~Juillard, et~al., J. Low Temp. Phys. \textbf{199}(3-4), 798 (2019).
\newblock \doi{10.1007/s10909-019-02269-5}

\bibitem{Tosi19}
L.~Tosi, D.~Vion, H.~le~Sueur, Phys. Rev. Applied \textbf{11}, 054072 (2019).
\newblock \doi{10.1103/PhysRevApplied.11.054072}.
\newblock
  \urlprefix\url{https://link.aps.org/doi/10.1103/PhysRevApplied.11.054072}

\bibitem{Brock21}
B.~Brock, J.~Li, S.~Kanhirathingal, B.~Thyagarajan, W.F. Braasch, M.~Blencowe,
  A.~Rimberg, Physical Review Applied \textbf{15}(4) (2021).
\newblock \doi{10.1103/physrevapplied.15.044009}.
\newblock \urlprefix\url{http://dx.doi.org/10.1103/PhysRevApplied.15.044009}

\bibitem{HOCHBERG2017239}
Y.~Hochberg, Y.~Kahn, M.~Lisanti, C.G. Tully, K.M. Zurek, Physics Letters B
  \textbf{772}, 239 (2017).
\newblock \doi{https://doi.org/10.1016/j.physletb.2017.06.051}.
\newblock
  \urlprefix\url{https://www.sciencedirect.com/science/article/pii/S0370269317305270}

\bibitem{hochberg2018detection}
Y.~Hochberg, Y.~Kahn, M.~Lisanti, K.M. Zurek, A.G. Grushin, R.~Ilan, S.M.
  Griffin, Z.F. Liu, S.F. Weber, J.B. Neaton, Phys. Rev. D \textbf{97}, 015004
  (2018).
\newblock \doi{10.1103/PhysRevD.97.015004}.
\newblock \urlprefix\url{https://link.aps.org/doi/10.1103/PhysRevD.97.015004}

\bibitem{bunting2017magnetic}
P.C. Bunting, G.~Gratta, T.~Melia, S.~Rajendran, Phys. Rev. D \textbf{95},
  095001 (2017).
\newblock \doi{10.1103/PhysRevD.95.095001}.
\newblock \urlprefix\url{https://link.aps.org/doi/10.1103/PhysRevD.95.095001}

\bibitem{Essig:2018tss}
R.~Essig, M.~Sholapurkar, T.T. Yu, Phys. Rev. \textbf{D97}(9), 095029 (2018).
\newblock \doi{10.1103/PhysRevD.97.095029}

\bibitem{Schwemberger:2022fjl}
T.~Schwemberger, T.T. Yu, Phys. Rev. D \textbf{106}(1), 015002 (2022).
\newblock \doi{10.1103/PhysRevD.106.015002}

\bibitem{Donchenko:2021fnf}
G.~Donchenko, K.~Kouzakov, A.~Studenikin, J. Phys. Conf. Ser. \textbf{2156}(1),
  012231 (2021).
\newblock \doi{10.1088/1742-6596/2156/1/012231}

\bibitem{stein2018analysis}
M.~Stein, An analysis of frenkel defects and backgrounds modeling for supercdms
  dark matter searches.
\newblock Ph.D. thesis, Southern Methodist University (2018)

\bibitem{ramanathan:2017compton}
K.~Ramanathan, A.~Kavner, A.E. Chavarria, P.~Privitera, D.~Amidei, T.L. Chou,
  A.~Matalon, R.~Thomas, J.~Estrada, J.~Tiffenberg, J.~Molina, Phys. Rev. D
  \textbf{96}, 042002 (2017).
\newblock \doi{10.1103/PhysRevD.96.042002}.
\newblock \urlprefix\url{https://link.aps.org/doi/10.1103/PhysRevD.96.042002}

\bibitem{Botti:2022lkm}
A.M. Botti, et~al., Phys. Rev. D \textbf{106}(7), 072005 (2022)

\bibitem{Araujo:2022wjh}
H.M. Ara\'ujo, et~al.,   (2022)

\bibitem{Adams:2022zvg}
D.~Adams, D.~Baxter, H.~Day, R.~Essig, Y.~Kahn, Phys. Rev. D \textbf{107}(4),
  L041303 (2023)

\end{thebibliography}

\end{document}